%% file: survey2018_pdflatex.tex
\documentclass[twocolumn,trackchanges]{aastex61}
\usepackage{rotating,epsfig}
\pdfoutput=1
\def\msun{\rm M_{\sun}}

\submitjournal{Astronomical Journal}

\begin{document}


\shortauthors{Brice\~no et al.}
\shorttitle{The \emph{CVSO} II: demographics of Ori OB1}

\title{The CIDA Variability Survey of Orion OB1. II:
demographics of the young, low-mass stellar populations\footnotetext{Based on observations obtained at the Llano del Hato National Astronomical Observatory of Venezuela, operated by CIDA for the Ministerio del Poder Popular para la Educaci\'on Universitaria, Ciencia, Tecnolog{\'\i}a, the Fred Lawrence Whipple Observatory of the Smithsonian Institution, the WIYN 3.5m telescope, the 6.5m Magellan Telescopes at Las Campanas Observatory, and the Southern Astrophysical Research (SOAR) telescope.}}

\author[0000-0001-7124-4094]{C\'esar Brice\~{n}o}
\affiliation{Cerro Tololo Inter-American Observatory, National Optical Astronomical Observatory, Casilla 603, La Serena, Chile}

\author{Nuria Calvet}
\affiliation{Department of Astronomy, University of Michigan, 
	311 West Hall, 1085 S University Av, Ann Arbor, MI 48109, USA}

\author[0000-0001-9797-5661]{Jes\'us Hern\'andez}
\affiliation{Instituto  de  Astronom{\'\i}a,  UNAM,  Unidad
	Acad\'emica en Ensenada, Ensenada, 22860, M\'exico.}

\author[0000-0003-4341-6172]{A. Katherina Vivas}
\affiliation{Cerro Tololo Inter-American Observatory, National Optical Astronomical Observatory, Casilla 603, La Serena, Chile}

\author{Cecilia Mateu} 
\affiliation{Centro  de  Investigaciones  de  Astronom{\'\i}a  (CIDA),
	Apdo. Postal 264, M\'erida 5101-A, Venezuela}

\author{Juan Jos\'e Downes}
\affiliation{Centro  de  Investigaciones  de  Astronom{\'\i}a  (CIDA),
	Apdo. Postal 264, M\'erida 5101-A, Venezuela}

\author{Jaqueline Loerincs}
\affiliation{School of Mines, University of Colorado, Boulder, CO, USA}

\author{Alice P\'erez-Blanco}
\affil{University of Leeds, School of Physics and Astronomy, LS29JT, Leeds, UK.}

\author{Perry Berlind} 
\affiliation{Harvard-Smithsonian Center for Astrophysics, 60 Cambridge, MA 02138, USA}

\author{Catherine Espaillat},
\affiliation{Boston University, Astronomy Department, 725 Commonwealth Avenue, Boston, MA 02215, USA}

\author{Lori Allen}
\affiliation{National Optical Astronomical Observatory, 950 North Cherry Avenue, Tucson, AZ 85719, USA}

\author{Lee Hartmann} 
\affiliation{Department of Astronomy, University of Michigan, 
	311 West Hall, 1085 S University Av, Ann Arbor, MI 48109, USA}
    
\author{Mario Mateo} 
\affiliation{Department of Astronomy, University of Michigan, 
	311 West Hall, 1085 S University Av, Ann Arbor, MI 48109, USA}

\author{John I. Bailey III}
\affiliation{Leiden University, Niels Bohrweg 2, 2333 CA Leiden, Netherlands}

\email{cbriceno@ctio.noao.edu}
\correspondingauthor{C\'esar Brice\~no}

\begin{abstract}
 
We present results of our large scale, optical, multi-epoch photometric survey of $\sim 180$ square degrees
across the Orion OB1 association, complemented with extensive follow up spectroscopy. Our focus is mapping
and characterizing in an uniform way the off-cloud, low-mass, pre-main sequence populations.
We report 2064, mostly K and M-type, confirmed T Tauri members.
Most (59\%) are located in the OB1a subassociation, 27\% are projected onto the OB1b subassociation,
and the remaining 14\% are located within the confines of the A and B molecular clouds. There is significant
structure in the spatial distribution of the young stars. We characterize two new clusterings of T Tauri stars,
the HD 35762 and HR 1833 groups, both located in the OB1a subassociation, not far form the 25 Ori cluster.
We also confirm two stellar overdensities in OB1b, I and II, containing a total of 231 T Tauri stars.
A $\sim 2$ deg wide halo of young stars surrounds the Orion Nebula Cluster; the north and south parts
corresponding to the low-mass populations of NGC 1977 and NGC 1980, respectively.
There is indication of two populations of young stars in the OB1b region, located at two different distances,
which may be due to the OB1a subassociation overlapping on front of OB1b.
The various groups and regions can be ordered in an age sequence that agrees with the long standing picture
of star formation starting in Orion OB1a some 10-15 Myr ago.

We use the strength of the H$\alpha$ line in emission, combined with characteristics of IR excesses and optical
variability, to define a new type of T Tauri star, the C/W class, stars we propose may be nearing the end
of their accretion phase, in an evolutionary state between that of Classical T Tauri and Weak-lined T Tauri stars.
The evolution of the ensemble-wide equivalent width of the Li I$\lambda 6707$  line shows the depletion of Li
with a timescale of 8.5 Myr. The decline of the accretion fraction, from $\sim 2 - 10$ Myr, implies an
accretion e-folding timescale of 2.1 Myr, consistent with previous studies.  Finally, we use the median amplitude
of the $V$-band variability in each type of star, to show the decline of stellar activity,
from the accreting Classical T Tauri stars to the population of least active field dwarfs.

\end{abstract}

\keywords{stars: formation, pre-main sequence --- 
open clusters and associations: individual (Orion OB1 association) --- 
surveys}

\section{Introduction}
\label{sec:int}

Large star-forming complexes containing early spectral type stars, 
also known as OB associations, are the prime sites for star formation in our Galaxy
\citep{briceno07a}. 
These regions can extend on scales of tens up to hundreds of parsecs, and span a rich
diversity of environments and evolutionary stages in the early life of stars,
ranging from young stars still embedded in their natal 
molecular clouds (ages $\lesssim 1$ Myr), 
up to somewhat more evolved populations (ages $\sim 10$ Myr) in areas largely devoid
of gas, where the parent gas clouds have already dissipated.
Though the most conspicuous members are the few massive O and B stars,
the bulk, by number and mass, of the stellar population is 
composed of solar-like and lower mass
pre-main sequence (PMS) stars, also known as T Tauri stars \cite[TTS -][]{joy1945,herbig1962},
whose defining characteristics in the optical wavelength regime are, among others, 
photometric variability, 
late spectral type (K-M), and emission lines.
Because for any reasonable Initial Mass Function (IMF), 
the TTS are far more numerous than the O and B stars, the low-mass young stars 
are the best tracers of
the spatial extent, structure and star-forming history of any association.
Moreover, they are the only way to study how the early Sun and its planetary system
may have evolved.
Therefore, building a {\sl complete} census of the TTS population in an OB association  
is an essential first step 
to investigate issues like
the degree of clustering, cluster sizes and dispersal timescales,
the IMF,
disk evolution, and the role of the environment on protoplanetary disks.

However, our knowledge of the full stellar content of most nearby OB associations is still far from complete. 
This is largely because the majority of existing studies have focused on the most easily 
recognizable components, that is, the youngest PMS stars, in particular those densely packed 
in clusters projected on their natal molecular clouds (e.g. the Orion Nebula 
Cluster - ONC, $\sigma$ Ori, the NGC 2071, NGC 2068 and NGC 2024 clusters in the 
Orion B cloud, Tr 37 in Cepheus OB2, IC 348 in Perseus, NGC 2264 in Monoceros, among others). Recent work at optical \citep{hsu2013,bouy2014,kounkel2017b,kubiak2017} and near-IR wavelengths \citep{megeath2016} have mapped the youngest populations over larger areas in Orion, but mostly still limited to the molecular clouds. In regions like Scorpius-Centaurus and Orion, extensive spectroscopy has been done to characterize the young population \citep{rizzuto2015,pecaut_mamajek2016,da-rio2016,da-rio2017}.
However, the fact remains that in most regions the older off-cloud populations have been poorly studied.
Only recently, with the advent of large scale multi-wavelength surveys we have started to build, for the first time, complete pictures of the young stellar populations in these nearby OB associations.  The advent of GAIA will bring about the exploration of the full extent of OB associations, well beyond the confines of the molecular clouds \citep{zari2017,galli2018,wright2018}, though extensive ground based spectroscopy will still be essential to fully confirm and characterize the young stellar populations, specially for the solar and lower mass stars.

The Orion OB1 association
\citep[for reviews see the various chapters on Orion in][]{reipurth08},
located well below the Galactic plane
($-11^\circ \ga b \ga -20^\circ$), at a distance of roughly 400 pc
\citep{genzel89,briceno08,kounkel2017a},
and spanning over $\rm 200 \> deg^2$ on the sky,
is one of the largest and nearest OB associations.
\cite{blaauw64} 
counted 56 massive stars with spectral types
earlier than B2, more than Scorpius-Centaurus and Lacerta OB1,
and only slightly less than Cepheus. He 
estimated a total mass
for Orion OB1 of $\sim 8\times 10^3 \msun$, though this number is probably
best interpreted as a lower limit, because it
does not include the lower mass stars.
Orion OB1
exhibits all stages of the star formation process, from very young, embedded
clusters, to older, fully exposed OB associations, as well as both
clustered and distributed populations.
Therefore, this region
is an ideal laboratory for investigating fundamental questions
related to the birth of stars and planetary systems.

 \begin{figure*}[ht]
\centering
 \includegraphics[angle=0,width=5.5in]{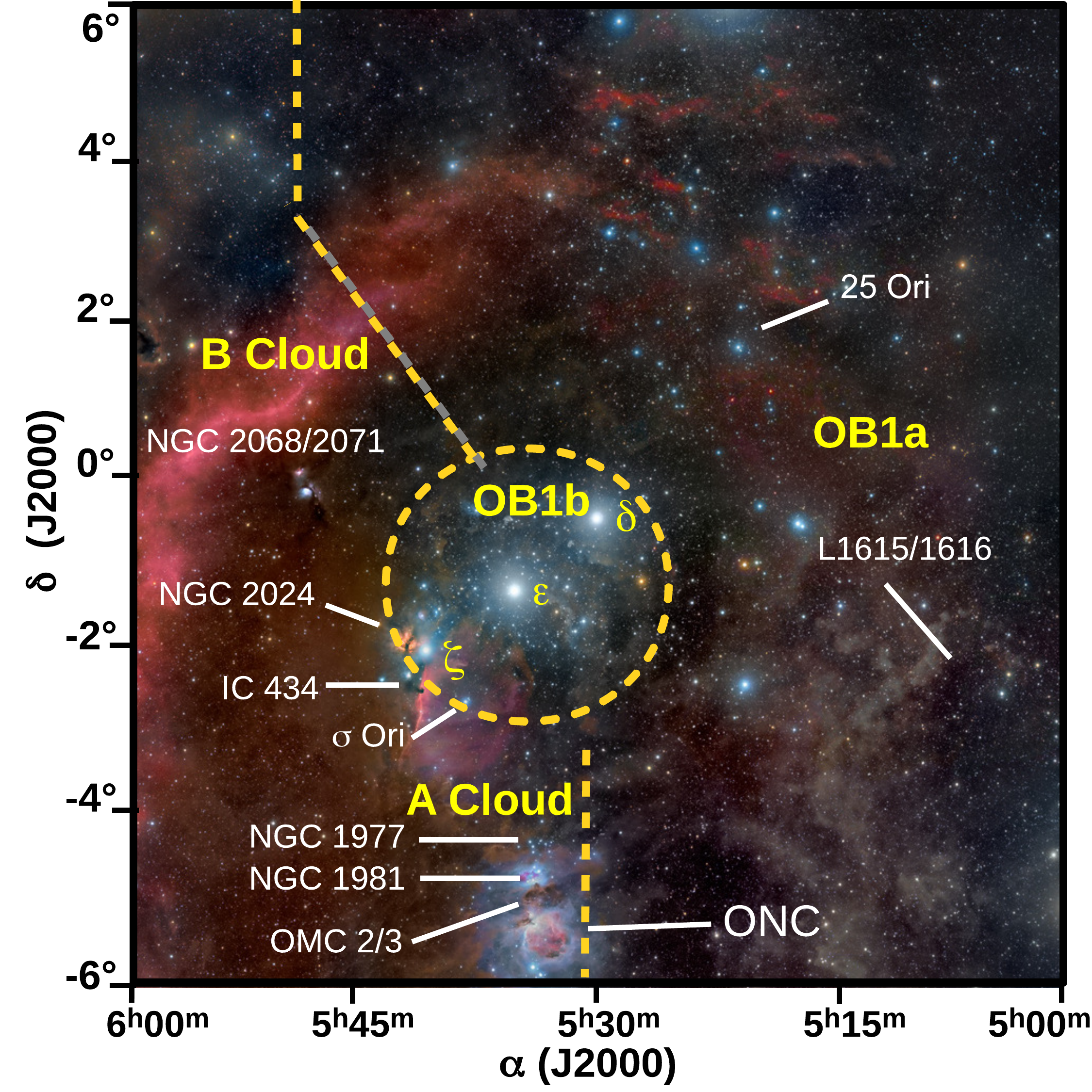}
 \caption{The Orion OB1 association as studied by the \emph{CVSO}.
 The most prominent and well known stellar groups and nebulosity found across
 this extended region are labeled.
 Throughout this work we assume the OB1b subassociation to be the area contained within the dashed lined circle roughly centered on $\epsilon$ Ori, as defined in \cite{briceno05}. The OB1a subassociation is all the area west of the straight dashed lines and of OB1b. 
 The B Cloud is considered here as the region east of the boundary with OB1a and OB1b, 
 north of $-2^\circ$, while the A Cloud is roughly the region east of 5h28m, 
 south of $-2^\circ$ and excluding OB1b
 (optical image courtesy of Rogelio Bernal Andreo, DeepSkyColors.com).
 }
 \label{surveymap}
 \end{figure*}

From late 1998 to 2012 we carried out 
the CIDA Variability Survey of Orion (\emph{CVSO}),
a large-scale photometric variability survey
(in the optical V,R and I-bands), complemented with an extensive spectroscopic study,
encompassing $\sim$ 180 $\rm deg^2$ in the Orion OB1 association,
(Fig. \ref{surveymap}), 
with the goal of identifying the low mass ($0.1~\msun
\la M \la 1~\msun$) stellar populations with ages $\la$
12 Myr \citep{briceno01,briceno05,briceno07b,briceno08}. 
The \emph{CVSO} has pioneered the use of large scale optical synoptic surveys 
to find and characterize populations of young, low-mass stars, 
a technique which has been successfully applied in several other studies
\citep{mcgehee05,mcgehee06,caballero2010, vaneyken2011,covey2011}. 
Though the \emph{CVSO} goes over the Orion A and B  clouds \citep{maddalena1986}, 
the real strength of our survey is the detection of the slightly extincted, optically 
visible low-mass PMS populations located in the extended areas devoid of cloud material; 
our completeness in the on-cloud regions is limited to members with low reddening 
($A_V$ less than a few magnitudes).

In this work we present a comprehensive large scale census of 
the off-cloud Orion OB1 low-mass 
young stellar population. We expand the initial results
presented in \citet[][- hereafter B05]{briceno05} by covering a much larger area, 
spanning all the region between $\alpha_{J2000}=$5-6h,  and $\delta_{J2000}=$-6 to +6 deg). 
This corresponds to the entire $\rm \sim 117\, deg^2$ encompassed by the Orion OB1a subassociation and the $\rm \sim 10.2 \, deg^2$ spanned by the
OB1b subassociation, both defined as shown in Figure \ref{surveymap}, 
plus $\rm \sim 31 \, deg^2$  on the A and B molecular clouds.
We do not consider here other parts of Orion OB1 which fall outside our survey boundaries,
such as the L1641 cloud \citep{allendavis2008}, located south of the ONC,  or the  $\lambda$ 
Orionis region \citep{mathieu2008}. Though located within our survey area, the embedded NGC 2024 \citep{meyer2008}, 2068 and 2071 clusters
\citep{gibb2008}, and the Orion Nebula Cluster \citep[ONC -]{muench2008} 
are not discussed here. Neither is the $\sigma$ Ori cluster \citep{walter2008}, 
which has been the subject of a separate study by \cite{hernandez2014}.

In \S \ref{sec:obs} we describe the optical variability survey,
the processing of the observations, the normalization and photometric calibration of the 
instrumental magnitudes, the \emph{CVSO} photometric catalog,
the selection of candidate, low-mass young stars,
and our follow up spectroscopy program. 
In \S \ref{sec:results} we describe our results,
and in 
\S \ref{sec:conclusions} 
we present a summary and conclusions.

\section{Observations}
\label{sec:obs}

Here we describe the two-stage methodology of our Orion OB1 large scale survey, starting with the multi-epoch, photometric variability survey for selecting candidate low-mass, young stars, followed by spectroscopic observations in order to provide membership confirmation and derive parameters like spectral type, reddening and type of object.

\subsection{The photometric survey}
\label{sec:surv}

The \emph{CVSO} consists of multi-epoch optical quasi-simultaneous V, R, and I-band
observations across the entire
Orion OB1 association,
obtained with the $8000 \times 8000$ pixel QUEST CCD Mosaic camera \citep{baltay2002},
installed on the 1m aperture J\"urgen Stock Schmidt-type telescope
at the National Astronomical Observatory of Venezuela. 
The system is optimized to operate in drift-scan mode: 
the telescope remains fixed at a given hour angle and declination, 
and because of the sidereal motion stars move across the length
of each CCD.
Up to four separate filters can be fitted at any given time in a special
filter holder. 
A single observation produces 4 stripes of the sky, since there are 4 columns in the array of CCDs in the QUEST camera. During drift scanning, stars go consecutively across the 4 CCDs in the same row, each fitted with a different filter. The total width of the scan area is 2.3 degrees, including small gaps between columns of CCDs.
At a survey rate of
roughly $\rm 34\, deg^2$ per hour per filter, large areas can be imaged
efficiently.
Many of the observations reported here were obtained with 
two V or two I filters in the filter holder; therefore, 
in those cases 
two frames in the V-band and/or I-band were produced for every star 
in that particular strip of the sky.
The typical minimum time between observations in two adjacent 
filters ($\Delta t_{min}$) is set by how long it takes a star to go 
from one detector to the next in the same row
of CCDs during a drift-scan observation. 
In the QUEST Camera, at the equator, $\Delta t_{min} \sim 140$ sec,
smaller than the time scale of most brightness variations
seen in TTS, which range from $\sim 1$ hour for flare-like events, up to days for
the rotational modulation produced by dark or bright spots in the stellar surface or
changes in accretion flows,
to weeks in the case of variations due to obscuration by features in a circumstellar disk.
Other variations take place over even longer time scales,
like those observed in eruptive young variables like EX Ori and FU Ori objects \citep{briceno2004}, or the
many year cycles observed in some TTS \citep{herbst1994,grankin2007,grankin2008,herbst2012}.
\cite{flaherty2013} discuss the various causes of variability in TTS.

For every Orion season (rougly from October to March), 
we observed as many nights as possible, limited essentially by
weather and instrument availability. Usually, on a given night, 
we concentrated on a particular declination, and made as many 
drift-scan observations of that stripe of sky as
possible during the $\sim 6$h when Orion is accesible at airmasses $\la 2$
from our equatorial location. Our temporal sampling is very heterogeneous,
spanning a wide range of time baselines,
from several minutes, to a few hours, days, weeks, months and years.

\input{tabobs_small.tex}

\begin{figure}[htb!]
\centering
\epsscale{1.20}
\plotone{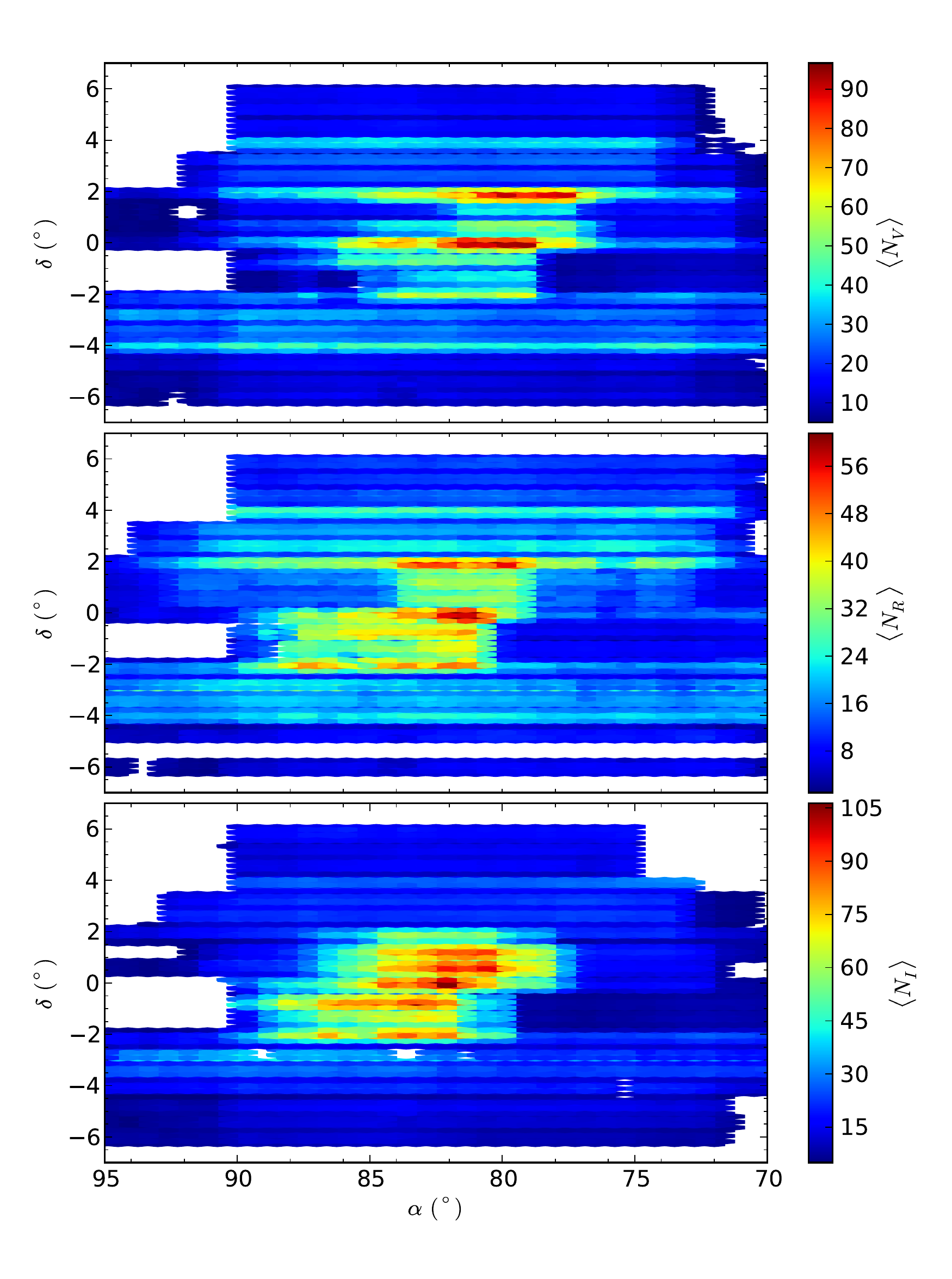}
\caption{Spatial coverage of the \emph{CVSO} in the $V$ (top), $R$ (middle)
and $I_C$ (bottom) filters. The color scale bars indicate the number of epochs per filter. Adapted from \cite{mateu2012}.
}
\label{nscans}
\end{figure}

The \emph{CVSO} comprises $337$ drift-scans, listed in Table \ref{tabobs}, 
performed along strips of right ascension
centered at  declinations $-5^\circ$, $-3^\circ$, $-1^\circ$, $+1^\circ$, $+3^\circ$ and $+5^\circ$.
We included observations obtained specifically for our program, 
but also for other programs that had targeted the same part of the sky
with the QUEST camera \citep[e.g.][]{vivas04,rengstorf2004,downes2008,downes2014}.
 The 4.7 Tb dataset corresponds 
to $\sim 15000$ hours of observations in the V, R, I
filters, obtained during 190 nights between 1999 and 2008.
The nature of the final dataset is quite heterogeneous, first, because filters 
changed position depending on the particular project being executed, which combined 
with the fact that not all detectors were functional over this many-year period, 
resulted in some regions having different multi-epoch coverage
in any of the V, R or I-band filters. Second, not all declinations were observed
as many times in each filter; though for the Orion project we planned as uniform coverage
as possible, including additional data from other projects meant that some
regions were more densely sampled. However, overall, the entire 
survey area was observed at least 10 or more times in each band (Figure \ref{nscans}),
with roughly 50\% of stars having in excess of 10 measurements in at least two photometric bands;
on average a given star has $\sim 20$ measurements.   
The average seeing measured in our data is $3.01\arcsec$, with a $\sigma = 0.64$,
which, given our plate scale of 1.02 arcsec/pixel,
means that the stellar point spread function is well sampled.

\begin{figure}[ht]
\includegraphics[angle=-90,scale=0.35]{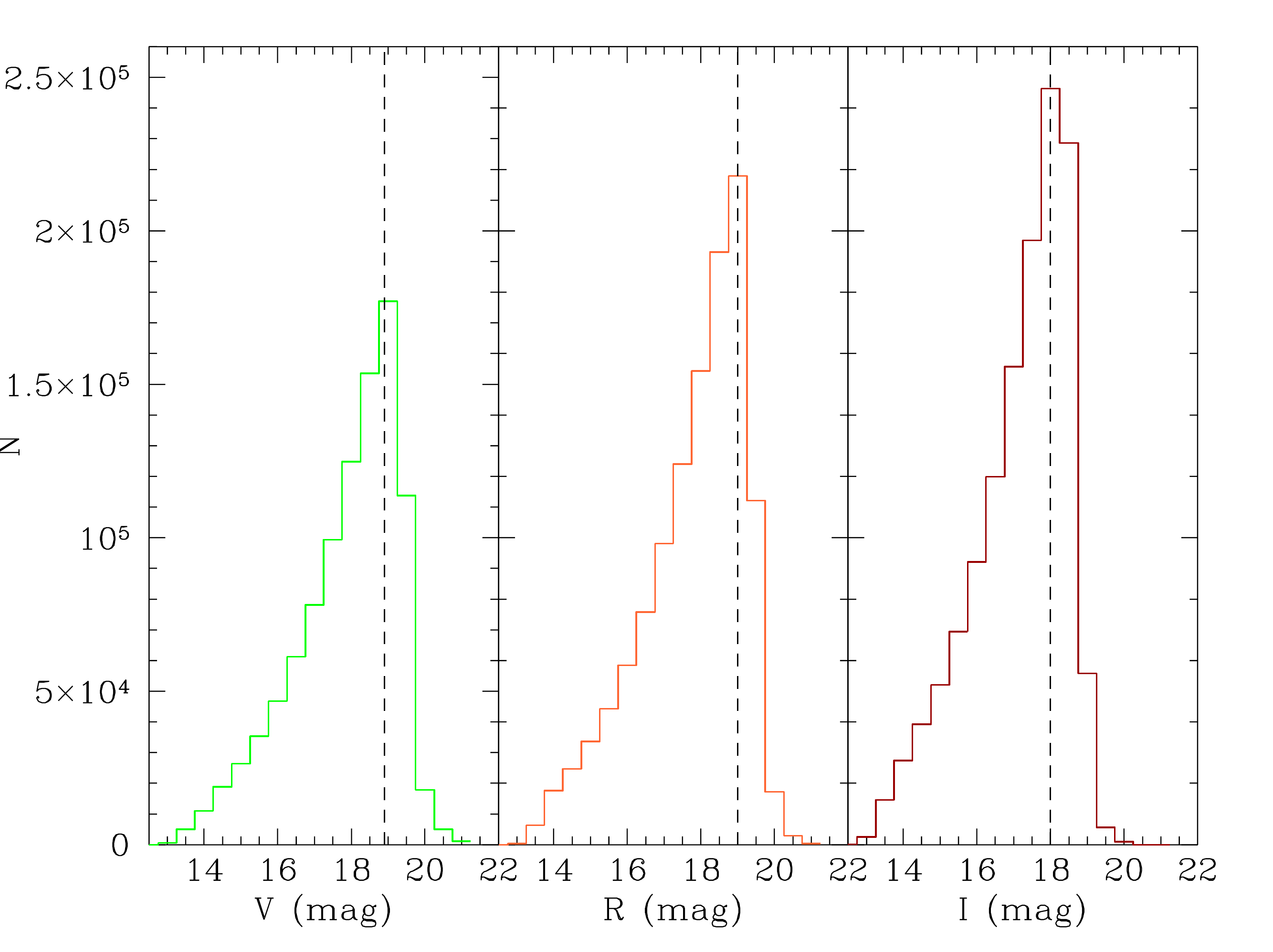}
\caption{Histogram of magnitudes in the \emph{CVSO} photometric catalog.
The vertical dashed lines indicate the completeness magnitudes:
V=18.5, R=18.5, I=18.0
}
\label{vrihisto}
\end{figure}

Because the exposure time in the drift-scan mode observations is
fixed, this defines the usable magnitude range for each individual
observation. At the bright end, our data saturate at magnitude $\sim 13.5$ in $V$, $R$ and $I_C$;
at the faint end, the $3\sigma$ limiting magnitudes for individual scans
are $\rm V_{lim}\sim 20.5$, $\rm R_{lim}\sim 20.5$, $\rm I_C{lim} \sim 20$,
with completeness (magnitude at which the distribution of sources reaches
its peak) of  
$\rm V_{com}\sim 18.9\pm 0.06$, $\rm R_{com}\sim 19.0\pm 0.07$, $\rm I_C{com} \sim 18.0\pm 0.08$
\citep[Figure \ref{vrihisto}; also, see][for more details on the photometry]{mateu2012}.

\subsection{Data Processing and optical photometry}
\label{sec:dataproc}

We used an updated, newer version of 
the automated QUEST data pipeline described in B05 and \cite{vivas04} 
to automatically process every single drift-scan. 
This software corrects the raw images by bias, dark current and flat field,
masks bad columns and pixels,
and then goes on to perform the detection of point sources, aperture photometry, 
and determination of detector coordinates for each object, 
independently in each of the 16 devices.
The software then solves the world coordinate system by
computing the astrometric transformation matrices for each CCD of the mosaic, 
based on the USNOA-2.0 astrometric catalog \citep{monet1998}, typically with an
accuracy of $\sim 0.14\arcsec$, sufficient for follow up work with multi-object
fiber spectrographs and cross-identification with other large scale catalogs. 
As a result, for every drift-scan observation there are 16 output catalogs produced,
containing for every object the following information: 
X and Y detector coordinates, J2000 equatorial coordinates,
instrumental magnitude and its corresponding $1\sigma$ error, and various
photometric parameters, like the Full-Width at Half-Maximum (FWHM) of
the image profile, ellipticity, the average sky background value and its associated error, 
and various flags (bad columns, edges, etc.).
Because Orion OB1 is well below the galactic plane
($b \ga -12^\circ$), stellar crowding is not an issue, so we could safely
perform aperture photometry; 
in fact, the average spatial density across our entire Orion survey area is 
1 point source every 1334 arcsec$^2$, which translates into a mean distance between
stars of $\sim 36\arcsec$. We used an aperture equal to the mean FWHM of our images.
Following B05,
for each filter and declination strip, we constructed reference catalogs of instrumental photometry, obtained under the best possible atmospheric
conditions. All the other catalogs are normalized to these reference catalogs,
such that
differing sky conditions like variations in transparency from night to night,
or a passing cloud causing an extinction of up to 
1 magnitude were accounted for, as shown in Figure 3 of
\cite{vivas04}, where more details of this method are provided.

In order to calibrate our instrumental magnitudes
in the Johnson-Cousins system, we followed a several step process
\citep[see also B05;][]{vivas04,mateu2012}.
Instead of observing Landolt fields every night at the Venezuela National Observatory, 
we collected a set
of secondary standard star fields evenly distributed in declination, located
so that each scan obtained at a given declination would go over 4 secondary
standard fields, one per each row of four CCDs in the QUEST mosaic camera.
We refined the calibration done in B05 by defining
new $\rm 23\, arcmin \times 23\, arcmin$ secondary standard fields obtained with 
the Keplercam 4k $\times$ 4k CCD instrument on the 1.2m telescope 
at the SAO Whipple Observatory in Arizona.
We observed the secondary standard fields 
during several photometric nights in Feb. 2008,
together with several Landolt standard star fields
\cite{landolt83} at various airmasses.
Then, using our \emph{CVSO} observations of the stars in the Keplercam
fields, we selected $\sim 125$ secondary standard stars
per field in the range $13 \le V\le 17$,
determined to be non-variable at the $\le 0.02$ mag level 
\citep[the complete V,R, I photometry for the secondary standards 
is published in][]{mateu2012}.
With such a large number of standards we were able to derive 
a robust photometric calibration, with an
average RMS  $\sim 0.021$ in the V, R and I bands.


\begin{figure}[ht]
\includegraphics[angle=0,scale=0.45]{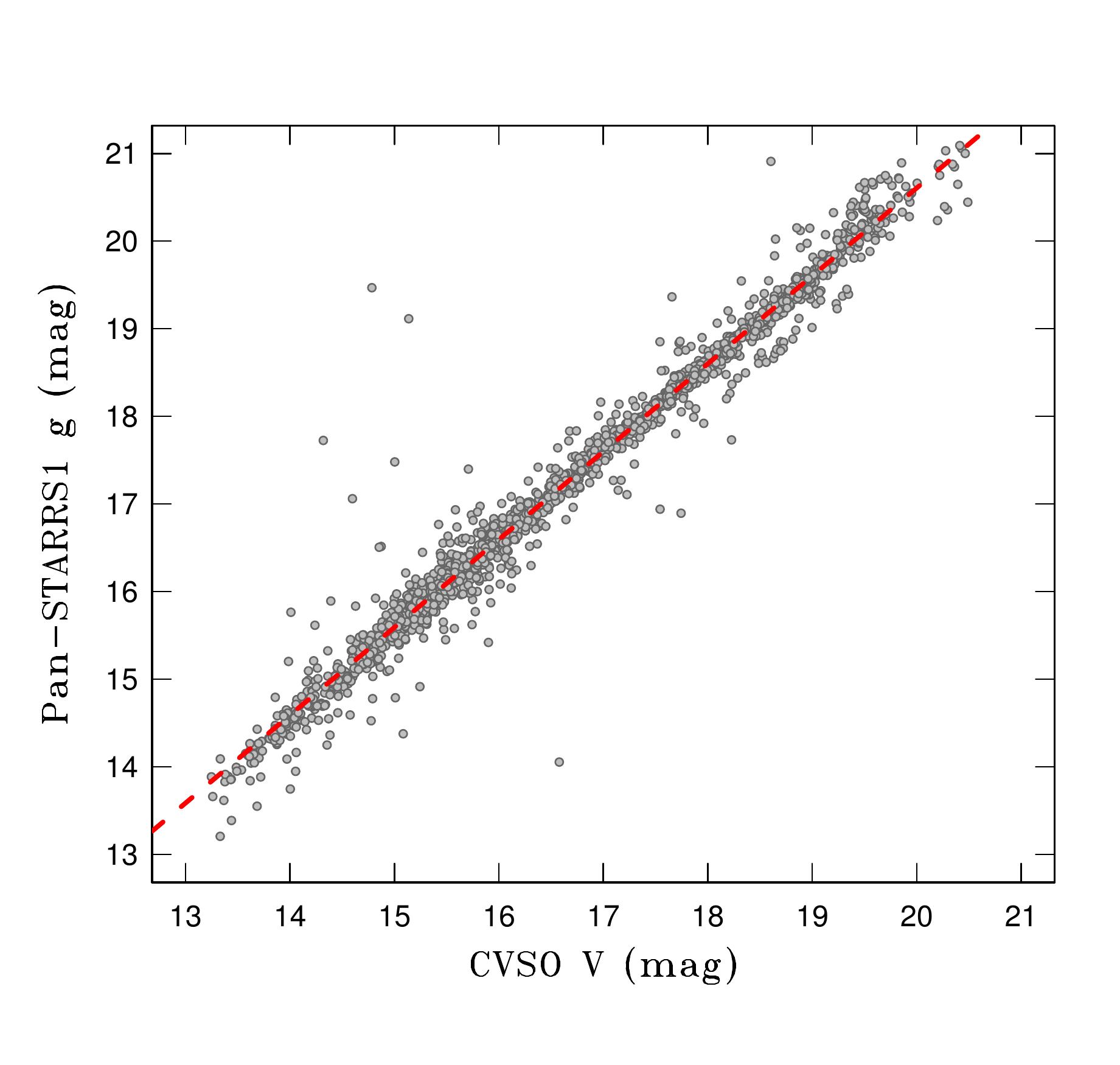}
\caption{Comparison of our \emph{CVSO} $V$ magnitudes with the Pan-STARRS1 DR1 $g_{PS1}$ photometry for the TTS sample described in this work.  The red dashed line is a least square fit to the data.
}
\label{cvso_vs_ps1}
\end{figure}

As an external check of our photometry, in Figure \ref{cvso_vs_ps1} we compare our $V$-band magnitudes with the Pan-STARRS1\footnote{Panoramic Survey Telescope \& Rapid Response System; 
http://pan-starrs.ifa.hawaii.edu/public/home.html} (PS1) first data release (DR1) $g_{PS1}$ magnitudes for the full sample of TTS presented in this work. The agreement is very good. The least square fit is $g_{PS1} = 1.003 \times V + 0.548$, which is consistent within 0.06 magnitudes with the relationship provided by \cite{tonry2012}, for the median $g_{PS1} - r_{PS1}=1.2$ color of our TTS sample.
The scatter observed for some stars is likely due to their variability between
the \textit{CVSO} and the PS1 observations.

\subsection{The \emph{CVSO} photometric catalog and the selection of candidate 
            pre-main sequence stars}
\label{sec:catalog_cand_selec}

\subsubsection{The Catalog}
\label{sec:cvso_catalog}

Our final \emph{CVSO} catalog contains $1702231$ sources 
located in the region $\rm \alpha_{J2000} \sim 5^h$ to $\rm 6^h$ and
$\delta_{J2000}= \sim +6\arcdeg $ to $-6\arcdeg$.
Each source has an arbitrary ID number, $\rm \alpha_{J2000}$ and $\delta_{J2000}$ 
coordinates, at least one measurement in either one of the 
V, R and I-bands ($640639$ objects have photometry in all three bands), 
with their corresponding errors
(the sum in
quadrature of the photometric calibration error and the
standard deviation in that magnitude bin of all stars determined to be non-variable),
the number of measurements in each filter, the maximum measured amplitude of photometric
variations in each filter, the standard deviation of photometric variations in each filter,
the probability that the object is variable in a given photometric band based on a
$\chi^2$ test, the actual value of $\chi$ in each filter, 
and the \cite{stetson96} $L_{VR}$, $L_{VI}$ and $L_{RI}$ variability indices with their 
corresponding weights, and the number of pairs of measurements
involved in the computation of each index \cite{mateu2012}. 

Though the Stetson variability indices in the \emph{CVSO} catalog are
a valuable new addition to our original B05 photometric variability dataset, in order to be able
to compare the new results presented here with our previous studies, we used the
same variability selection criterion as in B05, namely, we
flagged as variable (at a 99.9\% confidence level) 
those objects for which the probability in the V-band $\chi^2$ test 
that the dispersion of measured magnitudes is due to random errors,
is very low ($\le 0.001$). This criterion yielded $85011$ variable sources 
(5\% of the $1702231$ objects in the full \emph{CVSO} catalog).
With the adopted confidence level, formally only $85$ of the $85011$ objects 
flagged as variables are expected to be false-positives. 
But in reality, such a sample of variable objects 
can still potentially contain fake detections due to cosmic rays, 
bad pixels or columns not properly corrected for during the data processing, 
or other artifacts.
We dealt with this by requiring either that an object had a
counterpart in more than one band, that it had 3 or more measurements in a
single band, or that it has a counterpart in the 2MASS catalog
\citep[][ see below]{skrutskie06}. 
Our variability selection also 
sets the minimum $\Delta {\rm mag}$ that we can detect as a function of
magnitude: 0.08 for $V=15$, 0.12 for $V=17$ and
0.3 for $V=19$.
This means that for fainter stars we detect only those which vary the most.
We cannot decrease the confidence level too much without increasing the
contamination from non-variable stars to unacceptable levels, because we are
dealing with such large numbers of stars.

Comparing directly the overall fraction of variable stars 
found in the {\sl CVSO} with other similar surveys is far from straightforward.
The number of variables detected will depend on, among
other parameters, the criterion for declaring an object as variable,
the wavelength range considered, the time sampling,
the general direction on the sky,
the survey brightness limits, plate scale, seeing conditions, and 
methods for source extraction and photometry.
Nevertheless, it is useful to place the {\sl CVSO} results in context with other 
surveys spanning a similar magnitude range. 
The Catalina Sky Survey found 2-4\% of variable objects
on timescales spanning several years 
\citep{drake2013,drake2014}, and the PS1 $3\pi$ survey 
detects 6.6\% variables 
among $3.8 \times 10^8$ sources, in multi-epoch data spanning $\sim 3.7$ years 
\citep{hernitschek2016}.  
In Orion, the only other significant variability survey carried out so far is that by \cite{carpenter01},
which targeted a $0.84^\circ \times 6^\circ$ region centered roughly on the Orion Nebula cluster,
where they found a $\sim 7\%$ fraction of near-IR variables, similar to our result.

The availability of large scale astronomical surveys, together with new data mining tools
have made feasible the combination and analysis of multiple datasets containing information
across a wide range of wavelengths. 
We performed a spatial match of the full \emph{CVSO} catalog against 
the 2MASS Point Source Catalog \citep[PSC][]{skrutskie06}, that helped us
weed out artifacts that may still affect the optical catalog,
but more importantly, added near-infrared photometry that combined with
our optical magnitudes and variability information, 
provided us with a longer wavelength base for a refined color selection of candidate young stars, 
and later enabled the determination of fundamental parameters like the stellar luminosity.
We used the Tool for OPerations on Catalogues And Tables package \citep[TOPCAT - ][]{taylor2005}
and the Starlink Tables Infrastructure Library Tool Set \citep[STILTS - ][]{taylor2006}
to do a positional match, with a $2\arcsec$ search radius,
between the \emph{CVSO} catalog and the 2MASS Point Source Catalog (PSC); 
the mode of the distribution of separations between our catalog positions
and 2MASS is $0.18\arcsec$ (a significant improvement over the result obtained in B05),
of which $0.06\arcsec$ comes from a systematic offset 
between the USNOA-2 catalog and the 2MASS PSC.  
The resulting combined VRIJHK catalog contained $946934$ sources;
because of the reduced sensitivity of the 2MASS PSC compared to the \emph{CVSO},
this match effectively set the limiting depth of our survey, 
such that the completeness magnitudes 
dropped to $\rm V_{com}=18.2$, $\rm R_{com}=17.7$ and $\rm I_{com}=17.2$.
We could have used the much deeper YZJHK dataset obtained for the \emph{VISTA} telescope Galactic Science Verification \citep{petr-gotzens2011}, 
but then we would be limited to a much smaller area ($\sim 30$ sq.deg, or $\lesssim 20$\% of
our total photometric survey area); the \emph{VISTA} observations were used by \cite{downes2014,downes2015} and \cite{suarez2017} in their study of the 25 Ori cluster .
Since our purpose was to create a spatially complete map of the low-mass young populations 
in Orion OB1, we opted to sacrifice depth in favor of a combined optical/near-IR catalog 
that spanned the full area covered by the \emph{CVSO}.
Even with this modestly deep completeness level we still are sensitive to 
pre-main sequence stars down 
to $\sim 0.15\msun$ at 10 Myr \citep{siess00}, which means that we could still expect to
create a rather complete map of the stellar population to very low masses. Also,
from a purely observational point of view, this was a reasonable low-mass limit for a feasible
spectroscopic follow-up program using existing multi-fiber spectrographs on 4-6.5m class telescopes (see \S \ref{sec:lowres_spec}).

\subsubsection{Pre-main sequence candidates selection}
\label{sec:cand_selection}

In selecting our photometric pre-main sequence (PMS) candidates we followed a two step process that produced high and low priority targets for follow up spectroscopy: 1) Select objects located above the main sequence in optical and optical-near IR color-magnitude diagrams; 2) among the PMS candidates from step one, select those identified as variable.

Following the procedure outlined in \cite{briceno05}, of the $946934$ sources
in our combined CVSO-2MASS catalog, we selected $115071$
PMS candidate stars located {\sl above} the main sequence \citep{siess00}, 
set at a distance of 440 pc, 
in both V vs V-J and V vs $\rm V-I_c$ color-magnitude diagrams constructed using our robust mean $V$ magnitudes; this is
what we call here Candidate Sample 1 (CS1).
Formally, Orion spans a range of distances, from $\sim 360$ pc for the closer
OB1a subassociation  to $\sim 400$ pc for OB1b and the molecular clouds, 
as shown by recent accurate distance determinations from Very Long Baseline Interferometry by \cite{kounkel2017a}, and parallaxes from the \textit{Gaia} Second Data Release \citep[DR2; ][see \S \ref{sec:distances}]{gaia_dr2}. 
However, assuming a slightly farther distance gave us a more relaxed selection criterion, placing the main sequence lower (fainter) in 
the color-magnitude diagram (CMD), and therefore allowing us to gather a more
complete candidate sample among the more distant and in the older regions. 
The disadvantage of this approach is that a fainter main sequence allows more field contaminants, specially for the regions thought to be nearest to us. 

Among the objects in CS1, we selected $12928$ stars (11\% of the total CS1 sample)
as variable objects; 
this is Candidate Sample 2 (CS2), which by definition is a subset of CS1. 
For convenience, we define here as Candidate Sample 3 (CS3), those objects in sample CS1 
not flagged as variable, and therefore not included in CS2; this is either because they are
non-variable, or more likely, have variability below our detection threshold for that magnitude.
Candidates in CS2 were considered our highest priority targets for follow-up 
spectroscopy (see \S\ref{sec:lowres_spec}).
In order to have a general idea of the effectiveness of our PMS candidate selection scheme, we
looked up in the SIMBAD database all previously known T Tauri stars inside our entire survey region.
Though strictly this cannot be considered a quantitative 
test of our photometric search technique for young low-mass stars,
because the PMS objects in SIMBAD constitute a very heterogeneous set, 
containing objects from a variety of studies with differing biases, 
techniques and spatial coverage, it still provides a rough estimate  of how much of 
the low-mass PMS population we can expect to find.  
Out of 275 SIMBAD objects with a ``TT\*'' or ``Y\*O'' type 
(excluding sources in the ONC, in $\sigma$ Ori, NGC 2024, NGC 2068, 
and those published in B05 and B07),
in the magnitude range V=13.5-19 and located in the PMS locus in the 
V vs V-J CMD, we recovered 85\% in CS2. 
The ones we did not recover were because they coincided with bad columns, 
or fell in gaps between adjacent rows of detectors, in the master reference 
drift-scans used to calibrate the final \emph{CVSO} catalog (\S\ref{sec:dataproc}).

\subsection{Spectroscopy}
\label{sec:lowres_spec}

\input{speclog_short.tex}

\begin{figure*}[htb!]
	\centering
	\includegraphics[angle=270,scale=0.70]{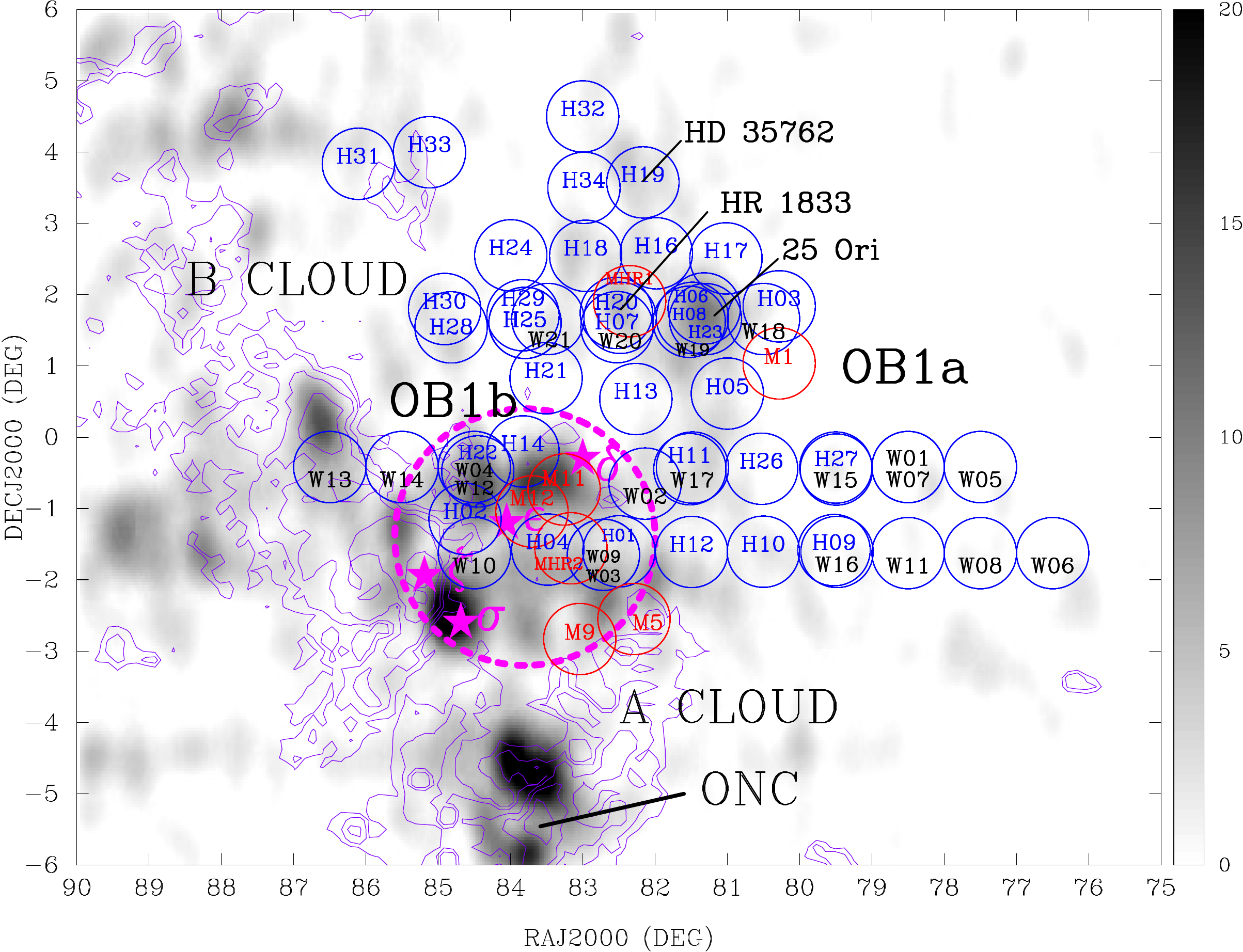}
	\caption{Schematic distribution of our Hectospec, Hydra and M2FS observations.
	Circles labeled ``H'' in blue correspond to the Hectospec fields, 
	``W'' black labels to Hydra, and ``M'' red labels to the M2FS fields (see Table \ref{speclog}).
	The gray scale map is the surface density of candidate TTS in sample CS2, per $1.6\arcmin \times 2.6\arcmin$. 
	Contours show the integrated  $^{13}$CO emissivity of the Orion
	A and B molecular clouds \citep{bally87}, covering the range from
	from 0.5 to 20 ${\rm K \, km \, s^{-1}}$.
	The circular dashed region indicates our definition of the Orion OB1b association,
	as described in B05.
	The solid starred symbols mark the positions of the Orion belt stars and $\sigma$ Ori.
	The location of the 25 Ori cluster and the new TTS clusterings HR1833 and HD35762	(see \S \ref{sec:ob1_offcloud_pop}) are also indicated. 
	}
	\label{hsp_hydra}
\end{figure*}

\begin{figure}[hbt!]
\epsscale{1.2}
\centering
    \plotone{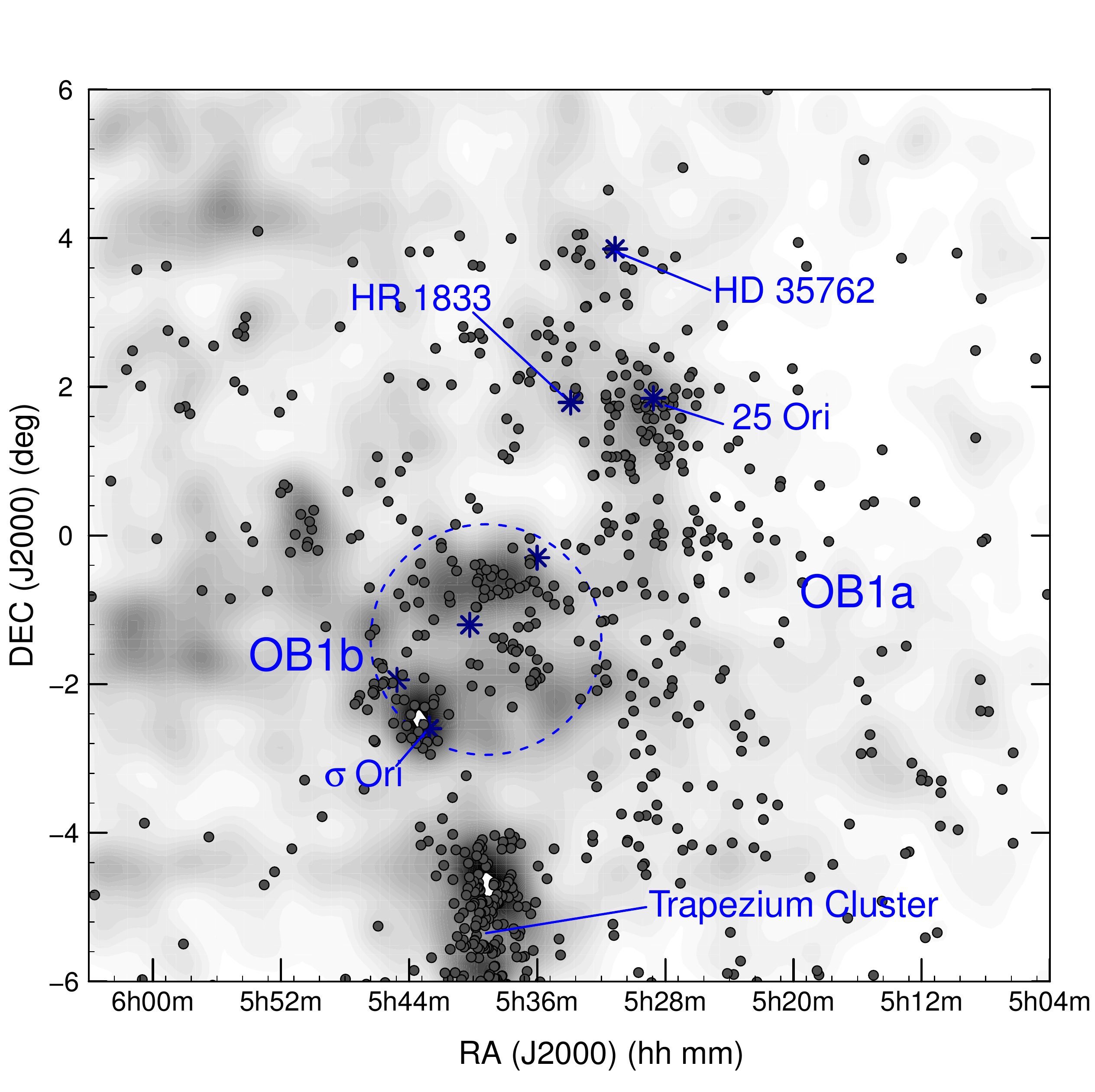}
\caption{Spatial distribution of TTS confirmed through FAST spectroscopy. 
Individual stars are indicated with the dark dots. The gray scale background is the surface density of photometric candidate PMS stars in our sample CS2.
The main stellar groups and regions are labeled, both the known groups and the new ones discussed in this work.
}
\label{spatial_fast}
\end{figure}

A sensitive photometric survey capable of 
identifying reliable and large samples of candidate PMS stars across 
the entire Orion OB1 association is the first step in mapping the full 
young low-mass population.
However, follow up, low-resolution optical spectroscopy is paramount for three main reasons: 
first, to confirm membership, because even the best candidate samples are inevitably affected 
by contamination from field stars; second, 
to determine basic quantities for each star like its luminosity and $\rm T_{eff}$, that
can then be compared with evolutionary models to estimate masses and ages,
and third, to distinguish between non-accreting Weak line T Tauri stars (WTTS) 
and accreting Classical T Tauri stars (CTTS), an important diagnostic for characterizing the disk accretion properties across the full stellar population.

Our spectroscopic low-resolution follow up program has been carried out with the following facilities:

1) The Hydra multi-fiber spectrograph \citep{barden94} on the WIYN 3.5m telescope at Kitt Peak.

2) The Hectospec multi-fiber spectrograph \citep{fabricant05} on the 6.5m MMT.

3) The Michigan/Magellan Fiber System \cite[M2FS;][]{mateo2012} on the 6,5m Magellan Clay telescope at Las Campanas Observatory.

4) The FAst Spectrograph for the Tillinghast Telescope (FAST) \citep{fabricant98} on the 1.5m telescope
of the Smithsonian Astrophysical Observatory.

5) The Goodman High Throughput Spectrograph \cite[GHTS;][]{clemens2004} on the 4.1m Southern Astrophysical Research (SOAR) telescope at Cerro Pach\'on, Chile.

We obtained spectra of a total of $11201$ candidate PMS stars among all instruments. 
 As a general strategy we selected our highest priority targets from the CS2 sample (PMS variables),
and then added targets from sample CS3 (PMS non-variables). For Hectospec and Hydra we also included as third priority, whenever there were available fibers, additional candidates selected from the PMS locus in J vs J-H color-magnitude diagrams made from 2MASS data.
In Figures \ref{hsp_hydra} and \ref{spatial_fast} we plot the spatial distribution of all sources observed with the multi-fiber spectrographs, and with FAST, respectively.

\subsubsection{Multi-fiber Spectroscopy}

A total of $7796$ candidates fainter than $V\sim 16$ were observed 
in our combined multi-fiber spectrograph campaigns.
Both Hectospec and Hydra have 1 deg diameter field of view, while M2FS has a 29.5 arcmin diameter field.
Hectospec has 300 fibers, each $1\farcs5$ on the sky. 
Hydra with the Red Channel has 90 fibers each
with a projected diameter of $2\farcs0$, and M2FS offers up to 256 fibers, 
128 for each of its twin Littrow spectrographs,
each fiber with a projected diameter of $1\farcs2$ on the sky.

In Table \ref{speclog} we show the full log of all the multi-fiber spectrograph observations for
Hydra, Hectospec and the Michigan/Magellan Fiber System (M2FS),
including those discussed in B05;

\begin{figure*}[htb!]
\epsscale{1.15}
\plottwo{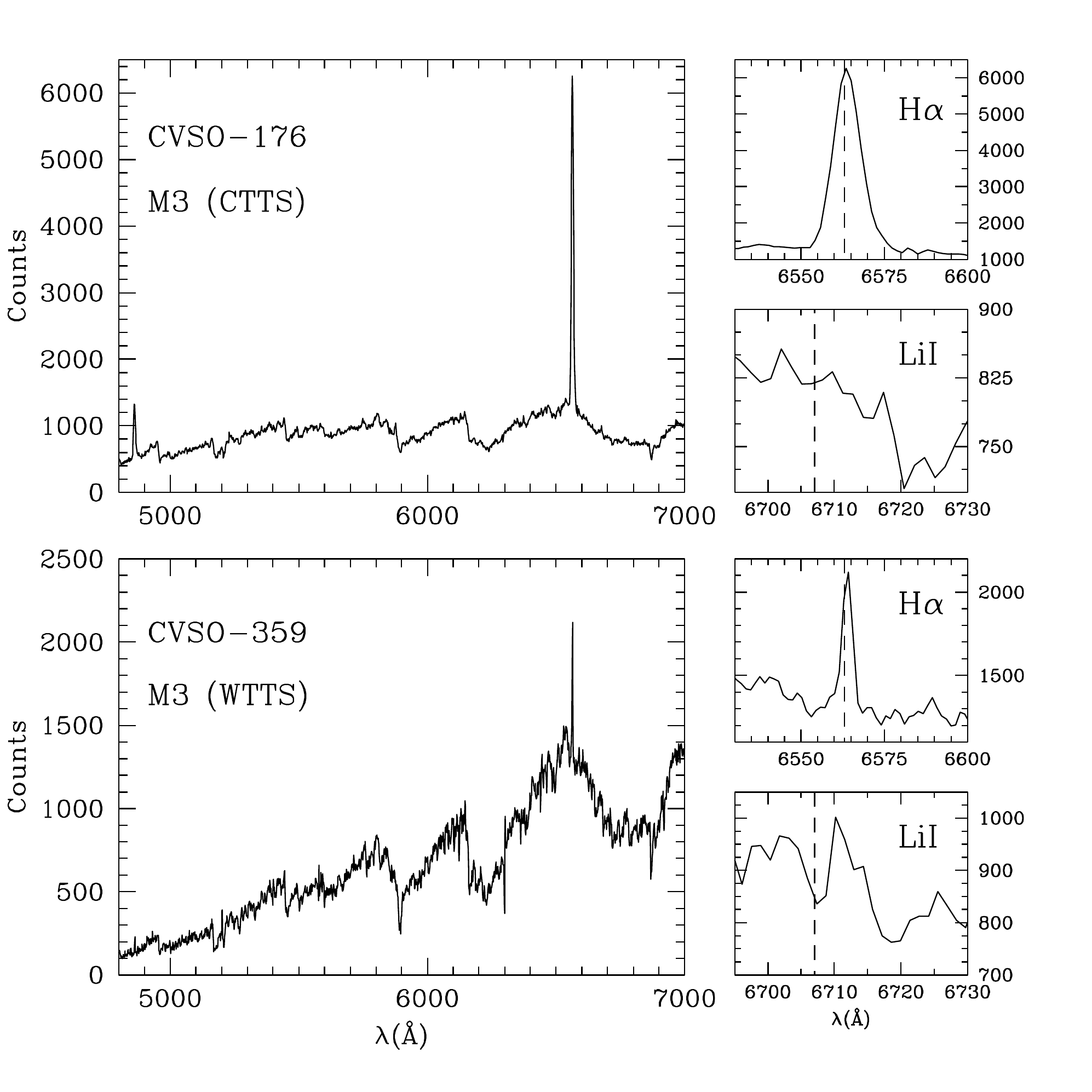}{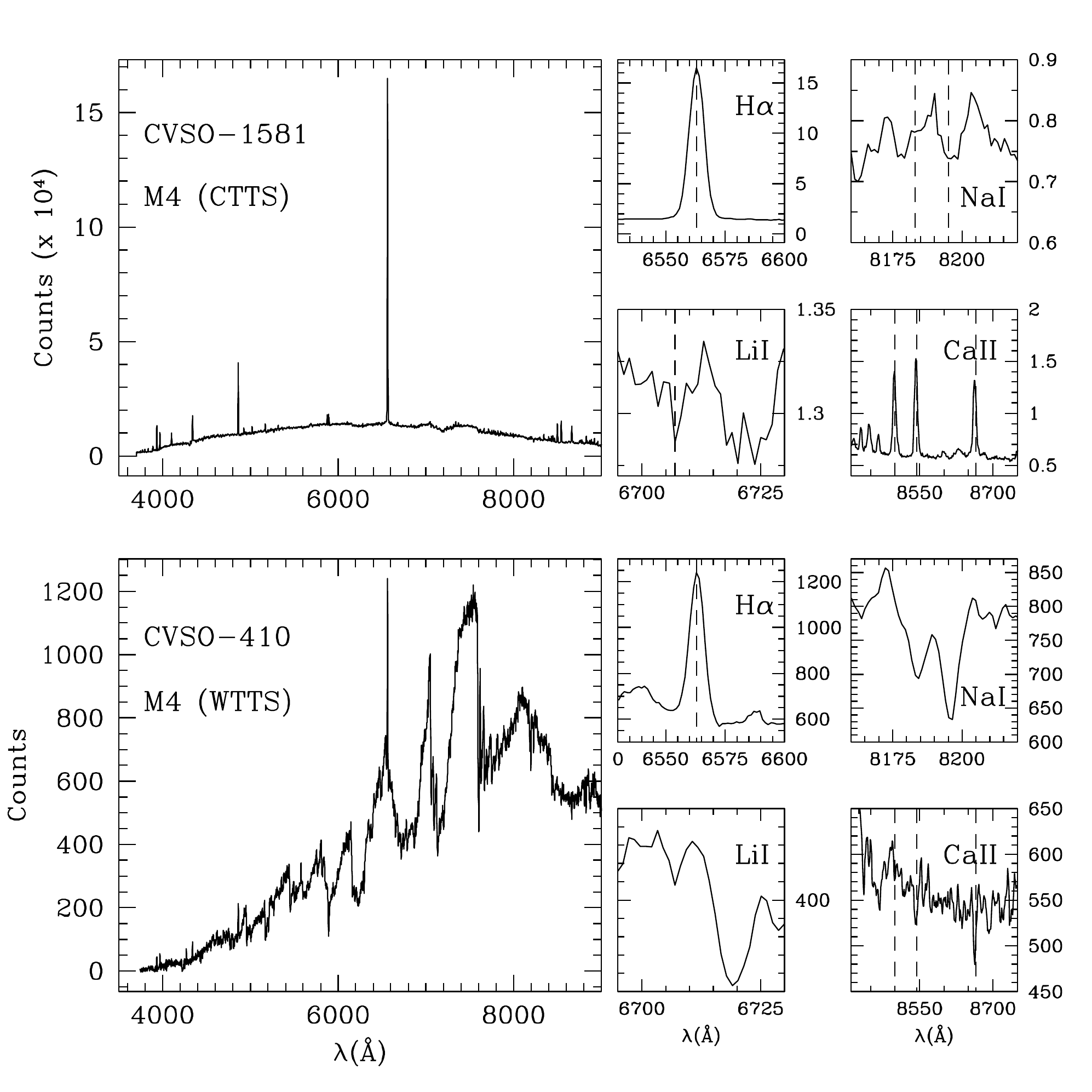}
\caption{{\bf Left:} Spectra of two M3 type TTS obtained with the Hydra spectrograph on the WIYN 3.5m telescope.
In the upper panel, the  CTTS \emph{CVSO}-176 has a W(H$\alpha$)$=-41.4${\AA},
and  W(LiI)$=0.2$\AA. 
The H$\beta$ Balmer line is also in emission, as are the He I lines at 5876{\AA } and 6876\AA.
In the lower panels, the WTTS star CVSO-359 has a W(H$\alpha$)$=-2.5${\AA} and W(LiI)$=0.6$\AA.
Other than H$\alpha$, no other lines are seen in emission.
In both stars the S/N ratio in the region between H$\alpha$ and Li~I is $\sim 35$.\\
{\bf Right:} Spectra of two candidates confirmed as new M4 TTS members of Orion,
obtained with the Hectospec spectrograph on the 6.5m MMT.
In the upper panel we plot an extreme CTTS, with W(H$\alpha$)$=-83.6${\AA} and W(LiI)$=0.1$\AA.
The entire Balmer series is clearly in emission, along with 
the Ca H and K lines (3933\AA,3968\AA), He I at 5876{\AA } and 6876\AA,
[OI] 6300{\AA } and [SII] 6716\AA,6732\AA.
In this star the Na~I 5890,5895{\AA } doublet and
the Ca~II triplet (8498\AA, 8542\AA, 8662\AA) are also strongly in emission
(W(CaII8498)= -10.2\AA, W(CaII8542)= -11.8{\AA } and W(CaII8662)= -10.5{\AA }).
The near-IR Na~I doublet has W(8183)=0.5{\AA } and W(8195)=1.1\AA.
Lower panel: The M4 WTTS shown here has W(H$\alpha$)$=-6.1${\AA} and W(LiI)$=0.4$\AA.
H$\beta$ and the Ca H and K lines are also in emission. 
The near-IR Na~I doublet has W(8183)=0.7{\AA } and W(8195)=1.0\AA, typical of
M-type stars with lower than main sequence gravities \citep{luhman2003,schlieder2012}.
}
\label{spec_hyd_hsp}
\end{figure*}

 \begin{figure*}[ht!]
 \centering
\includegraphics[width=3.5in,height=2.72in]{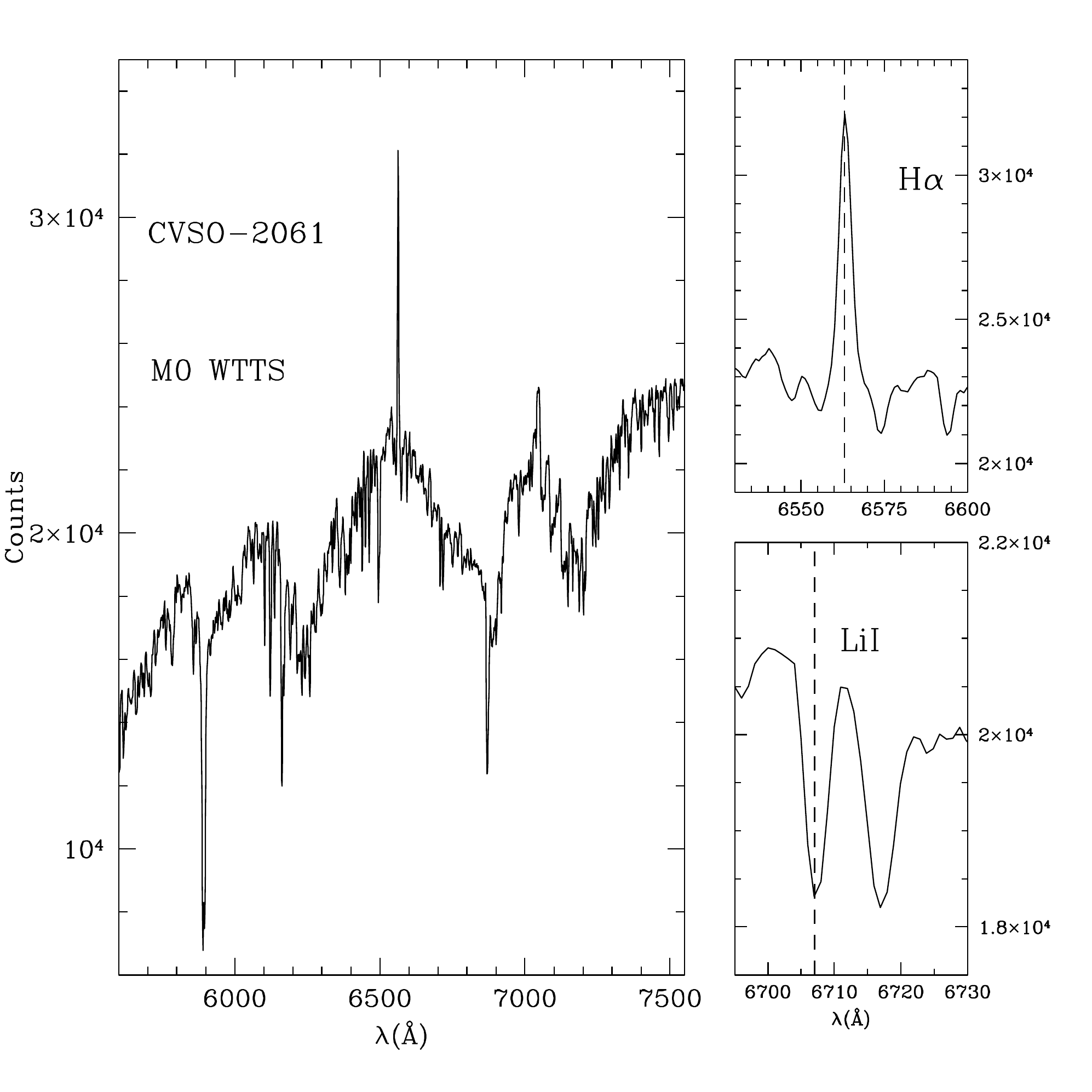}
\includegraphics[width=3.5in,height=2.72in]{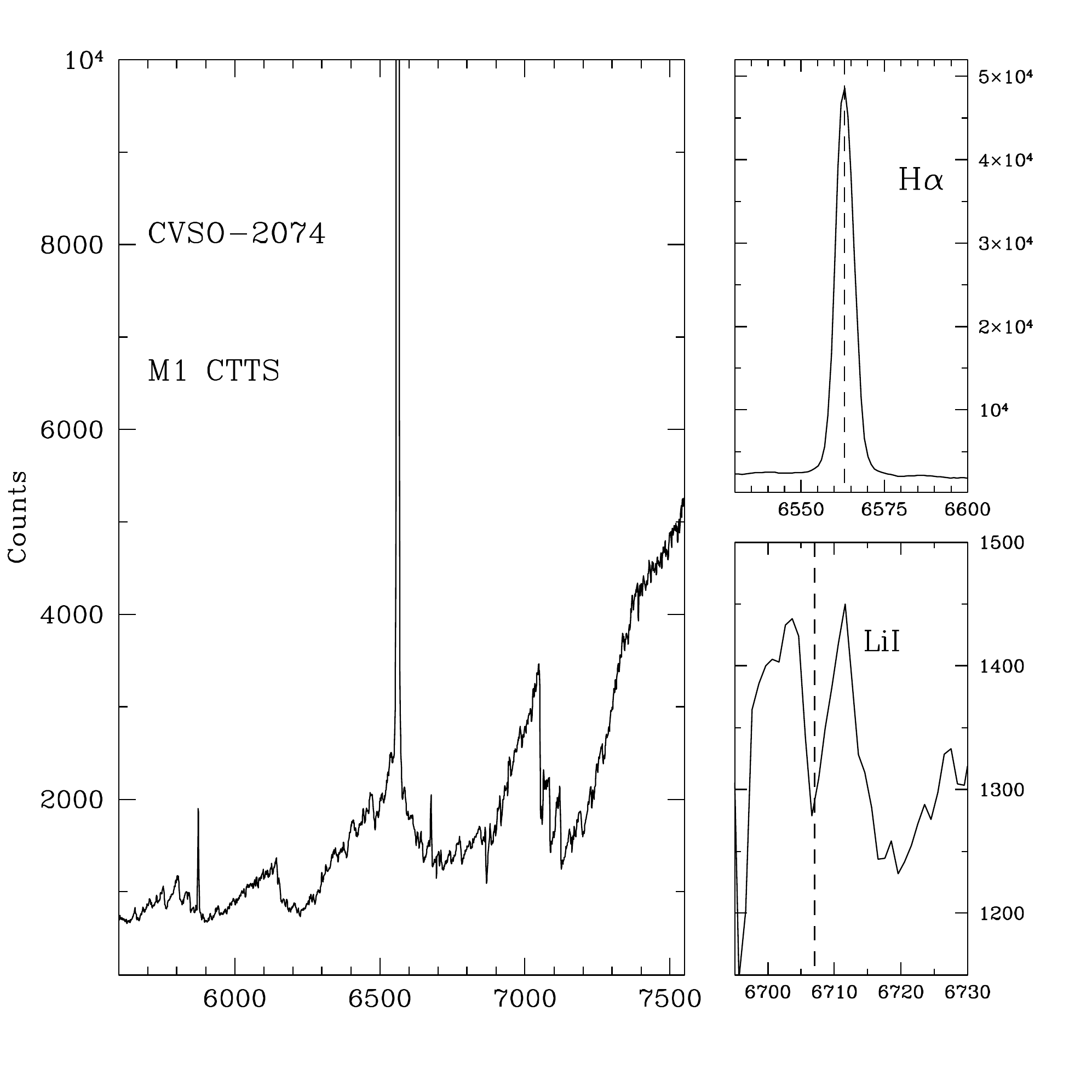}
\caption{Spectra of two newly confirmed early M-type TTS members of Orion,
obtained with the M2FS spectrograph on the 6.5m Magellan Clay telescope.
\textbf{Left:} A M0 WTTS, which has W(H$\alpha$)$=-2.2${\AA} and W(LiI)$=0.4$\AA.
In both spectra the S/N ratio is $\ga 50$.
\textbf{Right:} an accreting CTTS, with W(H$\alpha$)$=-139${\AA} and W(LiI)$=0.4$\AA.
In addition to the strong H$\alpha$ line, the He I line can also be seen 
in emission at 5876{\AA } and 6876{\AA}, as well as [OI] 6300\AA. 
	}
	\label{spec_m2fs}
\end{figure*}

\paragraph{WIYN+Hydra}

In B05 we reported on $320$ targets observed with Hydra in 5 fields 
(6 fiber configurations) on November 26 and 27, 2000
(W02, W03, W04, W05, W07 and W09 in Table \ref{speclog}).
Here we consider an additional $932$ objects for which we obtained WIYN-Hydra spectra
during the nights of November 26, 27 and 28, 2000, February 02, and November 13, 14 and 15, 2002. 
These candidates were distributed in 15 fields (see Figure \ref{hsp_hydra}), spanning an area of $\sim 11\> deg^2$;
$138$ (15\%) objects listed in CS2 were selected as priority 1 targets.
We used the Red Channel fibers ($2 \arcsec $ diameter),
the Bench Camera with the T2KC CCD, the 600@10.1 grating,
yielding a wavelength range $\sim 4700 - 7500${\AA } with a resolution
of 3.4\AA. 
All fields were observed with airmasses=1.0-1.5, and integration
times for individual exposures were 1800 s. When weather allowed we obtained
two or three exposures per field.
Comparison CuAr lamps were obtained between each target field.
In each Hydra field we assigned fibers to all candidates from CS2
with V=$16 - 18.5$, then to candidates from CS3, and with the lowest priority, to objects from the 2MASS near-IR CMD.
Typically we assigned 10-12 fibers to empty sky positions and
5-6 fibers to guide stars.
We used standard IRAF routines to remove the 
bias level from the two dimensional Hydra images.
Then the {\sl dohydra} package was used to extract
individual spectra, derive the wavelength calibration,
and do the sky background subtraction.
Since the majority of our fields are located in regions
with little nebulosity, background subtraction was in
general easily accomplished.

\paragraph{MMT+Hectospec}

With Hectospec we observed $6110$ targets distributed in 28 fields
(34 fiber configurations), covering $\rm 23\,deg^2$,
during the period Nov 2004 to Feb 2010 (Table \ref{speclog} and Figure \ref{hsp_hydra}). We excluded the 124 members of 25 Ori from B07,
and the 77 very low-mass PMS members in that same cluster already 
reported by us in \cite{downes2008} and \cite{downes2014}.
We assigned the highest priority in the fiber configuration software
to the $813$ candidates flagged as PMS variables (CS2).
The spectrograph setup used
the 270 groove $mm^{-1}$ grating, yielding spectra in the range  
$\lambda 3700 - 9000 $\AA, with a resolution of 6.2 \AA. 
As for Hydra, objects in CS2 had the highest priority,
followed by objects from CS3, and filling remaining fibers 
with 2MASS JH-selected PMS candidates. 
On average we assigned 50 fibers per field to empty sky positions;
the majority of our fields are located in regions with little or no 
extinction, so nebulosity was not an issue for sky subtraction.
 All the Hectospec spectra were processed, extracted, corrected for sky lines,
 and wavelength calibrated by 
S. Tokarz at the CfA Telescope Data Center, using customized IRAF routines 
and scripts developed by the Hectospec team \citep[see][]{fabricant05}. 

\paragraph{Magellan+M2FS}

We used the M2FS instrument to obtain spectra of 434 targets distributed in 5 fields,
spanning a total area of $1\,deg^2$, during several runs between November 2013 and February 2015 
(Table \ref{speclog} and Figure \ref{hsp_hydra}).
As we did for Hectospec, we assigned the highest priority
to candidates flagged as PMS variables (CS2).
The spectrograph setup used
the 600 lines/$mm^{-1}$ grating and $125\mu$m slit, yielding spectra in the range  
$\lambda 5670 - 7330 $\AA, with a resolution of 1.3 \AA. 
Sky fibers were assigned as for Hydra and Hectospec.
The raw data were processed using custom Python scripts developed by John Bailey, 
that apply a bias correction, merge the four files per image produced by each of the four
amplifiers, and correct cosmic rays  \citep[see ][ for a more detailed description of the software]{bailey2016}.  
Extraction of spectra, wavelength calibration and
correction for sky lines was done using the routines in the {\sl twodspec} and {\sl onedspec} packages in IRAF.

\subsubsection{Single Slit Spectroscopy}

\begin{figure*}[htb!]
\epsscale{1.15}
\plottwo{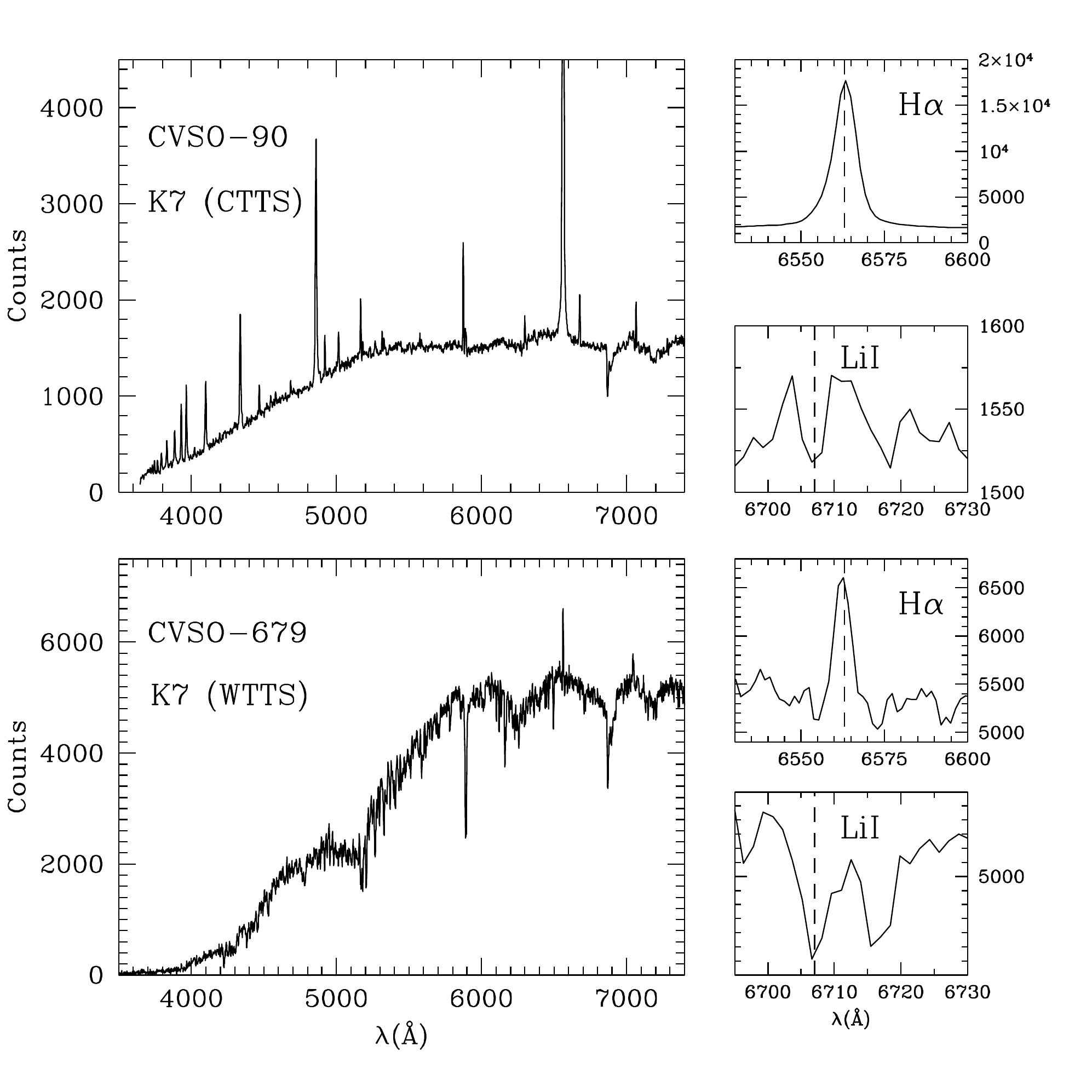}{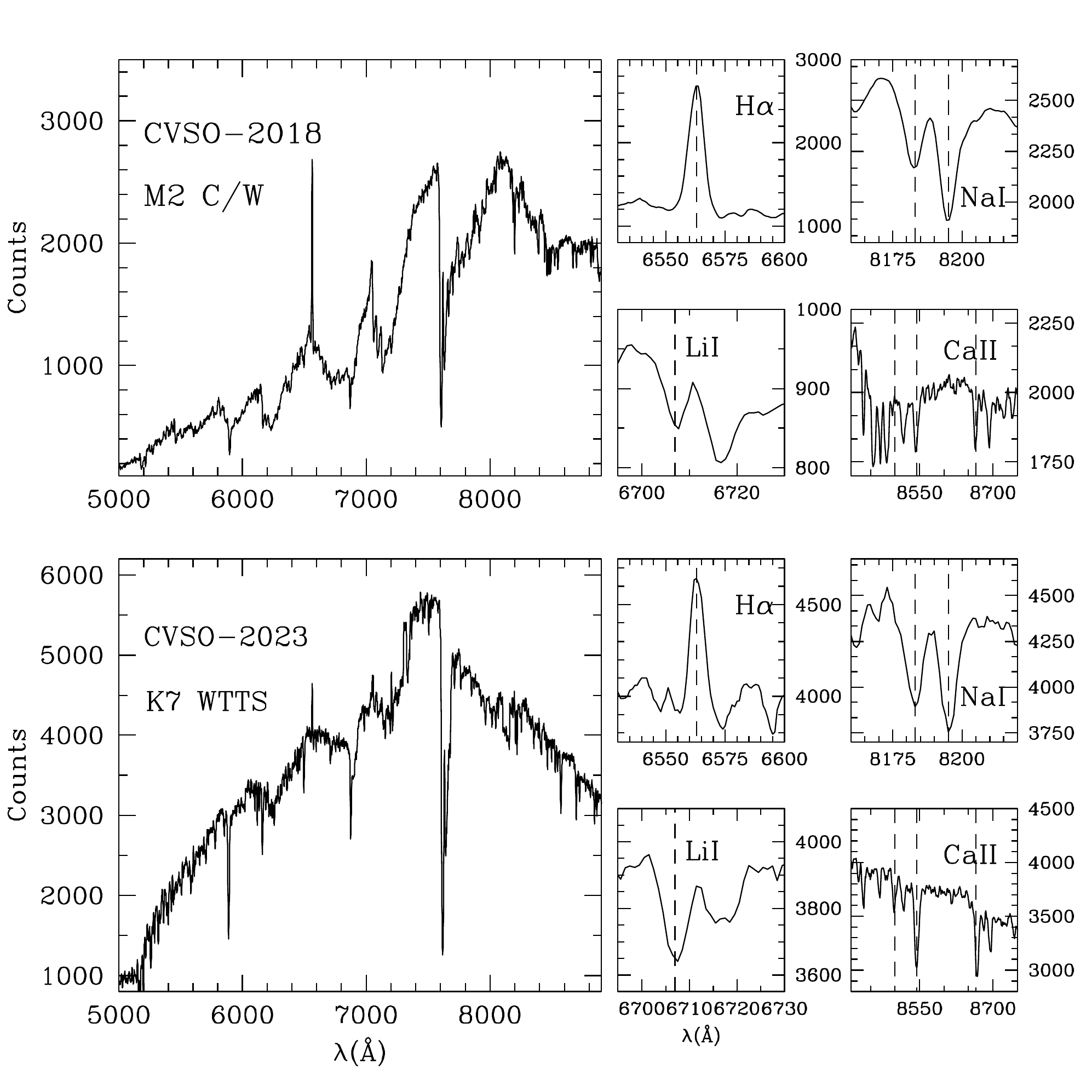}
\caption{{\bf Left:} Spectra of K7 type TTS obtained with the FAST spectrograph on the SAO 1.5m telescope.
In the upper panels the CTTS \textit{CVSO}-90 shows the characteristic extreme emission line spectrum of a strongly
accreting TTS; the equivalent width of the H$\alpha$ emission line is W(H$\alpha$)$=-97${\AA}. 
The equivalent width of Li~I is W(LiI)$=0.2$\AA, and the absorption line next to it is the Ca~I line at 6716\AA.
The entire Balmer series is seen clearly in emission; also in emission are the
the Ca H and K lines (3933\AA,3968\AA), [OI] at 6300\AA, and He I at 5876{\AA } and 6876\AA.
In contrast, the WTTS \textit{CVSO}-679 in the lower panels has a weak W(H$\alpha$)$=-1.6${\AA}, and 
no other emission lines; in this star W(LiI)$=0.4$\AA.
{\bf Right:} Spectra of two newly identified TTS obtained with the GHTS spectrograph on the SOAR 4.1m telescope.
In the upper panels the C/W type (see \S \ref{sec:ctts_classif}) \textit{CVSO}-2018 shows the characteristic strong emission line spectrum of a moderately accreting TTS; the equivalent width of the H$\alpha$ emission line is W(H$\alpha$)$=-9.9${\AA}. 
The equivalent width of Li~I is W(LiI)$=0.2$\AA, and the absorption line next to it is the Ca~I line at 6716\AA.
The K7 WTTS \textit{CVSO}-2023 in the lower panels has a weak W(H$\alpha$)$=-1.4${\AA}; in this star W(LiI)$=0.5$\AA.
In all spectra the S/N ratio in the region between H$\alpha$ and Li~I is $\ga 20$.
}
\label{spec_fast_ghts}
\end{figure*}

\paragraph{FLWO 1.5m+FAST}

We obtained spectra for a total of $3383$ bright candidates ($V < 16$)
in queue mode at the FAST spectrograph.
Out of these $3383$ FAST targets, $1235$ (37\%) objects from subset CS2
were observed as highest priority. We then continued with the remaining
$2148$ objects from set CS3.

In B05 we presented results for the first $1083$ FAST candidates
observed from January 1999 through January 2002. Here we discuss the
additional $2300$ candidates for which we obtained FAST spectra up to April 2013.
In Figure \ref{spatial_fast} we show the spatial distribution of the Orion TTS members confirmed with FAST. We obtained a relatively uniform spatial coverage across the entire survey area, except for the declination band $+4\deg \la \delta \la +6\deg$.
The FAST Spectrograph was equipped with the Loral $512 \times 2688$ CCD,
in the standard configuration used for ``FAST COMBO'' projects: 
a 300 groove $mm^{-1}$ grating and a $3\arcsec$ wide slit, 
producing spectra spanning the range from 4000 to 7400{\AA } with a resolution of 6 {\AA }.
The spectra were reduced at the CfA using software
developed specifically for FAST COMBO observations.
All individual spectra
were wavelength calibrated using standard IRAF routines.
The effective exposure times ranged from 60 s for the $V\sim 13$ stars
to $\sim 1500$ seconds for objects with $V\sim 16$.

\paragraph{SOAR+Goodman}

We used the Goodman High Throughput Spectrograph (GHTS), installed on the SOAR 4.1m telescope on
Cerro Pach\'on, Chile, to obtain slit spectra of 23 candidate TTS that needed confirmation, and had not been observed with any other spectrograph.
We used available time slots during the engineering nights of  Mar 19, 2014,  Sep 6, 2017 and Dec 5, 2017. 
The GTHS is a highly configurable imaging spectrograph that employs all-transmissive optics
and Volume Phase Holographic Gratings, that result in high throughput for
low to moderate resolution spectroscopy over the 320-850 nm wavelength range.
The 2014-03-03 and 2017-09-06 observations both used the 400 l/mm grating in its 400M2 preset mode combined with the GG 455 order-sorting filter. This configuration provides a wavelength range
$\sim 5000 \la \, \lambda \, \la 9000$\AA, that
combined with the $1\arcsec$ wide slit results in a Full Width at Half Maximum (FWHM) resolution of 6.7{\AA } (equivalent to $R\sim 800$).
The 2017-12-05 observations were done with the 600 l/mm grating in the "mid" setup, with the GG385 order-blocking filter, and the $1\arcsec$ slit. This setup results in a FWHM spectral resolution of 4.4\AA (equivalent to R$\sim 1300$) in the wavelength range of $\sim 4450 \la \, \lambda \la \, 7050$\AA.
In all cases we used $1\times 1$ binning, keeping the native pixel scale of $0.15\arcsec$/pixel.
For each object we obtained three integrations, which were median combined after correcting for
bias, and spatially registering the second and third exposure to the first one, which we used
as reference. Integration times ranged from 300s for the brightest targets (V $\sim 15$), to 900s for the fainter ones (V$\sim 18$). The basic image reduction was performed using standard IRAF packages: CCDPROC and IMSHIFT.
The one-dimensional spectrum was extracted
using routines in the IRAF TWODSPEC and ONEDSPEC packages.
For wavelength calibration we used a HgArNe lamp. 

We did not perform flux calibration in any of our spectra, since 
the main purpose of our follow up spectroscopy is membership identification
and spectral type classification. 
We measured H$\alpha$ and Li~I equivalent widths in all our low resolution spectra from the various instruments, 
using the {\sl splot} routine in IRAF and the 
SPTCLASS tool \citep{hernandez2017}, an IRAF/IDL code based on the methods described in \cite{hernandez04}. 
Lines in the far red region of the spectrum were only available for TTS confirmed in Hectospec and SOAR spectra.
The S/N ratio of our spectra was typically $\ga 25$ at H$\alpha$,
sufficient for detecting equivalent widths down to a few $0.1${\AA }
 at our spectral resolution of $\sim 6-7${\AA } FWHM.
In Figures \ref{spec_hyd_hsp}, \ref{spec_m2fs} and \ref{spec_fast_ghts} 
we show sample spectra of Orion OB1 WTTS and CTTS observed with the Hydra and Hectospec, M2FS, and FAST and SOAR spectrographs, respectively.

\section{Results and Discussion}
\label{sec:results}

\subsection{Identification of T Tauri Stars} \label{sec:newtts}

\subsubsection{H$\alpha$ emission and Li I absorption}
\label{sec:membership}


We establish membership based on our low resolution spectra from 
FAST, Hydra, Hectospec, M2FS and SOAR. Our criteria to identify PMS 
low-mass stars are the following:

1) Spectral type between K to M-type, which corresponds to the range of
colors and magnitudes expected from our photometric survey candidate selection.

2) Presence of the Balmer hydrogen lines in emission, 
in particular H$\alpha$ 6563{\AA}, which are characteristic of 
active late spectral type young ($\la 1 $ Gyr) dwarfs
\citep[e.g.][]{stauffer_hartmann86,stauffer97}.

\begin{figure}[hbt!]
\centering
\epsscale{1.25}
\plotone{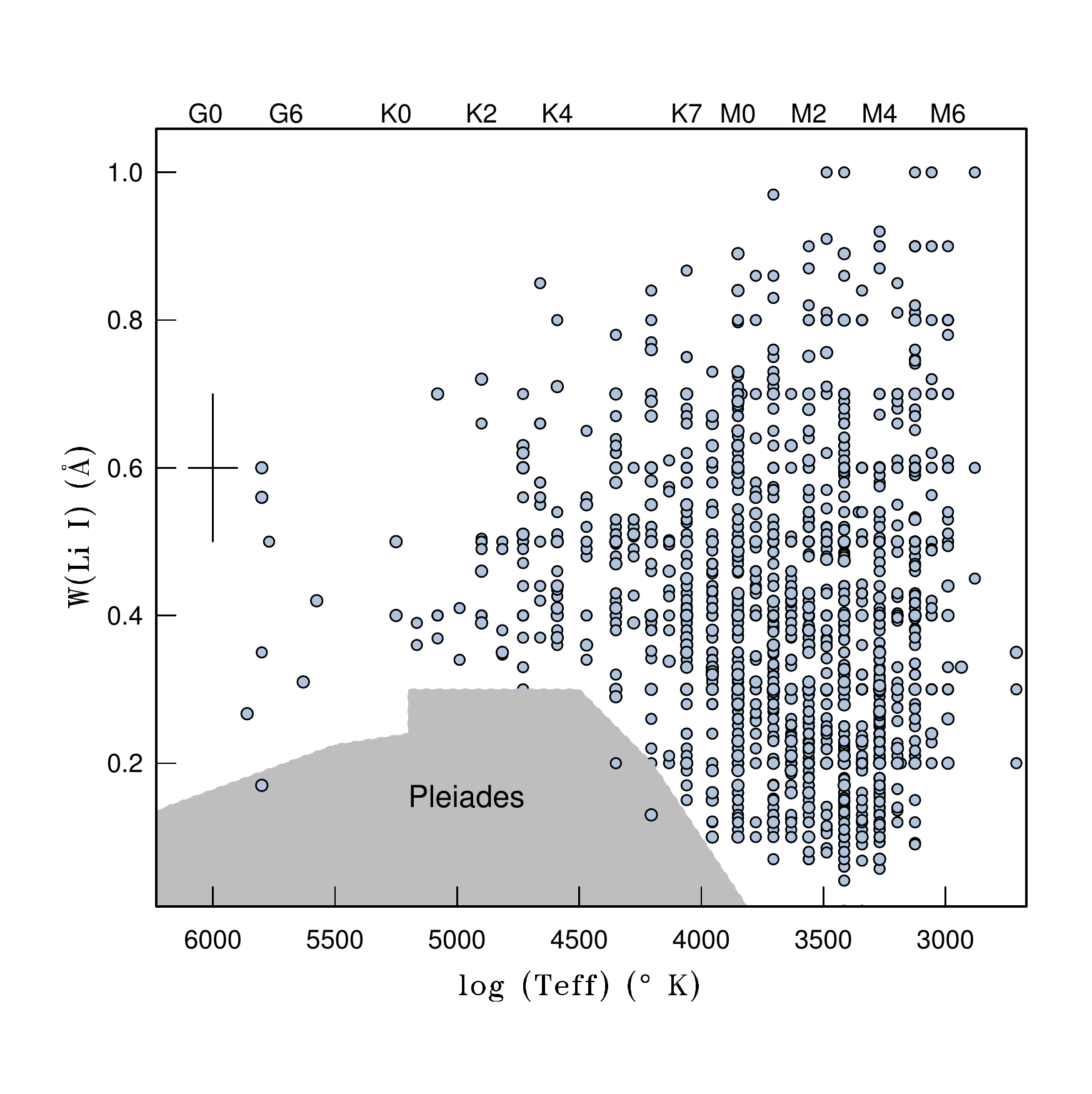}
\caption{Equivalent width of the Li I $\lambda 6707$ line for the 2064 Orion OB1 TTS. 
The locus of Pleiades stars is indicated by the shaded grey region
\citep{soderblom1993,garcia-lopez1994}.
A candidate young star is considered a TTS if it falls above this part of the diagram (see text). The typical errorbar is indicated.
}
\label{ewli_ob1ab}
\end{figure}

3) Presence of the Li I (6707\AA) line strongly in absorption \citep{briceno97,briceno98}. Li I is our main youth criterion. Since lithium is depleted during the PMS stage in stars in the deep convective interiors of
 K and M-type stars, we regarded a candidate object as a TTS if it had H$\alpha$ in emission, and Li~I(6707\AA) in absorption with equivalent 
 width larger than the upper value for a Pleiades star of the same spectral 
 type \citep{soderblom1993,garcia-lopez1994}, which represents the young main 
 sequence for late type stars (Figure \ref{ewli_ob1ab}).
 With a S/N ratio $\ga 25$ in our spectra we could detect Li~I$\lambda 6707$ 
absorption down to $\rm W(Li~ I) \sim 0.1${\AA }.
They may have a weak Li~I line if the CTTS spectrum is heavily veiled by the excess continuum emission from an accretion shock, 
created by material infalling from the circumstellar disk onto the star. 
In these cases, we still classified those stars as TTS, 
regardless of the small value for the W(Li~I).

4) Presence of additional youth signatures, in particular gravity-sensitive features like the Na~I 8173/8195{\AA } doublet weakly in absorption
\citep[e.g][]{martin1996,luhman2003,martin2004,slesnick06,downes2008,martin2010,lodieu2011,schlieder2012}. 
In strongly accreting young stars, other spectral features like He I at 
5876{\AA} and 6678{\AA }, forbidden [OI]6300{\AA} and 6346{\AA}, 
[SII] at 6716{\AA } and 6732{\AA }, and the 
Ca II triplet (8500\AA,8544{\AA } and 8664.5\AA) can 
also be seen in emission \citep{edwards87,hamann1990,hamann1992,hamann1994}, and we take these as features that confirm and reinforce the PMS nature of a star.


\begin {deluxetable*}{rll}
\tabletypesize{\scriptsize}
\tablewidth{0pt}
\tablecaption{T Tauri stars in the \emph{CVSO} \label{newttsinfo}}
\tablehead{
\colhead{Column number} & \colhead{Column Name} & \colhead{Description} 
}
\startdata
  1   &  CVSO          & \emph{CVSO} number  \\
  2   &  H07           & ID from \cite{hernandez07a}  \\
  3   &  Other ID      & Other designation, from SIMBAD \\
  4   &  RA(J2000)     & Right Ascension J2000.0 (hh:mm:ss.ss)  \\
  5   &  DEC(J2000     & Declination J2000.0 (dd:mm:ss.s)  \\
  8   &  SpT           & Spectral type \\
  9   &  WHa           & Equivalent width of H$\alpha$ 6563 (\AA) \\
 10   &  WLi           & Equivalent width of Li I 6707 (\AA) \\
 11   &  WNaI          & Total equivalent width of the Na~I 8183,8195 doublet (\AA) \\
 12   &  Type          & C=CTTS, CW=C/W, W=WTTS (\S \ref{sec:ctts_classif}) \\
 13   &  $\bar{V}$     & V filter robust mean \citep{stetson96} (mag) \\
 14   &  err($V$)      & 1-$\sigma$ error of $\bar{V}$, computed from err vs mag diagram (mag) \\
 15   &  $N(V)$        & Number of non-null $V$-band observations \\
 16   &  $\bar{R}$     & R filter robust mean  \citep{stetson96} (mag) \\
 17   &  err($R$)      & 1-$\sigma$ error of $\bar{R}$, computed from err vs mag diagram (mag) \\
 18   &  $N(R)$        & Number of non-null $R$-band observations \\
 19   &  $\bar{I_C}$   & I filter robust mean \citep{stetson96} (mag) \\
 20   &  err($I_C$)    & 1-$\sigma$ error of $\bar{I}$, computed from err vs mag diagram (mag) \\
 21   &  $N(I_C)$      & Number of non-null $I$-band observations \\
 22   &  $\Delta(V)$   & $V$-band peak-to-peak amplitude (mag) \\
 23   &  $\Delta(R)$   & $R$-band peak-to-peak amplitude (mag) \\
 24   &  $\Delta(I_C)$ & $I$-band peak-to-peak amplitude (mag) \\
 25   &  $\sigma(V)$   & $V$-band standard deviation (mag) \\
 26   &  $\sigma(R)$   & $V$-band standard deviation (mag) \\
 27   &  $\sigma(I_C)$ & $V$-band standard deviation (mag) \\
 28   &  $V_prob$      & $\chi^2$ probability of variability in $V$  \\
 29   &  $R_prob$      & $\chi^2$ probability of variability in $R$  \\
 30   &  $I_prob$      & $\chi^2$ probability of variability in $I$  \\
 31   &  $L_{VR} $     & \cite{stetson96} variability index for $V$ \& $R$ magnitudes \\
 32   &  $L_{VI} $     & \cite{stetson96} variability index for $V$ \& $I$ magnitudes \\
 33   &  $L_{RI} $     & \cite{stetson96} variability index for $R$ \& $I$ magnitudes \\
 34   &  $J$           & 2MASS $J$ magnitude (mag) \\
 35   &  $err(J)$      & 2MASS $J$ 1-$\sigma$ combined error (mag) \\
 36   &  $H$           & 2MASS $H$ magnitude (mag) \\
 37   &  $err(H)$      & 2MASS $H$ 1-$\sigma$ combined error (mag) \\
 38   &  $K$           & 2MASS $Ks$ magnitude (mag) \\
 39   &  $err(K)$      & 2MASS $Ks$ 1-$\sigma$ combined error (mag) \\
 40   &  $T_{eff}$     & Effective temperature ($^\circ K$) (1) \\
 41   &  $A_V$         & Extinction in the $V$ band (mag) (2) \\
 42   &  Loc           & Location within Orion OB1: 1a, 1b, 25 Ori, HR 1833, A\_cloud, B\_cloud \\
\enddata
\tablenotetext{1}{Corresponding to the spectral type interpolated in Table A5 of \cite{kh95}}
\tablenotetext{2}{For 86\% of the sample $A_V$ was derived from the $V-I_c$ color. For an additional 8\% of stars which lacked an I-band measurement, 
we used the $V-J$ color. For the remainder of the stars we used either the $R-I$, $R-J$ or $I-J$ colors. We adopted the \cite{cardelli1989} extinction law, and intrinsic colors from \cite{kh95}.}
\end{deluxetable*}
\label{mastertable}

Following this approach, we classified $2064$ candidates as confirmed low-mass, PMS stars of
 K and M-type, based on our low resolution spectra (passed H$\alpha$, Li~I criterion); a few G-type TTS were also identified. 
 The properties of each TTS are provided in Table \ref{mastertable}: ID, designation in \cite[][; H07]{hernandez07b} and in \textit{SIMBAD} (when available), coordinates on the sky, spectral type, equivalent width of H$\alpha$, equivalent width of Li I 6707\AA, total equivalent width of Na I 8183\AA+8195\AA, type (WTTS, CTTS or C/W - see \ref{sec:ctts_classif}), \textit{CVSO} photometry and variability information, 2MASS JHKs photometry, {$\rm T_{eff}$}, $A_V$, location within the Orion association, and luminosity. 
 Because of the large amount of information available for each star, 
the complete optical/near-IR VRIJHK photometry, variability and and low-resolution spectroscopic data is provided in an electronic-only version. 
 There are a few very active CTTS that show many emission lines in their spectra, but we provide here measurements only for H$\alpha$, Li~I, and 
the Na~I 8183, 8195{\AA } doublet (this later spectral feature only for the $1025$ objects observed with Hectospec).

Of the $2064$ TTS, $245$ were identified in Hydra spectra, 
$1025$ in Hectospec spectra, $49$ in the M2FS spectra, 
$724$ in FAST spectra,
and $21$ in SOAR GHTS spectra. 
About 50\% of the CS2 candidates were confirmed as TTS, compared to 
a $\sim 9$\% success rate for the TTS confirmed from the CS3+JH-selected sample.  This result highlights the importance of optical variability as a tool for tracing young, low-mass populations of young stars in regions 
devoid of molecular gas.
Since our variability detection rate is highest for the
bright (V$\lesssim 16$) sample (because of the smaller measurement
errors), we can look at the success rate of
the FAST follow up spectroscopy as an indicator of the best-case
efficiency we can expect from our variability selection technique.
Of the $1235$ variable candidates observed with FAST, $650$ ($\sim 53$\%) 
were classified as TTS. In contrast, only $\sim 21$\% of the stars in the entire FAST sample were labeled as TTS, a number we would expect from 
a conventional single-epoch color-magnitude selection. 
In the combined Hydra and Hectospec sample, $428$ of $951$
PMS candidate variables (45\%) were classifed as TTS, 
a slightly lower success rate compared to the FAST sample, but
consistent with the fact that as we go to fainter magnitudes, we can
detect only those variables which have increasingly larger amplitudes.

\subsubsection{The Na~I 8200 Doublet as a youth indicator}
\label{sec:sodium}

The usefulness of the Na~I lines as surface gravity indicators
has been known since \cite{luyten1923} first showed that
the sodium D doublet (5890,5895\AA) was stronger in dwarfs than in giants.
The low ionization potential of alkali atoms like sodium, that
have a single valence electron, means that they are easily 
pressure-broadened, and therefore the
absorption line strength increases as the gas density gets larger.
However, the Na~I (5890,5895\AA) is strongly affected by TiO absorption
bands in stars later than $\sim$M2, 
and absorption by the interstellar medium may also affect the line strength. 
On the other hand, the 
Na~I subordinate doublet at 8183{\AA } and 8195{\AA } is located
in a region of high S/N in our Hectospec spectra and not significantly affected
by telluric absorption, or by TiO bands up to spectral types as late as M9.
This feature has been used many times to discriminate between
field dwarfs and younger, late type objects 
\citep{martin1996,martin2004,lawson2009,martin2010,lodieu2011,hillenbrand2013,hernandez2014,suarez2017}, though with samples of limited sizes.
With our spectroscopic follow up with Hectospec we have amassed a large number of spectra going out to $\sim 9000${\AA }.
Armed with such a large sample of TTS and field stars, all identified and measured with the same instrumental setup and uniform criteria, we can now explore the behavior of the Na~I(8183,8195\AA) doublet as a youth indicator across the K to M spectral type range in a consistent way.

\begin{figure}[htb!]
\epsscale{1.3}
\centering
\plotone{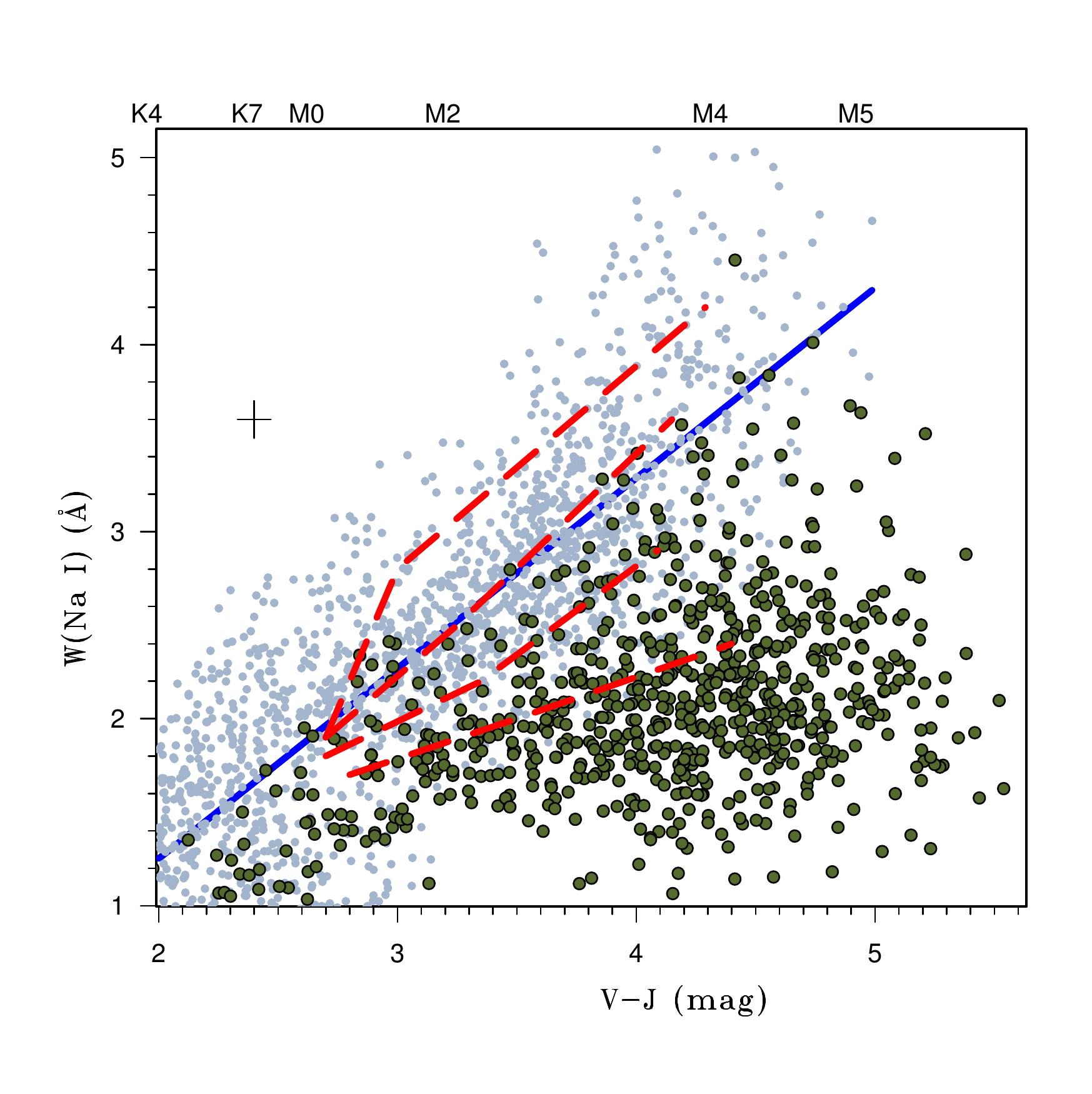}
\caption{Total equivalent width of the Na I 8183,8195{\AA } lines for the new Orion OB1 TTS confirmed with Hectospec spectra, shown as dark green dots.
The small light blue dots correspond to stars classified by us as field stars.
The blue straight line is the least square fit to the distribution of those field stars.
The dashed red lines are, from top to bottom, the 1 Gyr, 100 Myr, 50 Myr and 10 Myr 
isochrones from \cite{schlieder2012}.  A typical errorbar is indicated.
}
\label{ewna}
\end{figure}

In Figure \ref{ewna} we show the total equivalent width of
the Na~I(8183,8195\AA) doublet as a function of the observed V-J color\footnote{because the overall extinction of our sample is small, it is
appropriate to use V-J without correcting for reddening; see Table \ref{mastertable}.}
for $3441$ stars: $1025$ TTS TTS (dark green dots) and $2416$ field stars (light blue dots). For every object the equivalent widths of each of the Na I lines were measured both with SPTCLASS \citep{hernandez2017}
and interactively with the \textit{splot} utility in IRAF, which allowed us to
obtain an estimate of the measurement uncertainty, indicated by the errorbar in the figure. 
It is important to point out that the TTS were identified based solely on the presence of H$\alpha$ in emission \textit{and} Li I 6707{\AA } strongly in absorption, above the Pleiades level, as described in \ref{sec:membership}. Stars lacking Li I were classified as field stars. A least squares fit to the distribution of field stars is shown as a blue straight line.

As expected from the V$\sim 13$ saturation limit of the photometry, our
color selection, the relatively low extinction, and the nature of the stellar mass function 
in the thin disk of the Galaxy \citep{robin2003},
the majority (65\%) of the field stars in our sample are K and M type dwarfs \citep[see \ref{sec:spectra} below, and also ][]{downes2014}, with a median spectral type of M0 and median $A_V=0.64$ mag.
Among the late type field stars,  23\% were classified as active K and M type dwarfs with H$\alpha$ in emission (dKe and dMe stars);
this subset is characterized by a later median spectral type of M3.

Despite the large scatter of Na~I equivalent widths at any given V-J, it is readily apparent that the bulk of the TTS lie distinctly below the distribution of field stars.
{
To quantify this effect, we show
in Figure \ref{ewna} isochrones 
calculated by  \cite{schlieder2012}, using the \cite{siess00} evolutionary
tracks and the PHOENIX model atmospheres
\citep{hauschildt1999,rice2010}. The
mean value for the field stars in
Figure \ref{ewna} matches very well the
100Myr isochrone.
99.9\% of the TTS fall below this lines, and
}
most of the spread seen
in the field stars for spectral types later than $\sim $M0 is encompassed within the
1 Gyr and 50 Myr isochrones.
It is also noticeable that the TTS and field stars tend to separate in
the W(Na~I) vs Teff plane, only for V-J$\ga$ 3,
corresponding to spectral types later than $\sim$ M1.5.

Because most of the K and M-type field dwarfs are located 
in front of Orion, 
they fall above the main sequence when assumed at the distance of the star-forming region, thus mimicking PMS stars.
Late type WTTS differ from their main sequence K and M-type counterparts only in the presence of the Li I 6707{\AA } line in absorption, 
and the weaker Na~I(8183,8195\AA) doublet. Otherwise they are identical, 
having the same color, 
similar median amplitude of photometric variability (see \S \ref{sec:variability} and  Figure \ref{variability_AmpV}), 
same spectral types and weak to  modest emission in H$\alpha$. 
In the absence of a measurement of W(Li~I),
a star located in the PMS locus of an optical/near-IR CMD,
with a mid-K to M spectral type, H$\alpha$ in emission,
and a small W(Na~I), can be considered a strong TTS candidate. Of course, other characteristics like an IR excess, strong X-ray emission, proper motion and radial velocities consistent with membership to a given association, 
add to the classification of an object as a PMS star. 
However, an IR-excess will be
absent in most WTTS, which constitute the bulk of the population in
off-cloud regions of OB associations \citep{briceno08}, 
and strong X-ray emission decays slowly during the first several hundred million
years in G-M dwarfs \cite{briceno97,ingleby2011}. 
Proper motions and radial velocities
will be unavoidably contaminated by field stars that happen to share the same
space motion as a PMS member of a given association. Therefore, unless an X-ray emitting, or proper motion/radial velocity candidate member
source is found to be a dKe or dMe star with W(Na~I) well below the 50 Myr isochrone, it still could be a main sequence field star with age between 
$\sim 100$Myr to a few Gyr. 
In the end, the detection of Li I at 6707\AA, as shown in Figure \ref{ewli_ob1ab}, 
remains the crucial criterion to confirm the PMS nature of K and M-type dwarfs. This is why spectroscopy is essential for a reliable census of young stars, in particular for identifying the mostly diskless, more evolved WTTS found in regions devoid of molecular clouds.

\subsection{Determination of Spectral Types}
\label{sec:spectra}

Our low resolution spectra provide the large wavelength coverage necessary to
measure several temperature-sensitive features like the various TiO bands 
from $\sim 4500 - 8000${\AA }, which are characteristic of late K and M-type stars.

\begin{figure}[hbt!]
\centering
\includegraphics[scale=0.40]{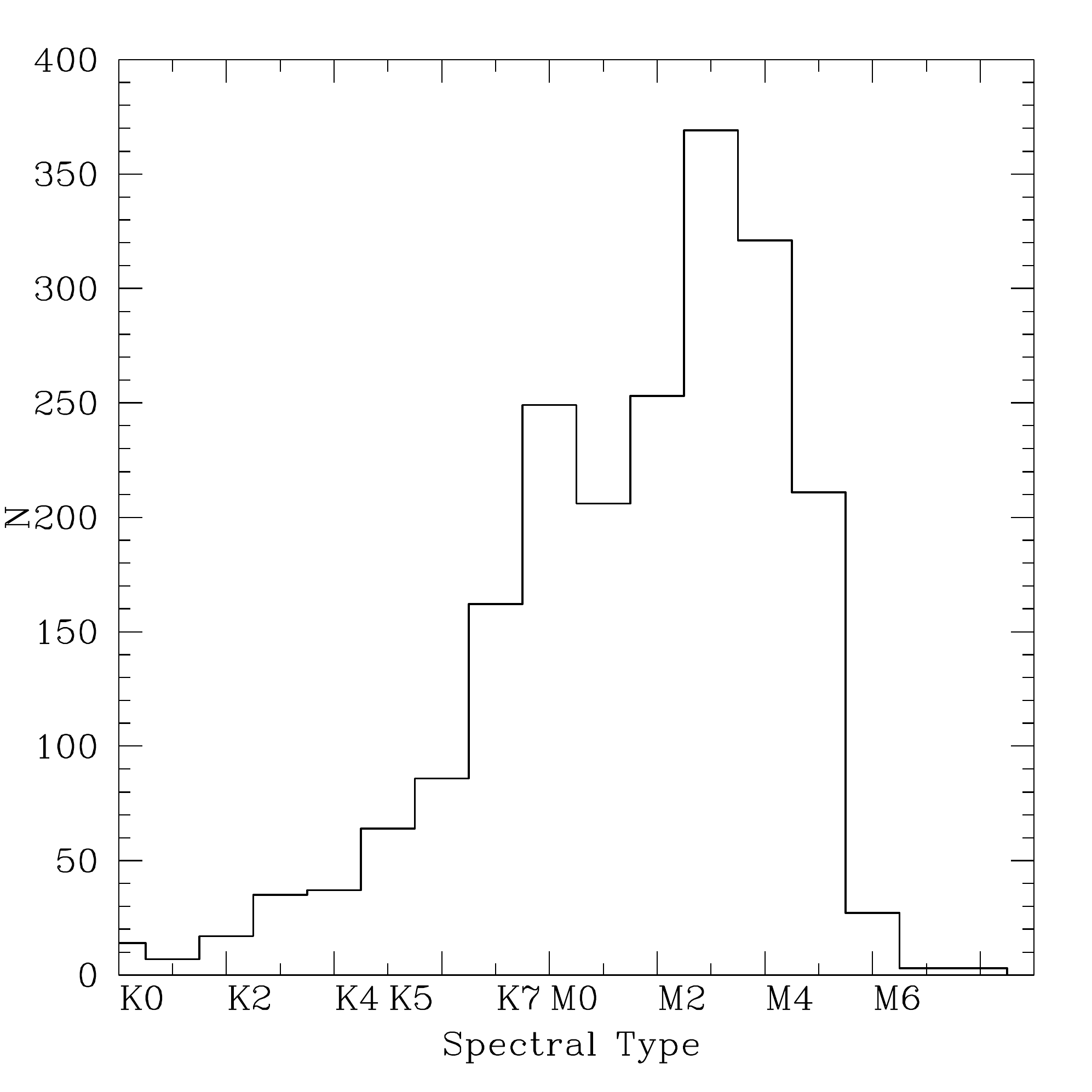} 
\caption{Distribution of spectral types of the 2064 \textit{CVSO} TTS.
}
\label{sptypes}
\end{figure}

We classified our sources with the SPTCLASS package \citep{hernandez2017}.
On average, the
uncertainty in spectral type is roughly 1 subclass, but this depends largely on the S/N of the spectrum. The only object in our final TTS table that lacks a spectral type is \emph{CVSO}-157, an extremely active CTTS located in 1b, that shows essentially no absorption features in its spectrum, and hence
no reliable spectral type could be determined, so
it was classified as ``C'' (a ``Continuum'' object).
In Figure \ref{sptypes} we show the distribution of spectral types of our full Orion TTS sample, plotted with the solid black line histogram. The distribution peaks at M3, and the decline at later spectral types is largely due to the lower completeness for increasingly fainter stars in our spectroscopic follow up. 

\subsection{Classification of accreting and non-accreting young stars: 
definition of the new C/W class.}
\label{sec:ctts_classif}

An important result from our extensive spectroscopic observations is the ability to classify in a systematic way a large number of Orion association young members according to the strength of the H$\alpha$ emission line. 
In this section we propose a new type of TTS, the C/W class, objects with
H$\alpha$ emission strength intermediate between that of CTTS and WTTS. We show that the C/W class, defined from a purely spectroscopic criterion, also exhibits
intermediate behavior between CTTS and WTTS in other properties like IR excesses and variability.

It has long been recognized that H$\alpha$ emission is a telltale signature of accretion in solar-like PMS stars (see \cite{hartmann2009}.
Objects showing very strong H$\alpha$ lines are classified as accreting CTTS, while WTTS exhibit weak emission, consistent with levels of chromospherically active young stars. The qualitative idea remains a useful classification scheme, specially because by estimating the fraction of accreting stars across different regions and over a range of ages, we can infer fundamental properties like circumstellar disks lifetimes and the effects of the environment on such disks. However, the quantitative criterion to separate the two classes of objects has evolved significantly. The original threshold of 10{\AA } proposed by \cite{herbig1988} was revised by \cite{white03} and \cite{barrado2003}, based on the fact that the equivalent width of H$\alpha$ due to chromospheric emission is a function of spectral type. 
This is caused by a contrast effect between the emission in the line and the underlying photosphere, such that for equally strong intrinsic line fluxes, the H$\alpha$ equivalent width would be larger in an 
M-type star than in a K-type star, because of the weaker photospheric continuum near 6500{\AA } in the M star. This had the implication of reducing the number of accreting TTS (or the number of CTTS) at later M spectral types.

\begin{figure}[htb!]
\epsscale{1.22}
\centering
\plotone{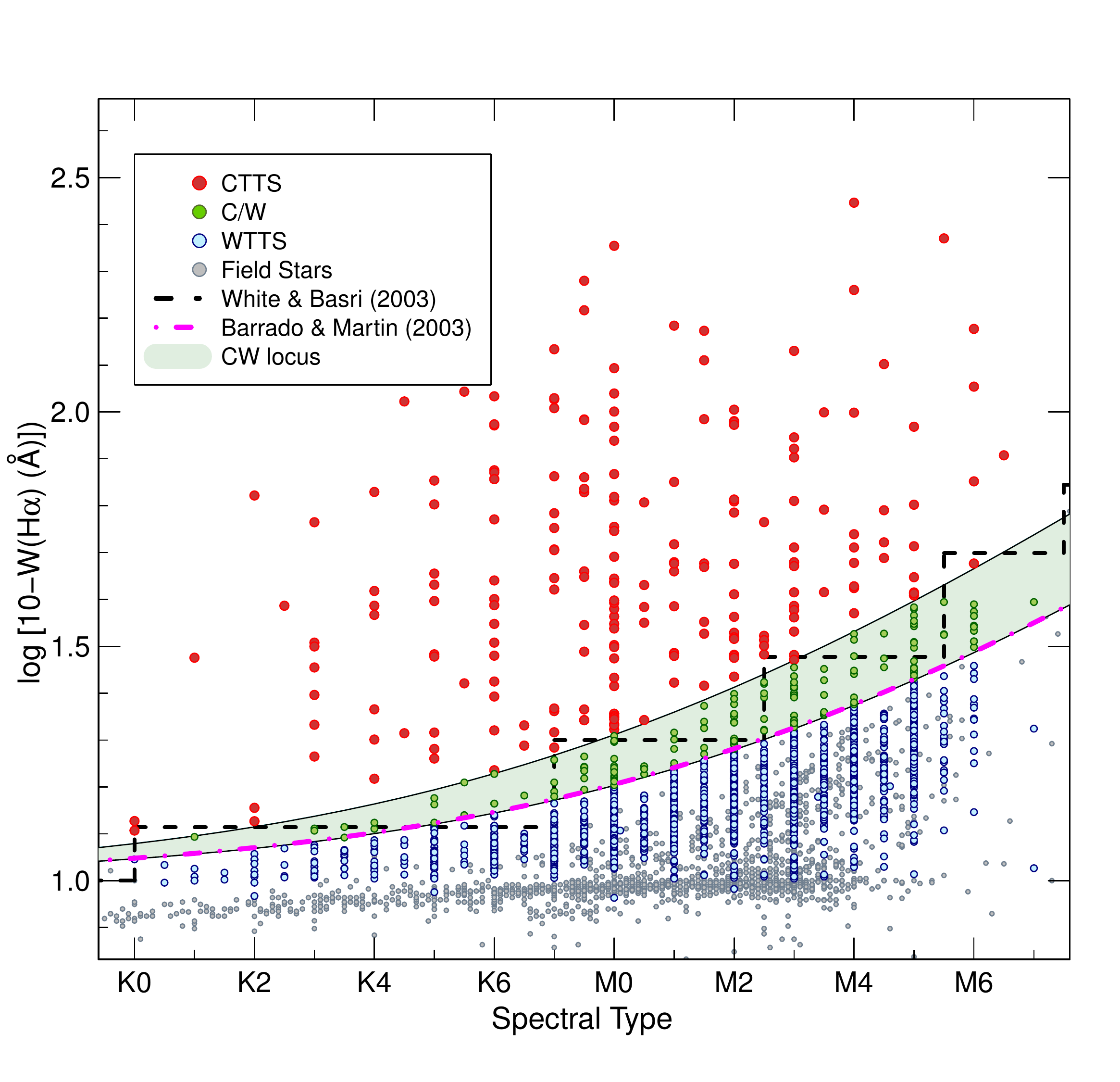}
\caption{Logarithm of the equivalent width of H$\alpha$ as a function of spectral type 
for all the  Orion OB1 members, shifted by 10\AA. 
We plot CTTS as large red dots and WTTS as smaller blue dots. 
For reference, we also plot as grey dots stars classified as field 
objects in our low-res spectra.
The separation between CTTS and WTTS as defined by \cite{white03} 
is shown with the dashed line, while the criterion adopted by \cite{barrado2010} is plotted as a dash-dot magenta line. 
The boundary we adopt here between CTTS and WTTS is shown by the pale green 
region, which we define as the transition between WTTS and CTTS. 
Objects falling within this area of the plot are classified as C/W objects, objects below are WTTS and objects above are CTTS. The height of this transition region was adopted by adding in quadrature the mean variability
of W(H$\alpha$) in CTTS and WTTS at each spectral type, to the measurement uncertainty.}
\label{wha_spn}
\end{figure}

\begin{figure}[htb!]
\epsscale{1.22}
\centering
\plotone{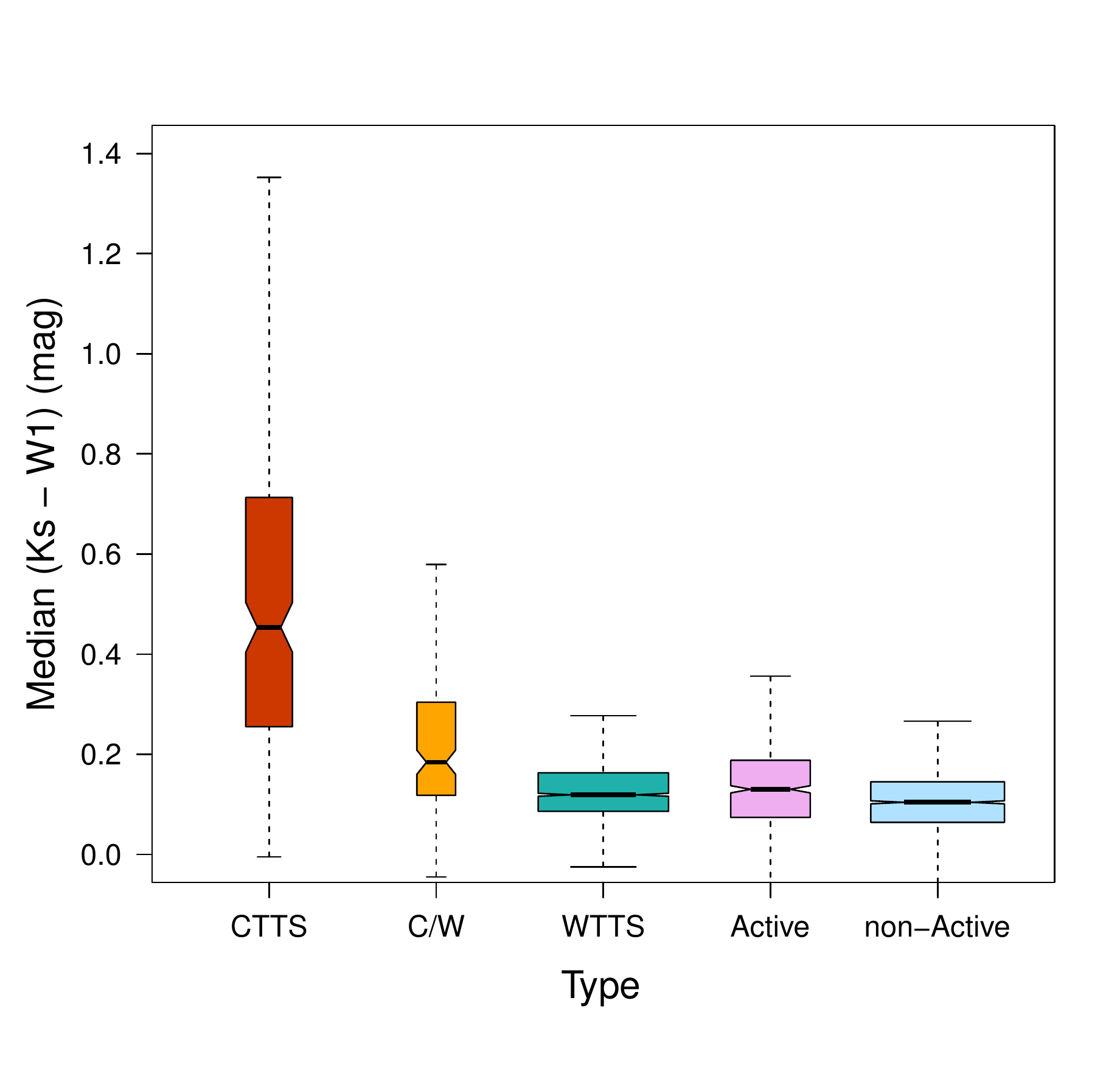}
\caption{Median $Ks-W1$ colors for each type of TTS, and for the field star sample. 
We subdivide the field stars between active (those with H$\alpha \lambda 6563$ in emission) and non-active (H$\alpha$ in absorption).
The first and third quartiles are indicated with the dashed vertical bars, and the width of each box is proportional to $\sqrt N$, where $N$ is the number of stars in each group. 
The width of the notches in each box correspond to the 95\% confidence interval, such that if the notches in two boxes do not overlap it is considered their medians differ \citep{chambers1983}.
}
\label{KW1_colors}
\end{figure}

However, the new criterion introduced by \cite{white03} and \cite{barrado2003} does not account for the fact that H$\alpha$ emission is variable among PMS stars. This variability is greater for the most active, strongly accreting stars, but is present even in the WTTS. The equivalent width of H$\alpha$ can vary by up to factors of a few in CTTS and up to $\times 2$ in WTTS \citep[e.g. ][]{rugel2018}. Therefore, a star classified as CTTS in one observation may be deemed as a WTTS at some other epoch. Some CTTS can go through quiescent phases in which they would be confused with a WTTS, unless high resolution (R$\ga 10000$) spectroscopy is used to resolve the profile of H$\alpha$ and look for
accretion signatures like broad line wings, or red-shifted absorption \citep{white03}. 
To account for such time-variable emission we introduce here a new class of object, the CTTS-WTTS stars or C/W, which are defined as TTS falling in the C/W locus in a H$\alpha$ equivalent width versus spectral type diagram, as shown in Figure \ref{wha_spn}.
The C/W locus is defined as the region of the diagram contained within the following expressions:

\begin{equation}
W(H\alpha)_{Upper}^{C/W} = 10^{(0.09*SpT-5.100)}
\end{equation}

\begin{equation}
W(H\alpha)_{Lower}^{C/W} = 10^{(0.09*SpT-5.345)}
\end{equation}        
        
where the spectral type (SpT) is defined numerically as 
G0=50, G1=51, G2=52 ... G9=59, K0=60, K1=61 ... 
K7=67, M0=68, M1=69, M2=70 ... M6=74. 

We derived the above expressions based on multiple measurements of $W(H\alpha)$ in a set of $95$ TTS with spectral types K4 to M6, 
for which we had two or more spectra obtained at different epochs.
At each spectral type we considered only stars with $W(H\alpha)$ 
values close to the limit between CTTS and WTTS defined by \cite[][; black dashed line in Figure \ref{wha_spn}]{white03}. We then adjusted the
width of the C/W locus to encompass the majority ($> 85$\%)
of the range of variation of $W(H\alpha)$ at each spectral type for this subset of TTS close to the CTTS/WTTS dividing line. 
The classification for each of the $2064$ TTS is indicated under the column ``Type'' in Table \ref{mastertable}: ``C'' for CTTS, ``W'' for WTTS, and ``CW'' for the newly defined C/W 
objects.
Stars in the C/W category may
represent objects
evolving from an active CTTS accretion phase to a non-accreting WTTS stage. We speculate that this group is likely composed of a mix of objects that are accreting at modest or low levels, constituting the weak tail of the CTTS, and a few objects in a quiescent stage between periods of enhanced accretion.
If the newly defined C/W class are indeed TTS at a stage intermediate between that of actively accreting CTTS and the WTTS, which are thought to have ceased accreting from a circumstellar disk \citep[though a fraction of WTTS likely retain passive, non-accreting disks; ][]{hernandez2014,natta2004,nguyen2009a,nguyen2009b}, we would expect this type of object to also show other properties intermediate between the CTTS and WTTS.

A well known indicator of the presence of a circumstellar disk is IR excess emission, originating in the warm dust heated by irradiation from the central star \citep{hartmann2008}.
In Figure \ref{KW1_colors}  we show the boxplot with the median ($Ks - W1$) color for each type of TTS, and first and third quartiles, where W1 is the \textit{WISE} [3.6] band magnitude, obtained by matching the \textit{CVSO} sample with the \textit{ALLWISE} catalog \citep{wright2010,mainzer2011}.
WTTS have an average $\overline{Ks-W1}= 0.14\pm 0.10$ with $median\, (Ks-W1)= 0.12$; C/W have $\overline{Ks-W1}= 0.24\pm 0.19$ with a $median\, (Ks-W1)= 0.19$; CTTS have $\overline{Ks-W1}= 0.50\pm0.32$ with a $median\,(Ks-W1) = 0.46$. We have also derived average and median $Ks-W1$ values for our field star sample, which contains 2814 photometric candidates classified as field stars in our spectra. Of these, 643 are stars with the H$\alpha \lambda 6563$ line in emission ($\sim 27$\%), almost all are dKe and dMe stars; we call these objects the ``active'' subset of field stars. These young main sequence field stars should have ages between $\sim 50-60$ Myr up to a few hundred Myr, as would be expected from the recent history of star formation in the solar vicinity \citep{briceno97}.
The active field stars have a $\overline{Ks-W1}=0.13\pm 0.16$ with a median value of 0.13. The ``non-active'' field stars have $\overline{Ks-W1}=0.11\pm 0.10$ with a median of 0.10.  Clearly, the C/W have an intermediate $Ks-W1$ color between that of WTTS and CTTS, while WTTS have a $Ks-W1$ color similar to the sample of active field stars, as would be expected from young stars which have largely lost their disks and exhibit only enhanced levels of chromospheric activity. The median $Ks-W1$ color for the ensemble of C/W stars is an independent measure that gives support to our suggestion of these stars being at an intermediate evolutionary stage between CTTS and WTTS;
in an upcoming paper we will discuss in further detail the IR properties of the C/W sample.
High resolution optical spectroscopy of C/W stars should provide further insight into their accretion state, allowing to look for accretion-related features in the H$\alpha$ profile. Also, intermediate to high resolution near-IR spectra will provide profiles for features like the He I $\lambda 10830$ line, which has been shown to be a sensitive probe of 
low levels of accretion (Thanathibodee et al. 2018, submitted).


\subsection{The \textit{CVSO} sample in the literature}
\label{sec:simbad}

\subsubsection{Crossmatch with SIMBAD}

Of our list of $2064$ spectroscopically confirmed TTS,
$1485$ (72\%) are characterized here for the first time, 
as newly identified members of the Orion OB1 association;
$787$ have a match within 3 arcsec in the 
SIMBAD database (the average separation is $0.31\arcsec \pm 0.33\arcsec$) and are classified as a star.
However, only $434$ of the matches (55\%) 
have a SIMBAD type ``TT\*'' (T Tauri star), ``Y\*O''
(Young stellar object), ``pr\*'' (pre-main sequence star), ``Or\*'' (Variable star of Orion type), 
``\*iC'' (Star in cluster) or ``\*iA'' (Star in association), {\sl and a spectral type}. 
These objects we adopt as previously known, confirmed young members of Orion OB1. In Table \ref{newttsinfo} we provide the SIMBAD identification and type for matching objects.

Among the $434$ TTS in SIMBAD, $226$ were published by us 
in B05, B07b, \cite{biazzo2011b}, another 77 in \citet{downes2014} 
and 53 by \cite{suarez2017}. 
The other $78$ are from various other studies (see \ref{sec:spec_studies}).
The remaining $483$ SIMBAD objects
are stars which lacked spectral type determination and Li I or Na I measurement. They are either candidate
PMS stars from other studies such as \citet[][212 objects]{hernandez07a}, \citet[][82 objects]{megeath2012},
or objects classified as an emission line star,
X-ray source, variable star, flare star, irregular variable, star in cluster, 
or star in association, but lacking membership information.
For these, we provide here, for the first time, spectroscopic confirmation of
their youth.

Combined with our previous work (B05,B07b), this constitutes the largest sample 
of low-mass PMS stars with both multi-band photometry and optical spectra, obtained in a consistent and systematic way in all Orion, even when compared to the intensively studied ONC \citep[H97;][]{carpenter01,slesnick04,tobin2009,hillenbrand2013}.

\subsubsection{Comparison with objective prism and photometric surveys}
\label{sec:surveys}

The Orion region has been studied extensively since it was recognized as
an OB association by \cite{blaauw64}. However, few studies have undertaken
the task of mapping in detail the young stellar population across the entire complex.  In the 90s, \cite{wiramihardja1989,kogure1989,wiramihardja1991} searched $\sim 100$ square degress for H$\alpha$ emission line stars using photographic plates on the 1m Kiso Schmidt telescope equipped with an objective prism. They published 759 candidate young stars, but without membership confirmation; 102 are confirmed as members here.
Newer objective prism studies like that of \cite{szegedi-elek2013} have focused on the ONC, finding 587 candidate young stars, of which 35 are in our \textit{CVSO} catalog.

Recent work has revealed large numbers of PMS stars in Orion, 
but concentrating on the youngest populations on the A and B molecular clouds.
Such is the case of the Spitzer study by \cite{megeath2012}, which covered the entire Orion A cloud and most of the B cloud. 
They identified nearly $3500$ young stars
based on their spectral energy distributions and infrared variability;
82 of the these sources match our \textit{CVSO} catalog, of which 80\% are classified by us as CTTS, a result consistent with a search for objects with IR-excess emission.
\cite{alves2012} presented a photometric study of the region surrounding the ONC, but their proposed population is based on photometric candidates only;
118 of their sources are in our TTS list. \cite{da-rio2012} carried out a deep photometric survey of $\rm 0.25\, deg^2$ in the ONC region, proposing 1750 objects as candidate young members; 26 match our \textit{CVSO} catalog, of which 20 are CTTS. \cite{pillitteri2013} conducted an X-ray study of the L1641 cloud; they found 716 candidate young stars, of which 16 have a counterpart in our \textit{CVSO} TTS list, with 12 being WTTS 3 CTTS and 1 C/W. 
\cite{sanchez2014} carried out an UV-based selection of candidate young stellar objects across $\rm \sim 400\, deg^2$ encompassing all Orion, using data from GALEX. They identified 111 candidate PMS objects, of which 11 are found in our list. As expected from the UV-excess selection, all but one are CTTS.
\cite{spezzi2015} used the \textit{VISTA} Orion Mini-Survey to identify 186 candidate
young objects in the B cloud, including the embedded clusters NGC 2068 and 2071. Ten of their targets match our \textit{CVSO} catalog, and all but one are CTTS, as would be expected from a selection biased toward IR-excess sources.
In the Orion OB1b sub-association \cite{kubiak2017} identified 789 candidate young members, based on optical/infrared photometry, with no spectroscopic confirmation. Unfortunately their target table is not yet available in either SIMBAD, Vizier or the CDS data services, so we could not cross match their list with our \textit{CVSO} sample. For all the objects in these studies in common with our \textit{CVSO} catalog, we provide here spectroscopic membership confirmation and characterization of spectral type, among other stellar parameters.

Among the few studies targeting the older populations, \cite{vaneyken2011} used the Palomar Transient Factory survey to look for eclipsing binaries in the 25 Ori cluster. They presented 16 candidate young stars, of which 6 have a counterpart in the \textit{CVSO}. One is \textit{CVSO}-35, already known from B05, and the other 5 are confirmed here as TTS. 

\subsubsection{Comparison with spectroscopic studies}
\label{sec:spec_studies}


Work based on optical low-resolution spectroscopic membership confirmation
has largely focused on the ONC region and its surroundings in the A cloud.
H97 carried the first landmark study of the ONC, producing spectral classification for 675 stars. Then \cite{hillenbrand2013} added 254 new stars
and reclassified many of the H97 stars;
27 of their sources are found in our TTS list.
\cite{hsu2012} surveyed $\rm \sim 3.25\, deg^2$ in  the L1641 dark cloud, located within the Orion A cloud, south of the ONC. They confirmed $864$ low-mass PMS members, $723$ of them in common with the  \cite{megeath2012} survey. A total of 22 objects have a match in our \textit{CVSO} catalog.

After the ONC, the $\sigma$ Orionis cluster is probably the most studied stellar aggregate in Orion. Among the most recent comprehensive spectroscopic studies is that of \cite{hernandez2014}. In our \textit{CVSO} catalog we have 16 sources in common with their list of 340 confirmed members.  Another cluster
that has been the subject of several spectroscopic studies is 25 Ori.
\cite{downes2014} used deep, coadded $I_C$-band photometry from the \textit{CVSO}, combined with \textit{VISTA} data and follow up spectroscopy, to search for very low-mass TTS and young brown dwarfs; 65 of their 77 new members are in our catalog. More recently, \cite{suarez2017} combined optical photometry from the \textit{CVSO} with spectroscopy from SDSS-III/BOSS to identify 53 new members in an area of $\rm \sim 7 \, deg ^2$ roughly centered on the cluster. We find 25 TTS in common with their study.

The proposed foreground population of NGC 1980 has been studied spectroscopically by \cite{kounkel2017b}, who obtained spectra for 148 young stars. Nine objects are in common with our work; spectral types determined in both studies compare well.
\cite{fang2017} also investigated the ``foreground'' population to NGC 1980 proposed by \cite{alves2012} and \cite{bouy2014}. They obtained spectra and stellar properties for 691 young stars in $\sim 12$ square degrees in this region. We found 24 stars in common with our list. 
In general, few objects in our \textit{CVSO} catalog match these various studies in the ONC and the A cloud. The main reason is that our south survey limit is close to $\delta \, 0 -6^\circ$, and we
avoided the central region of the ONC, so there is little spatial overlap with
these various studies.

After we take into account repeats between the aforementioned studies, there are only 78 objects with existing spectroscopic membership confirmation among the 2064 \textit{CVSO} TTS ($\sim 4$\%), excluding the $356$ sources already published in our previous work \citep[B05,B07b;][]{biazzo2011b,downes2014} and in \cite{suarez2017}. Aside from the stars in 25 Ori and $\sigma$ Ori, all the other sources are located south of $\delta = -6^\circ$. Therefore, our survey provides the basis for a consistent and comprehensive characterization of the young stellar population across the Orion OB1 association, north of $\delta = -6^\circ$ and south of $+6^\circ$.

With such a large number of confirmed TTS members we have 
the means to derive, for the first time, robust demographics of the populations of low-mass PMS stars in this star forming complex, across a range of ages and differing environments, with an emphasis on the largely unknown off-cloud component.

\subsection{The Large Scale Spatial Distribution of Young Stars across Orion OB1}
\label{sec:spatial}

With the \emph{CVSO} photometric catalog, complemented with spectroscopic confirmation of a substantial fraction of PMS photometric candidates, 
we can now examine the spatial distribution of both candidate and confirmed low-mass PMS members.
In this section we start with a qualitative exploration of the large scale spatial distribution of the photometrically-selected PMS population across Orion OB1, which allows
us to recognize the major groups and spatial features within each subassociation. Then, in the following sections, we use the properties of
the 2064 confirmed members to characterize the the general population in each region, and conduct demographic studies of the groups and clusters within.

\begin{figure*}[htb!]
\begin{center}
\includegraphics[angle=270,scale=0.65]{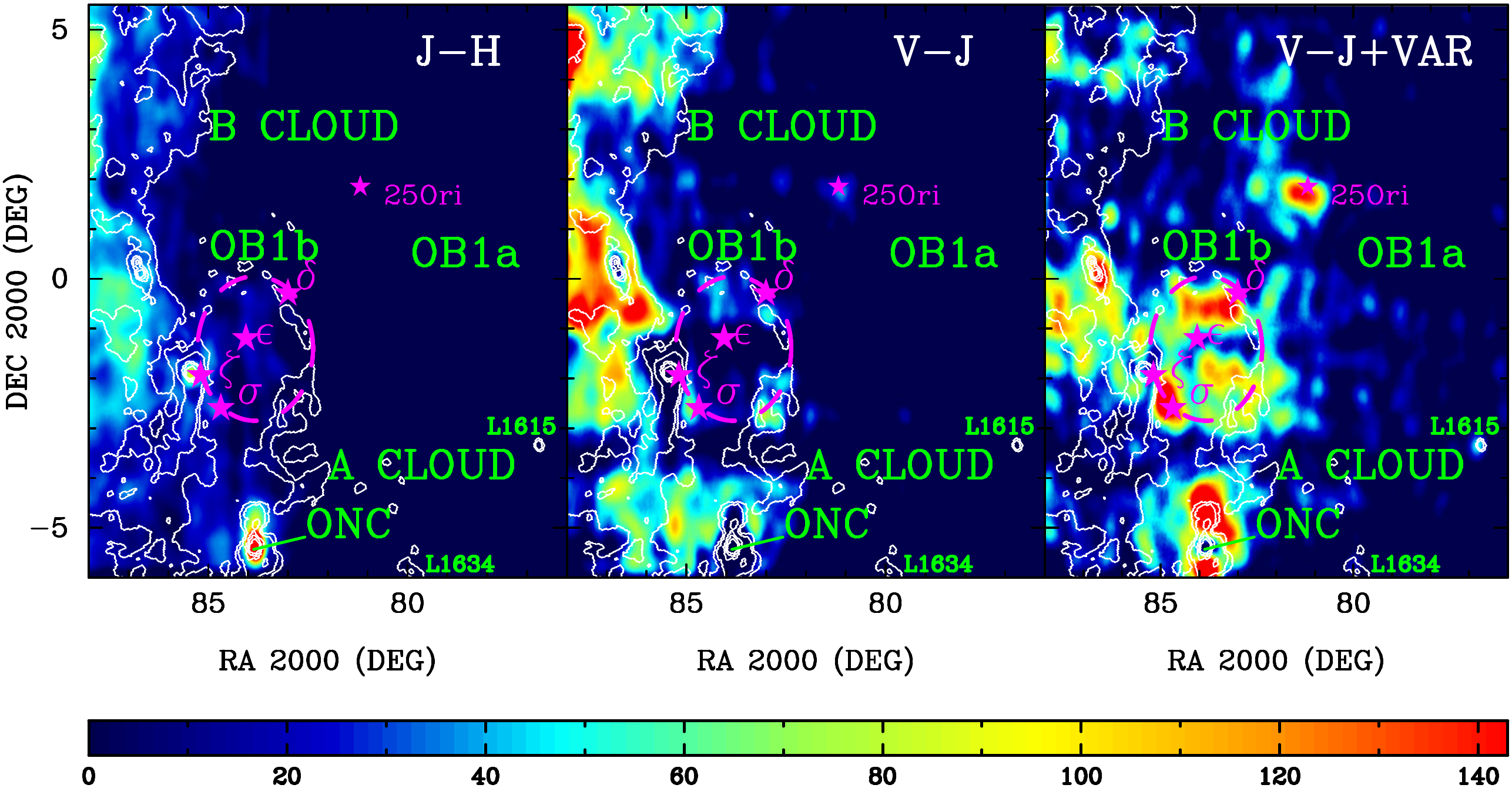}
\caption{Surface density of PMS candidates selected in different 
color-magnitude diagrams.
{\bf Left:} surface density ($\rm No. stars/4 arcmin^2$) of PMS candidates in Orion OB1, 
selected as objects located above the main sequence \citep[][at 440 pc]{siess00} in the 
2MASS J vs J-H diagram.
The maximum value corresponds to a density of $143\ stars/deg^2$.
The Orion A and B clouds are indicated, as well as the outlying clouds L1615 and L1634.
The three belt stars and $\sigma$ Ori are located within our definition of the OB1b subassociation,
indicated by the dashed line oval. Contours correspond to Av=0.5, 1, 5, 10, 20 from
the IRAS 100$\mu$m dust map of \cite{schlegel98}.
{\bf Middle:} surface density ($\rm No. stars/4 arcmin^2$) of PMS candidates 
selected as objects located above the main sequence \citep[][at 440 pc]{siess00} 
in the optical/near-IR V vs V-J and V vs $\rm V-I_c$ diagrams
(the ONC appears as a hole because we "cut out" this region of dense nebulosity in the \emph{CVSO} catalog).
{\bf Right}: surface density ($\rm No. stars/4 arcmin^2$) of PMS candidates 
selected as objects {\sl detected as  variable in the V-band and} 
located above the main sequence \citep[][at 440 pc]{siess00} 
in the optical/near-IR V vs V-J and V vs $\rm V-I_c$ diagrams.
}
\label{spatial1}
\end{center}
\end{figure*}

We start here by highlighting the importance of combining optical and 
near-infrared wavelengths, and particularly  the advantage of including
photometric variability, in searches for PMS populations in regions with
little or no extinction, areas devoid of molecular cloud material.
In the left panel of Fig. \ref{spatial1} we show the surface density of 
low-mass PMS candidates selected from the 2MASS J vs J-H CMD. In the middle and right panels respectively, we show 
the surface density of low-mass PMS candidates selected from the combined
optical/near-IR \emph{CVSO}-2MASS catalog, and the 
surface density of optical/near-IR PMS candidates selected as 
{\sl variables} in the V-band data of the multi-epoch VRI survey.
These maps were created from the two-dimensional histograms of the spatial distribution of PMS candidates in the 2MASS catalog, 
and for the combined PMS candidate sources in  the \textit{CVSO}-2MASS cross-matched catalog; 
these were then smoothed with a $2^\circ$ diameter Gaussian kernel.
Though its been long known that the near-IR bands are an ideal tool for
locating embedded populations of young low-mass stars, 
as evidenced by the strong peaks in the surface density of objects 
at the locations of the ONC and the NGC 2024 cluster (next to $\zeta$ Ori),
the JHK-only selection fails to detect any off-cloud population of PMS stars,
except for a faint trace of a distribution of young stars inside the relatively younger OB1b region; however, it is important to note that recent work has applied improved techniques for finding embedded young populations in near-IR data  \cite[e.g.][]{lombardi2017}, that can also detect the densest overdensities in the older OB1b and OB1a regions.
Still, when an optical band is added, the expanded color range offers an advantage for separating young stars in CMDs. The combined optical/IR selection (middle panel) starts to reveal, albeit still faintly, 
the existence of an off-cloud population of young low-mass stars, 
appearing in the form of overdensities such as the 25 Ori cluster \cite[B07b;][]{downes2014,suarez2017}, which is largely invisible in the 2MASS-only map.
The distribution of young stars within the OB1b subassociation is now more evident, and also shows hints of structure; south of OB1b 
an extended ``halo'' of PMS stars appears surrounding the ONC.

When the photometric variability is folded in with the VIJ color-magnitude diagrams
as a selection criterion, a new and completely different picture emerges. 
First, a widely spread low-mass, young stellar population is seen to extend 
well beyond the confines of the molecular clouds, as initially suggested by \cite{briceno01} and B05 within a smaller area, were only 25 Ori was apparent. Now, with the full survey area analyzed, it is clear that the young stellar population in the OB1a subassociation extends throughout most of the star forming complex.
The large scale spatial distribution of this low-mass PMS population can now be used to define the west edge of OB1a, at
roughly $\rm \alpha_{J2000}= 5^h16^m$, running from north to south the combined length of the A and B molecular clouds. 
We can portray this boundary as the fossil imprint of the general star formation event that formed Orion OB1a $\sim 10$ Myr ago.
Second, it is evident that in addition to the widely spread, 
low surface density population
of young stars (the ``field'' population of the OB1a subassociation),
there is a significant degree of substructure across all of Orion OB1. 

\begin{figure}
\includegraphics[angle=270,scale=0.45]{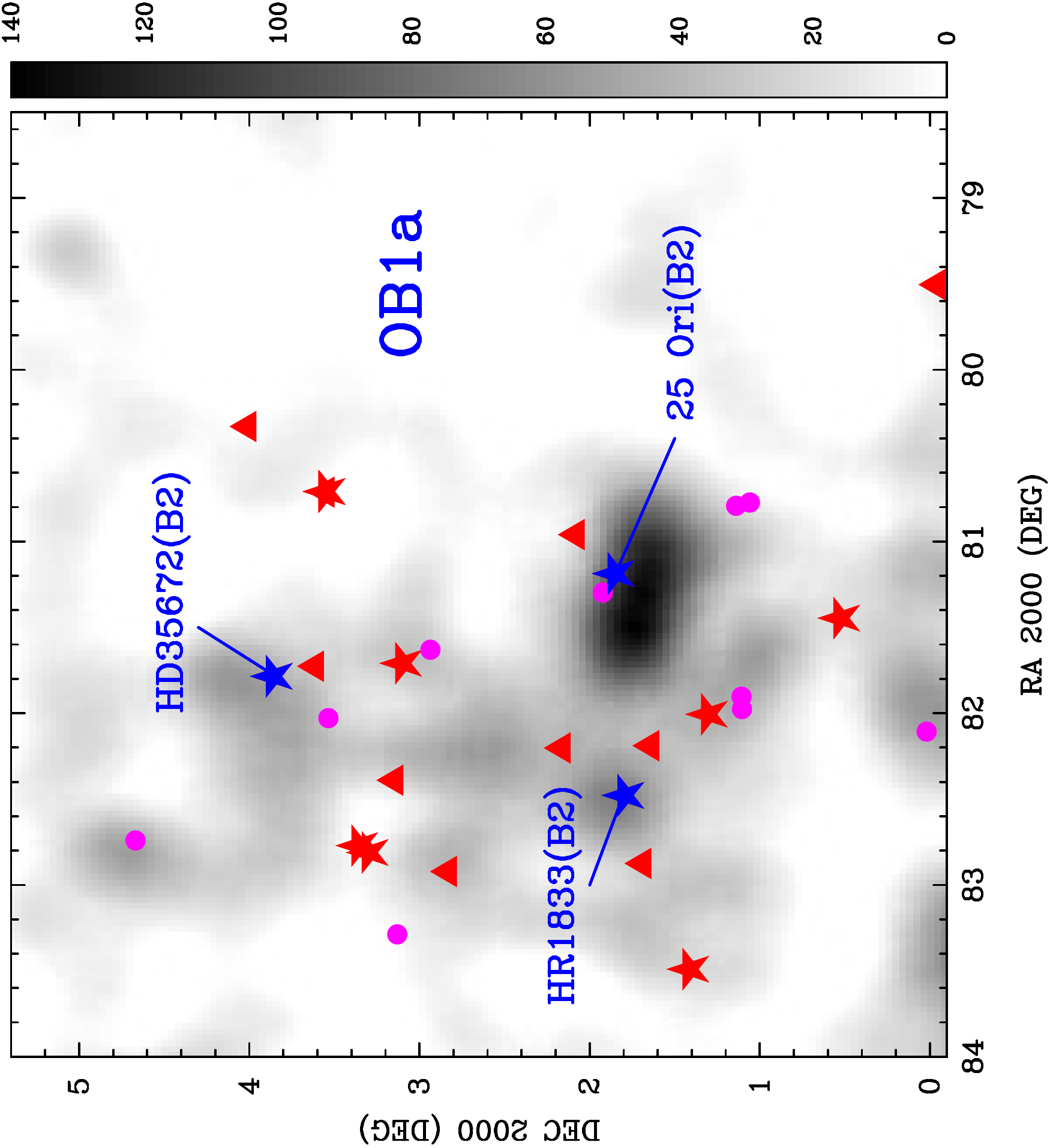}
\caption{Overdensities of candidate PMS sources in the OB1a region surrounding
the 25 Ori cluster.
The background greyscale map is the surface density of photometric
candidate PMS sources as described in Figure \ref{spatial1}. Red stars correspond to early type stars of
B1-B2 spectral type; red triangles to B3-B5 stars, 
and magenta dots to B6-B9 stars.  We have highlighted the
B2 stars 25 Ori, HR 1833 and HD 35672 with large blue stars.
}
\label{overdensities_1a}
\end{figure}

In the OB1a subassociation there are several evident stellar clumps
(Figure \ref{overdensities_1a}). 
Two were already shown in Figure 7 of \cite{briceno08} (despite the claim by \cite{lombardi2017} of having
discovered the overdensities of young stars in OB1a). 
The most prominent of these stellar density enhancements is
the 25 Ori cluster \citep[B05; B07b;][]{downes2014,suarez2017}, with a
clear elongation in the general E-W direction. This is consistent
with the findings by \cite{zari2017}, who noticed an extension in the
northward direction in galactic coordinates, corresponding roughly
to an eastward direction in equatorial coordinates.
The other overdensity is located just east of 25 Ori, at
$\rm \alpha_{J2000}= 5^h30^m$, $\rm \delta_{J2000}=+1^\circ54'$, 
($\rm \alpha_{J2000}= 82.5^\circ$, $\rm \delta_{J2000}=+1.9^\circ$), 
and $4\arcmin$ north of the B2 star HR 1833 (HD 36166). This 
structure may be present very faintly in Figure 15 of \cite{lombardi2017}, 
and in Figures 3, 8 and 9 of \cite{zari2017}, in both cases 
with much lower spatial resolution;
in fact, it is likely unresolved within the eastward extension of the
\cite{zari2017} maps.
However, \cite{lombardi2017} do not discuss it, because, 
in their own words, the analysis of the substructure in their 
density map of OB1a and OB1b was beyond the scope of their study, 
and \cite{zari2017} do not mention it, limiting themselves to 
recognizing the northern (in galactic coordinates) extension to 
the 25 Ori overdensity.
Our more detailed maps show that the HR1833 density enhancement is
a feature distinct from the distribution of sources that can be
associated with the 25 Ori cluster.
Both the 25 Ori cluster and what we will call henceforward the HR 1833
cluster, have been catalogued as clusters by \cite{kharchenko2005},
as ASCC 16 ($\rm \alpha_{J2000}= 81.15^\circ$, $\rm \delta_{J2000}=+1.80)$, 
and ASCC 20 ($\rm \alpha_{J2000}= 82.18^\circ$, $\rm \delta_{J2000}=+1.63^\circ$),
respectively \citep[see also Figure 1 in ][]{suarez2017}.
A third density enhancement in the OB1a region is found roughly located at 
$\rm \alpha_{J2000}= 5^h27^m12^s$, 
$\rm \delta=+4^\circ00'$, ($\rm \alpha_{J2000}= 81.8^\circ$, 
$\rm \delta_{J2000} +4^\circ$),
$8.8\arcmin$ north of the B2 star HD 35762. 
This structure can be seen in Figure 15 of \cite{lombardi2017}, and also
in Figure 4 of \cite{zari2017}. 
In both studies it appears as a northward extension
of the 25 Ori overdensity (westward in the galactic reference frame),
though neither comment further on this feature.  
Both the HR 1833 and HD 35762 overdensities
are shown more clearly in Figure \ref{overdensities_1a}. It can be noticed
that there is an almost continuous density enhancement going north
from HR 1833 up to HD 35762. Also, this last one seems to connect with 
an overdensity at $\rm \alpha_{J2000}= 82.8^\circ$, $\rm \delta_{J2000}= +4.8^\circ$,
roughly $8\arcmin$ north of the B8IV/V star HD 36310.

\begin{figure}
\includegraphics[angle=270,scale=0.44]{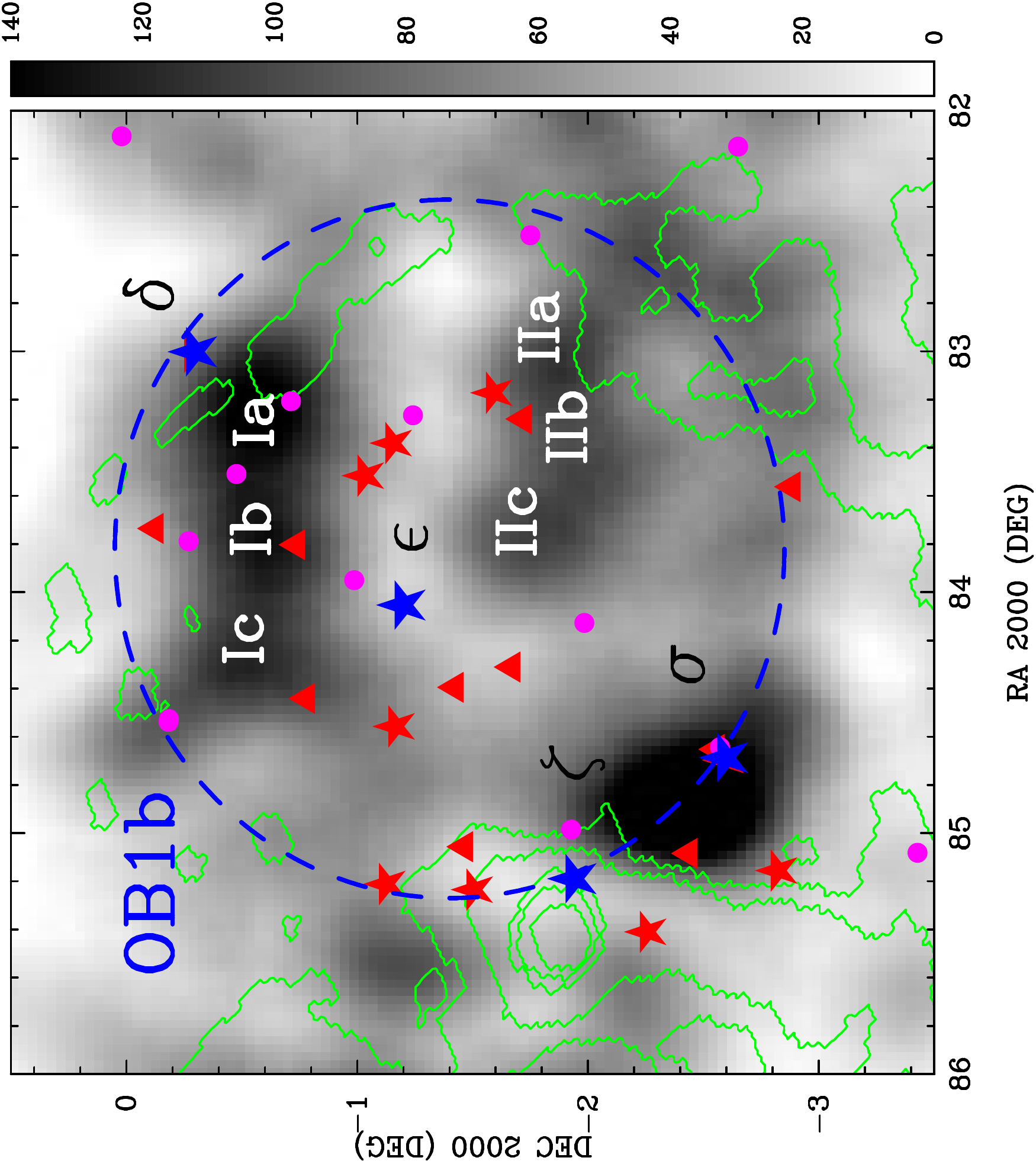}
\caption{Overdensities of candidate PMS sources in the OB1b region.
The background greyscale map is the surface density of photometric
candidate PMS sources as described in Figure \ref{spatial1}.
We have highlighted with large blue stars the three Orion belt stars:
$\delta$, $\epsilon$, and $\zeta$ Ori, and the massive O-type system
$\sigma $ Ori. The large blue dashed circle marks our adopted boundary for
the OB1b region, roughly following the contours of the dust extinction map
\citep{schlegel98}.
}
\label{overdensities_1b}
\end{figure}

Inside the OB1b subassociation, the distribution of stars is also 
quite clumpy
(Figure \ref{overdensities_1b}), and some overdensities are found near B-type stars.
The $\sigma$ Ori cluster is by far the densest grouping of stars, but
other significant groupings can be seen around $\epsilon$ Ori, 
with the densest being the feature labeled I in Figure \ref{overdensities_1b}, 
located approximately at 
$\rm \delta \sim -0^\circ36'$ ($\rm \delta \sim -0.6^\circ$)
and $\rm 05^h32^m \la \alpha_{J2000} \la 5^h37^m36^s$ ($\rm 83.0^\circ \la \alpha_{J2000} \la 84.5^\circ$), in which we recognize three distinct
structures that we label Ia, Ib and Ic, 
starting with the most westward one.
The other major density enhancement in OB1b is roughly located at
$\rm \delta \sim -1^\circ54'$ ($\rm \delta \sim -1.9^\circ$) and
$\rm 05^h31^m12^s \la \alpha_{J2000} \la 5^h36^m$ ($\rm 82.8^\circ \la \alpha_{J2000} \la 84.0^\circ$), $\sim 12'$ south-west of the B3V star HD 36646 and $\sim 22'$ south-west of the B2III HD 36591. 
This overdensity also shows three distinct clumps
that we label, from west to east, IIa, IIb and IIc.
These features have already been reported by \cite{kubiak2017}, who used a combination of optical data from SDSS and near-IR JHKs from 2MASS, to identify a $\sim 800$ candidate young objects within
Orion OB1b, which they call the ``Orion Belt Population''. In the surface density map shown in their Figure 5, aside from the obvious clump corresponding to $\sigma$ Ori cluster,  
they detect the same overdensities I and II, north-west and south-west 
respectively of the central belt star $\epsilon$ Ori. They also
mention that the density in these structures is clearly not homogeneous,
though their data did not allow them to draw further conclusions.

\begin{figure}
\includegraphics[angle=270,scale=0.44]{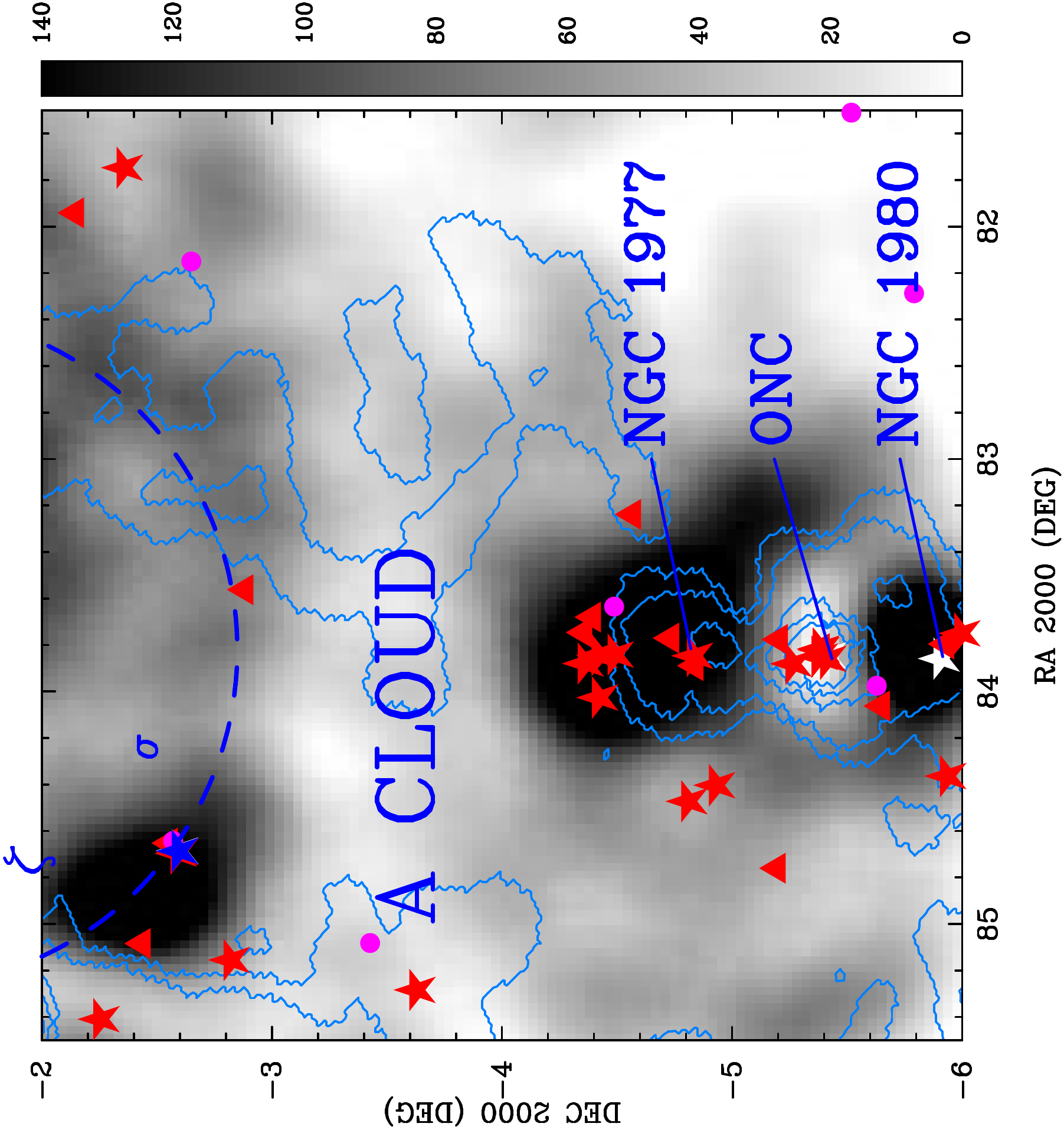}
\caption{Overdensities of candidate PMS sources in the area surrounding 
the ONC.
The surface density map and symbols are as in Figures \ref{overdensities_1a} 
and \ref{overdensities_1b}.
We indicate the location of the ONC, and the NGC 1977 and 1980 
stellar aggregates. The lowest blue contour  corresponds to the
Av=0.5 in the dust extinction map \citep{schlegel98}.
}
\label{overdensities_onc}
\end{figure}

In the Orion A cloud
(Figure \ref{overdensities_onc}), the most prominent feature is the strong
enhancement of candidate PMS stars surrounding the ONC,
encompassing to the north the NGC 1977 region and 
NGC 1981 to the south of the ONC, both known to contain
nearly a dozen O to B3 stars
\citep{rebull2000,peterson08,alves2012}.
The apparent ``hole'' in ONC region is an artifact from the \textit{CVSO} catalog,
which lacks photometry of sources in this region.
The ``halo'' of candidate PMS stars around the ONC
extends from  $\rm \alpha_{J2000} \sim 5^h31^m12^s$ 
($\rm \alpha_{J2000} \sim 82.8^\circ$)
to $\rm \alpha_{J2000} \sim 5^h38m24^s$ ($\rm \alpha_{J2000} \sim 84.6^\circ$), 
and from $\rm \delta_{J2000} \sim -4^\circ$ 
to further south than $\rm \delta_{J2000} \sim -6^\circ$; our data
do not allow us to probe beyond this limit.
This halo is not uniform; the strongest overdensities correspond to
NGC 1977 and 1980, while the areas directly east and west of the ONC 
have a much lower density of sources.
Because this extended structure spans $\rm \sim 4\,deg^2$ around
the ONC, some authors have called attention to the possible contamination of
existing censa of ONC members which extend beyond $\sim 20'$
from the cluster center, and indeed \cite{alves2012} and \cite{bouy2014}
propose an older, but still young population foreground to the NGC 1977/1980 region, called recently into question by \cite{fang2017}.
\cite{kounkel2017a} obtained spectra for 148 stars in NGC 1980, and they also find that the population is consistent with a younger $\sim 3$ Myr age,
also in disagreement with the \cite{alves2012} and \cite{bouy2014} studies.
What is clear is that these populations of young stars surrounding the ONC
will easily contaminate ONC samples, because these stars are essentially at the same distance \citep{kounkel2017a,kounkel2017b}, and share similar
radial velocities and proper motions close to zero, as the ONC, so they
will be difficult to distinguish.

\begin{figure*}[htb!]
\begin{center}
\includegraphics[angle=270,scale=0.80]{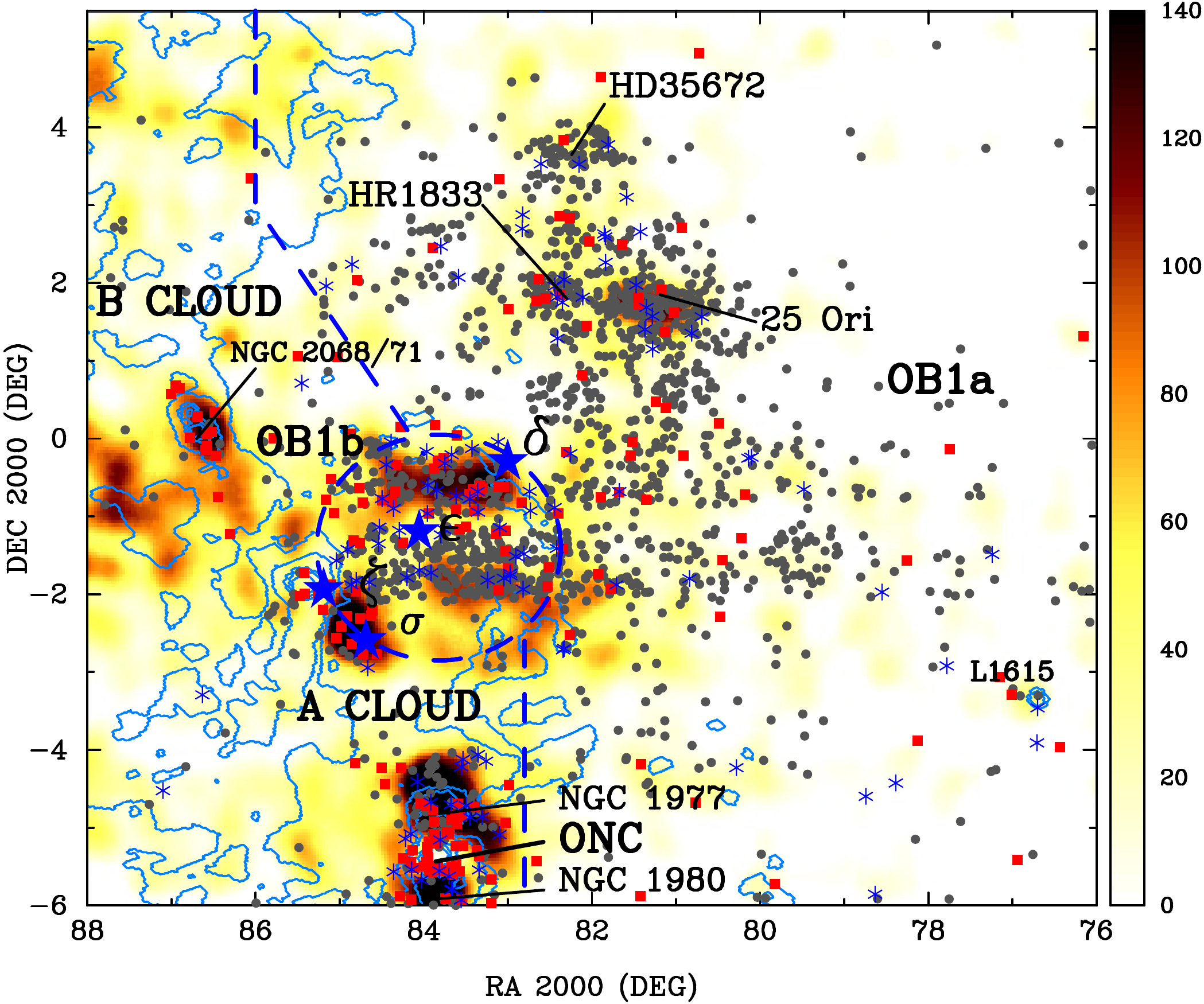}
\caption{Spatial distribution of the $2064$ confirmed TTS, 
projected against the surface density of PMS candidates 
selected as in Figures \ref{overdensities_1a}, \ref{overdensities_1b}
and \ref{overdensities_onc};
the density scale spans the range $0 - 143\ stars/deg^2$.
Gray dots are WTTS, blue starred symbols are C/W,
and red solid squares represent CTTS.
The outline of the A and B clouds is delineated by the Av=0.5 contour
from the \cite{schlegel98} dust map. Belt stars are indicated with
the large blue stars, and the dashed blue circle encompasses the area we assume as the OB1b subassociation. The dashed lines indicate our adopted
boundaries between the OB1a association and the A and B molecular clouds.
The off cloud TTS population stands out very clearly, 
following the complex structural traced by the surface density of
PMS candidates: a low-density component spread out throughout 
the region, starting at roughly $\rm \alpha_{J2000}=5^h16^m$, 
with a number of overdensitiesnthat include the 25 Ori cluster, 
and the HR 1833 and  HD 35762 groups in the OB1a subassociation, 
and the concentrations of TTS north-west and south-west of $\epsilon$ Ori
in OB1b.  The TTS surrounding the ONC trace well the underlying
density of PMS candidates.
}
\label{spatial2}
\end{center}
\end{figure*}

\subsection{The T Tauri population across Orion OB1}
\label{sec:ob1_offcloud_pop}

Now that we have gained a general picture of how the young, 
low-mass stars are distributed across the Orion OB1 association, and 
with the knowledge of the location of the various stellar groupings in
each region, we use our large dataset of spectroscopically confirmed members
to look at the overall characteristics and ensemble properties of the various groups, and compare
them in search of trends that can provide insight into the evolution 
of properties like Li I depletion and disk accretion.
In Figure \ref{spatial2} we show the location of the 2064 confirmed TTS
members of Orion OB1, projected against the surface density of photometric candidates. We plot WTTS as solid gray dots, intermediate C/W stars
as six starred blue symbols, and CTTS as red squares.
The low-mass young stars are distributed as follows: 
1218 in the 1a region, defined as the general area west of the straight dashed lines 
in Figure \ref{spatial2} and OB1b, 556 in the 1b region, defined within the dashed circle, 222 projected onto the A cloud, and 68 projected onto the B cloud.
Of the 1218 TTS in OB1a, 374 ($\sim 31$\%) are found within one of the three main stellar aggregates, 25 Ori, HR 1833 and HD 35762,
and 844 are what we call the OB1a ``field'' population, stars
not associated with any particular stellar overdensity.
In OB1b, 231 TTS ($\sim 42$\%) are associated with either the I or II
density enhancements shown in Figure \ref{overdensities_1b}.

\begin{figure}[hbt!]
\includegraphics[scale=0.4]{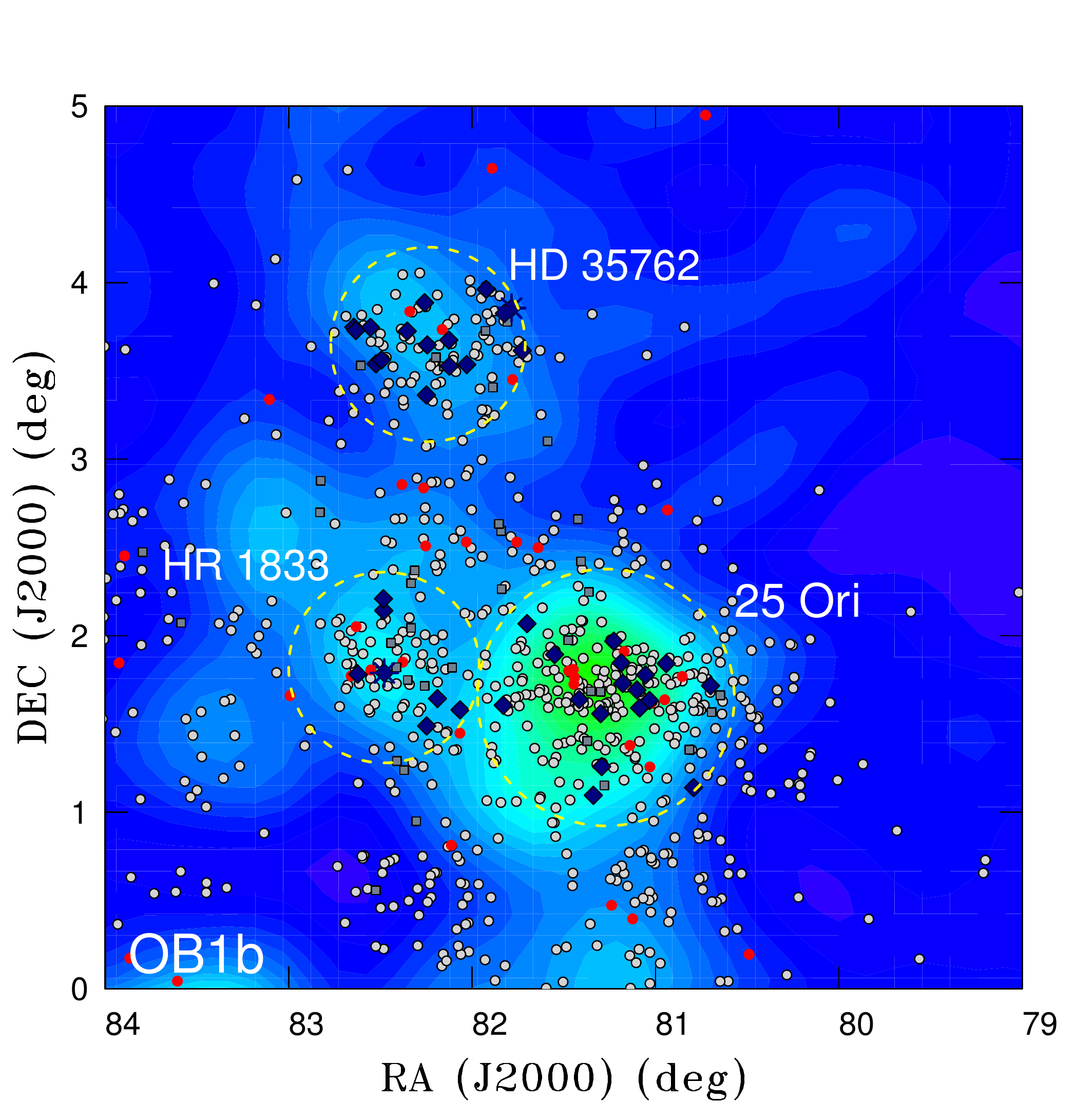} 
\caption{Spatial distribution of confirmed TTS in the 25 Ori cluster,
HR 1833 and HD 35762 groups, projected on the surface density map 
of PMS candidate stars.
Yellow dots are WTTS and red dots are CTTS. The B and A type stars in
each group are indicated with the dark blue diamonds. The 25 Ori, HR 1833
and HD 35762 stars are shown as large starred, dark blue, symbols.
Circles with our adopted radii of $0.7^\circ$ for 25 Ori, $0.5^\circ$ for HR 1833,
and $0.55^\circ$ for HD 35762 are shown with dashed lines. 
}
\label{clusters}
\end{figure}

In order to determine the stellar content and properties of each of
the stellar aggregates located in the OB1a subassociation, 
we defined as members of each group 
those spectroscopically confirmed 
TTS located within a circle encompassing most of the density contour at 
$\sim 3\sigma$ above the general background in the surface density map (Fig. \ref{clusters}).
This yielded radii of $0.7^\circ$ for 25 Ori,
$0.5^\circ$ for HR 1833 and $0.55^\circ$ for HD 35762.

With this selection criterion we assigned 223 members to the 25 Ori cluster,
65 to the HR 1833 group and 86 to the HD 35762 group. 
The 25 Ori members include 65 from \cite{downes2014} and 6 from \cite{suarez2017}. In HR 1833, there are 5 members in common with \cite{suarez2017}; another 13 TTS in their sample are assigned by us
to the OB1a field population.
In Figure \ref{clusters} we show the spatial distribution of all 
spectroscopically confirmed TTS in the region around the three stellar
aggregates in the OB1a subassociation, together with the location of
the early type stars in each group.

\subsubsection{Distances}
\label{sec:distances}

As is the case for 25 Ori (B07b), 
we expect these other groupings of TTS in 1a to be likely younger
than $\sim 20$ Myr. Because of this, we also expect the more massive B 
and A-type stars to be located on, or very close to the zero age 
main sequence (ZAMS) in an extinction-corrected color-magnitude diagram. 
This provides us with the opportunity to derive a photometric parallax for
each group, by applying the main sequence fitting technique instead 
of simply assuming a common distance for all groups. 

A practical advantage of these particular groups is that
they are located in an area of very low extinction, 
well removed from the Orion molecular clouds \citep[$\rm A_V \lesssim 0.3$ for the 25 Ori group][]{briceno07b,downes2014,suarez2017}; therefore,
reddening is not an issue when correcting the observed photometry.
Within each circular region we looked up the SIMBAD database and the \cite{kharchenko2005} catalog of 109 new open clusters for B and A-type stars.  
We then cross-referenced these lists with
\cite{hernandez05}, using their spectral types whenever available,
matched to the Spitzer Orion OB1 observations  
of early type stars by \cite{hernandez06}, and 
to the WISE catalog \citep{wright2010}.
In each group we determined which B and A-type stars showed infrared
excess emission typical of circumstellar dusty disks. 
In the 25 Ori cluster and the HR 1833 groups, 
where we have Spitzer $24\mu$m data, we
followed the criterion shown in Fig.4 of \cite{hernandez06}: 
we used the J-H vs Ks-[24] color-color diagram to select
objects with Ks-[24] $\ge 0.5$ as disk systems; this selection includes
those classified as Debris Disk systems ($\rm 0.5 < Ks-[24] < 5$), 
and those in the Herbig Ae/Be locus (Ks-[24]$>$5). 
In the HD 35762 group we used the WISE w4 band ($22\mu$m),
using only measurements with SNR $\ga 5$ and that had an actual
source visible in the W4 image.
We further assumed that the disk-bearing B and A-type stars located within the circular area defined for each group, are high probability early type members.
Once the disk systems were identified in each stellar aggregate, we narrowed
our selection to those
with spectral types roughly between B9 and A5 (equivalent to $-0.2 \lesssim (V-J)_0 \lesssim 0.25$).
This range corresponds to stars which should be on or very close 
to the ZAMS. At ages of several, up to $\sim 20$ Myr,
they are not so massive as to have started evolving off the ZAMS,
and they are not low-mass enough to be located significantly
above the ZAMS. In practice, this is equivalent to
these stars being located within $\sim \pm0.12$ mag of the main sequence 
in a V vs V-J CMD \citep{siess00}.  
We therefore assume that the bright end of
the cluster sequence is defined by the $\rm \sim B9$ to A5 stars with disks.
A least square fit of the \cite{siess00} ZAMS 
to these stars in each group yielded distances of $337\pm 35$ pc for 25 Ori (fit with 5 stars), $345\pm 35$ pc for HR 1833 (fit with 5 stars),
and $343 \pm 25$ pc for HD 35762 (fit with 8 stars). 
These distances are in agreement with the mean distances derived using
available {\sl Gaia} DR1 \cite{gaia_dr1} parallaxes for the B9 to A5 stars in each group, regardless of whether they were disk candidates or not, as long as they were located within $\pm 0.12$ mag of the main sequence in the CMD.
With the new \textit{Gaia} DR2 release \cite{gaia_dr2} we can now derive individual distances for a sizable subset of the 2064 TTS in the \textit{CVSO} sample. This provides a sensitive check on the above distances to the various Orion OB1 groups, with statistically robust samples of well characterized young stars, and enables us to derive for the first time, distances for the non-clustered populations in the OB1a and OB1b subassociations, and the populations projected on the molecular clouds.
In Table \ref{tab:distances} we compare the three distances determinations
for the groups, and add the \textit{Gaia} DR2 distances to the other regions.
We removed a few objects with negative parallaxes and used only stars with
parallax errors $< 20$\%.
Note that for the distances derived from the \textit{Gaia} parallaxes, we show the standard standard deviation of all measurements in each group. The actual typical measurement error in DR2 for our \textit{CVSO} TTS is $\sim \pm 25$ pc.

\begin {deluxetable}{lccccr}[ht!]
\tablewidth{0pt}
\tablecaption{Mean distances for each region\label{tabages}}
\tablehead{
\colhead{Region} & \colhead{MS fit$^{(1)}$} & \colhead{\textit{Gaia} DR1$^{(2)}$} &
\colhead{\textit{Gaia} DR2$^{(3)}$} & \colhead{Adopted} & \colhead{$\rm N_*^{(4)}$} \\ 
\colhead{}        & \colhead{(pc)}  &  \colhead{(pc)} & \colhead{pc} &
\colhead{pc} & \colhead{}\\
}
\startdata
Cloud B  &  \nodata & \nodata &  $395^{+68}_{-50}$ &  400 & 60 \\
Cloud A  &  \nodata & \nodata &  $398^{+39}_{-45}$ &  400 & 206 \\
1b       &  \nodata & \nodata &  $398{^{+69}_{-51}}^{(5)}$ &  400 & 510 \\
25 Ori   &  $337\pm 35$ & $347^{+58}_{-43}$ & $352^{+34}_{-28}$ & 350 & 211  \\
HD 35762 &  $343\pm 25$ & $361^{+44}_{-35}$ & $348^{+34}_{-28}$ & 350 &  78  \\
1a       &  \nodata & \nodata & $358^{+60}_{-43}$ & 360 & 797  \\
HR 1833  &  $345\pm 35$ & $361^{+45}_{-36}$ & $370^{+52}_{-40}$ & 360 &  64 \\
\enddata
\tablenotetext{}{(1): Main Sequence fitting for B9 to A5 stars, with the $\sigma$ of the fit.
(2): \textit{Gaia} mean distances derived from DR1 parallaxes of B9 to A5 stars.
(3): \textit{Gaia} mean distances derived from DR2 parallaxes for \textit{CVSO} TTS in each group.
(4): Number of \textit{CVSO} TTS involved in each mean \textit{Gaia} DR2 distance determination. (5): The distance distribution in OB1b is bimodal;
we find two distinct groups of stars, at two different distances.}
\end{deluxetable}
\label{tab:distances}


\begin{figure}[hbt!]
\includegraphics[scale=0.55]{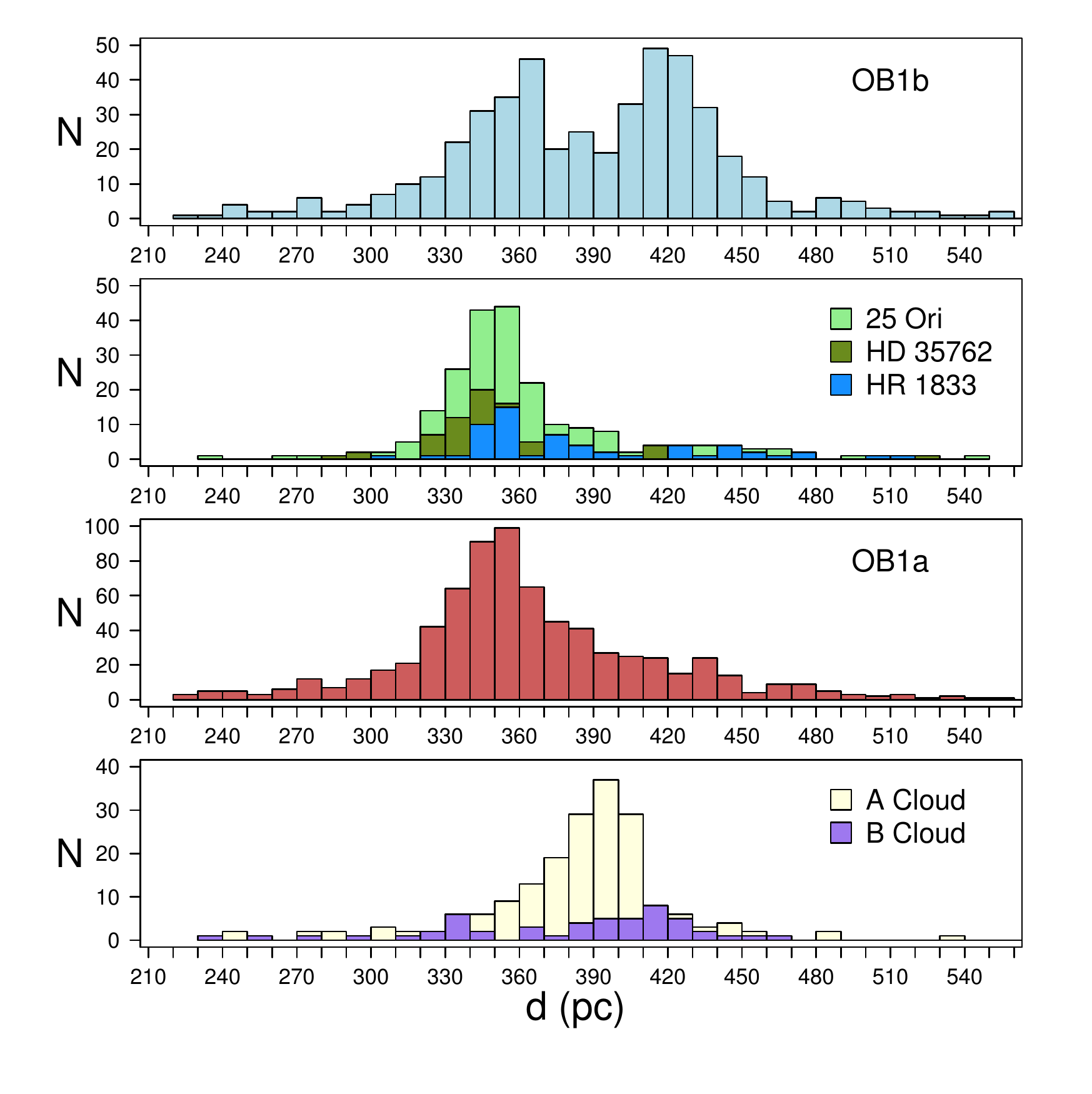} 
\caption{Distribution of distances derived from \textit{Gaia} DR2 parallaxes,
for confirmed TTS in the Orion OB1 regions.
}
\label{dist_histo}
\end{figure}

The distances derived from our main sequence fitting, from \textit{Gaia} DR1 parallaxes for the early type stars, and from the \textit{Gaia} DR2 data for the TTS populations, all agree within $1-\sigma$. In column 5 of Table \ref{tab:distances}
we provide our adopted distances for each region, and in Figure \ref{dist_histo} we show the distribution of distances derived from \textit{Gaia} DR2. 
An interesting result is that the OB1b region seems to contain two populations of TTS, a ``near'' one at a mean distance of $\sim 365$ pc and a ``far'' population located roughly at $420$ pc. With a sample of over 500 stars in this histogram, this is a rather robust result. The ``near'' subset distances peak at $\sim 365$ pc, 
essentially the same distance as the field population of OB1a. In retrospect this is not surprising, given that we would expect some amount of ``contamination'' of OB1b star samples by stars from the older, and closer OB1a subassociation (B05). In fact,
this result is consistent with the suggestion by \cite{jeffries2006} of an
older population of stars from OB1a, which have a mean radial velocity of 
$\rm 23.8 \, km \, s^{-1}$, offset about $7 \, km \, s^{-1}$ from the
$\rm 30 \, km \, s^{-1}$ mean velocity of the OB1b stars (see also B07b). 
Further study of the kinematics of these two populations, which is beyond the
scope of this work, will help disentangle their nature and origin.

\begin{figure}[hbt!]
\includegraphics[scale=0.47]{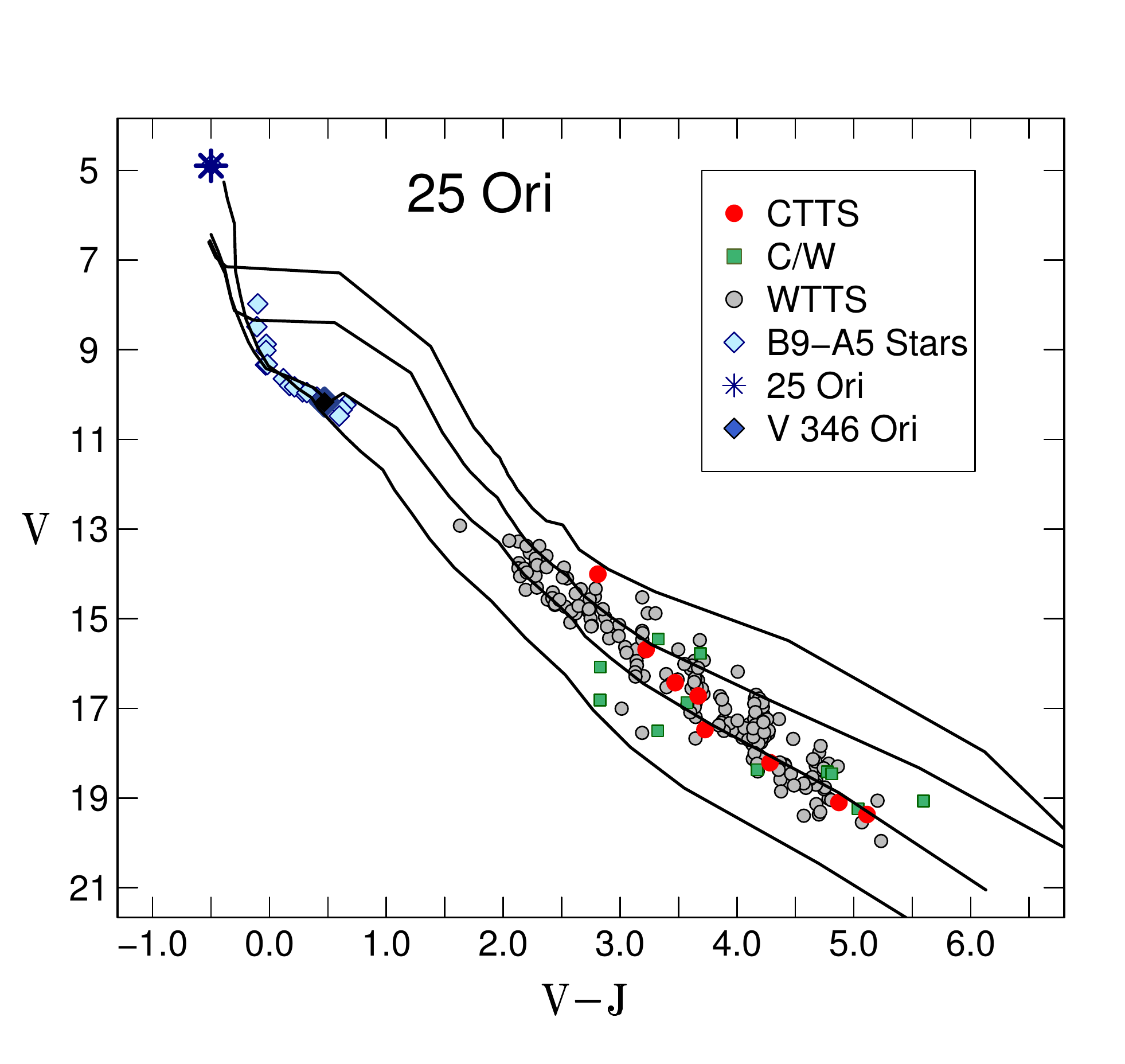} 
\caption{Extinction-corrected color-magnitude diagram for the 25 Ori cluster.
Solid lines represent, going down from the topmost one, 
the 1, 3, 10 and 100 Myr isochrones \citep{siess00}, set at 
the adopted distance of 350pc.
CTTS are plotted with red dots, WTTS with grey dots, C/W with green squares.
The early type stars are indicated with light blue diamond symbols.
The six-starred symbol marks 25 Ori itself, while the dark blue diamond
indicates the A9 Herbig Ae/Be stars V 346 Ori. An $\rm A_V=1$ reddening vector is also plotted.
}
\label{cmd_25ori}
\end{figure}

\begin{figure}[hbt!]
\includegraphics[scale=0.47]{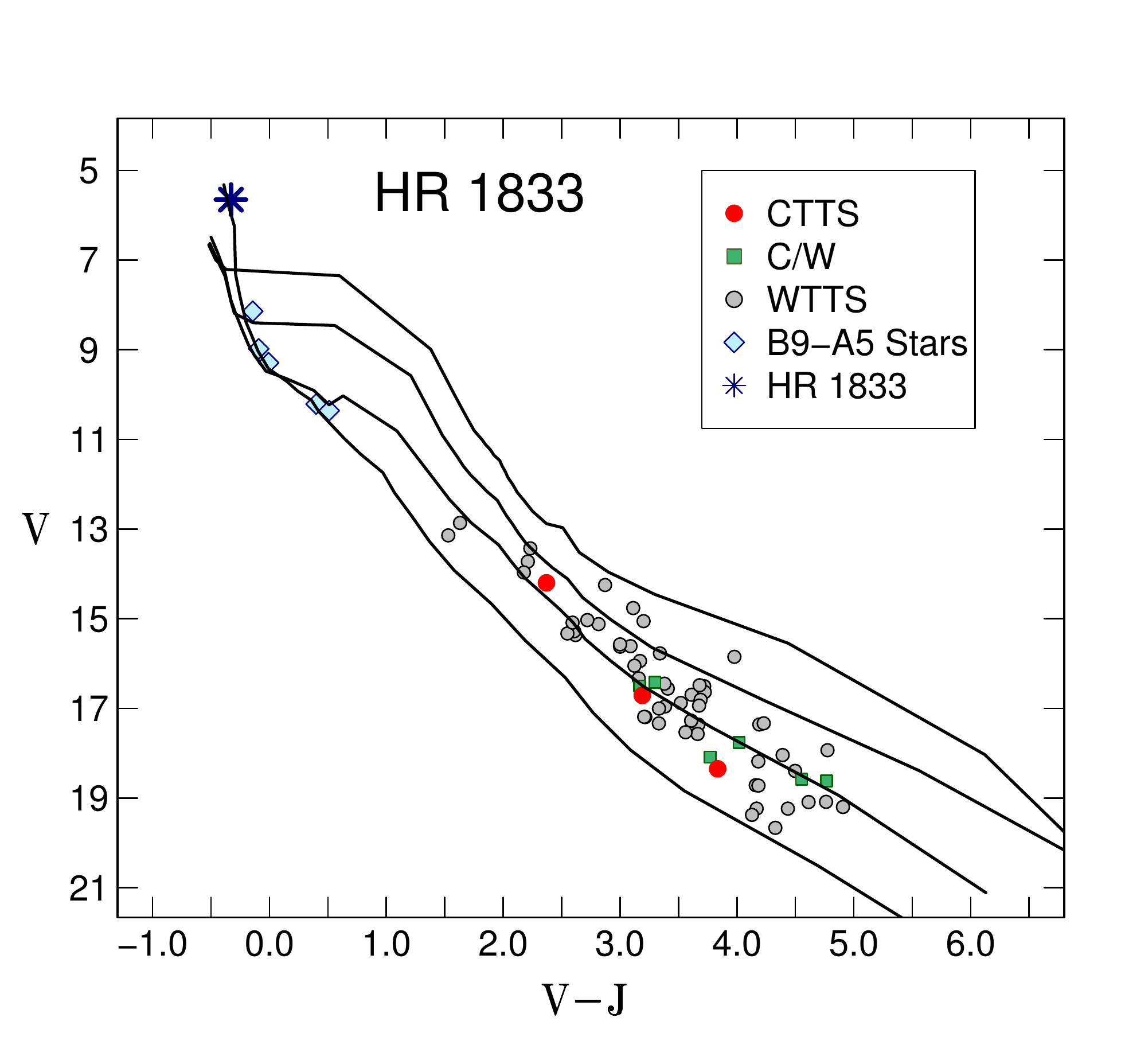} 
\caption{Extinction-corrected color-magnitude diagram for the HR 1833 group.
As in Figure \ref{cmd_25ori}, solid black lines correspond to the 
1, 3, 10 and 100 Myr isochrones \citep{siess00}, set at 
the adopted distance of 360 pc.
Symbols are as in Figure \ref{cmd_25ori}.
The six-starred symbol marks HR 1833.
}
\label{cmd_hr1833}
\end{figure}

\begin{figure}[hbt!]
\includegraphics[scale=0.47]{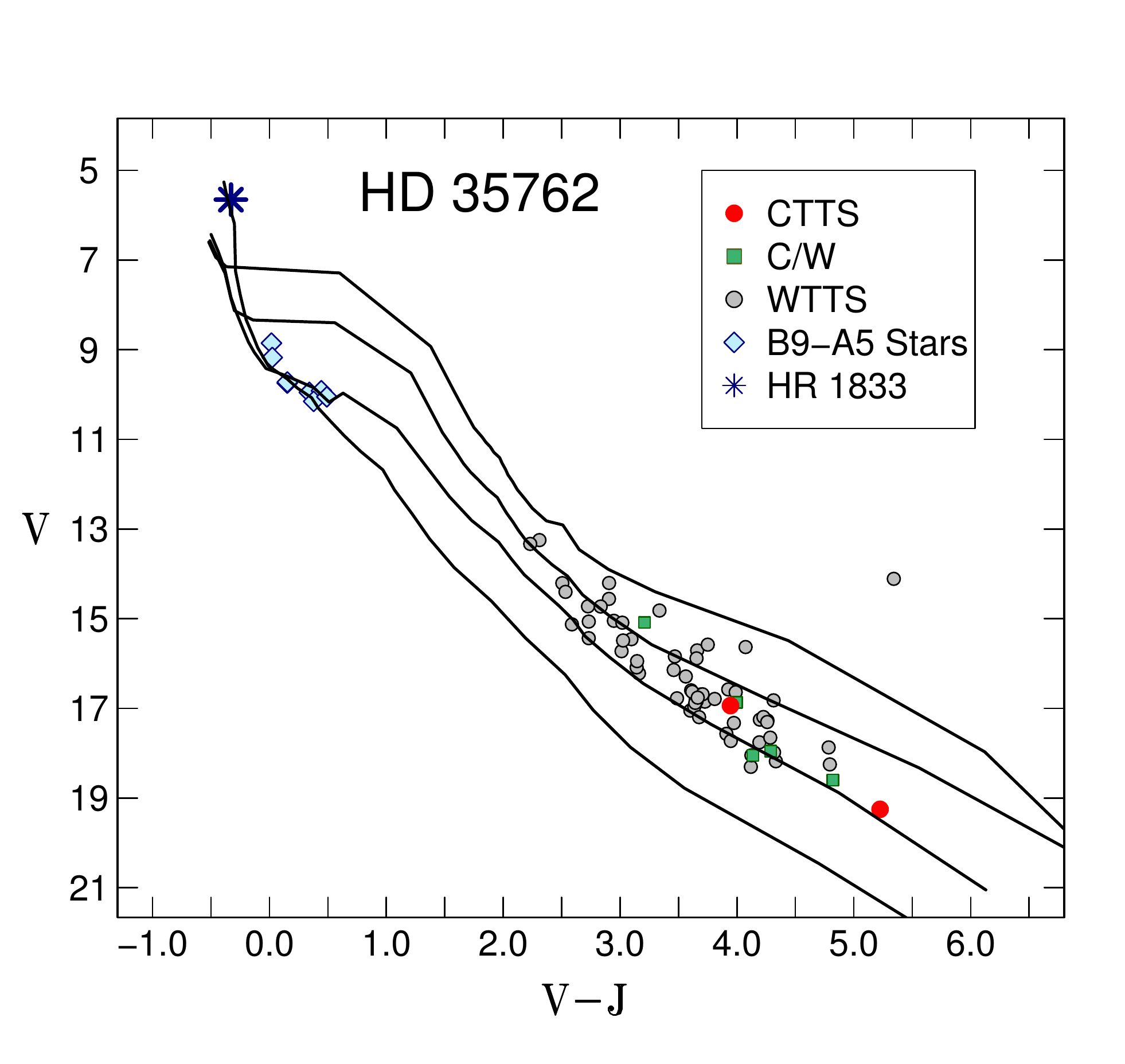} 
\caption{Extinction-corrected color-magnitude diagram for the HD 35762 group.
As in Figures \ref{cmd_25ori} and \ref{cmd_hr1833}, solid lines
correspond to the 1, 3, 10 and 100 Myr isochrones \citep{siess00}, set at 
the adopted distance of 350 pc. Symbols are as in Figure \ref{cmd_25ori}.
The six-starred symbol marks HD 35762.
}
\label{cmd_hd35762}
\end{figure}

\begin{figure}[hbt!]
\includegraphics[scale=0.47]{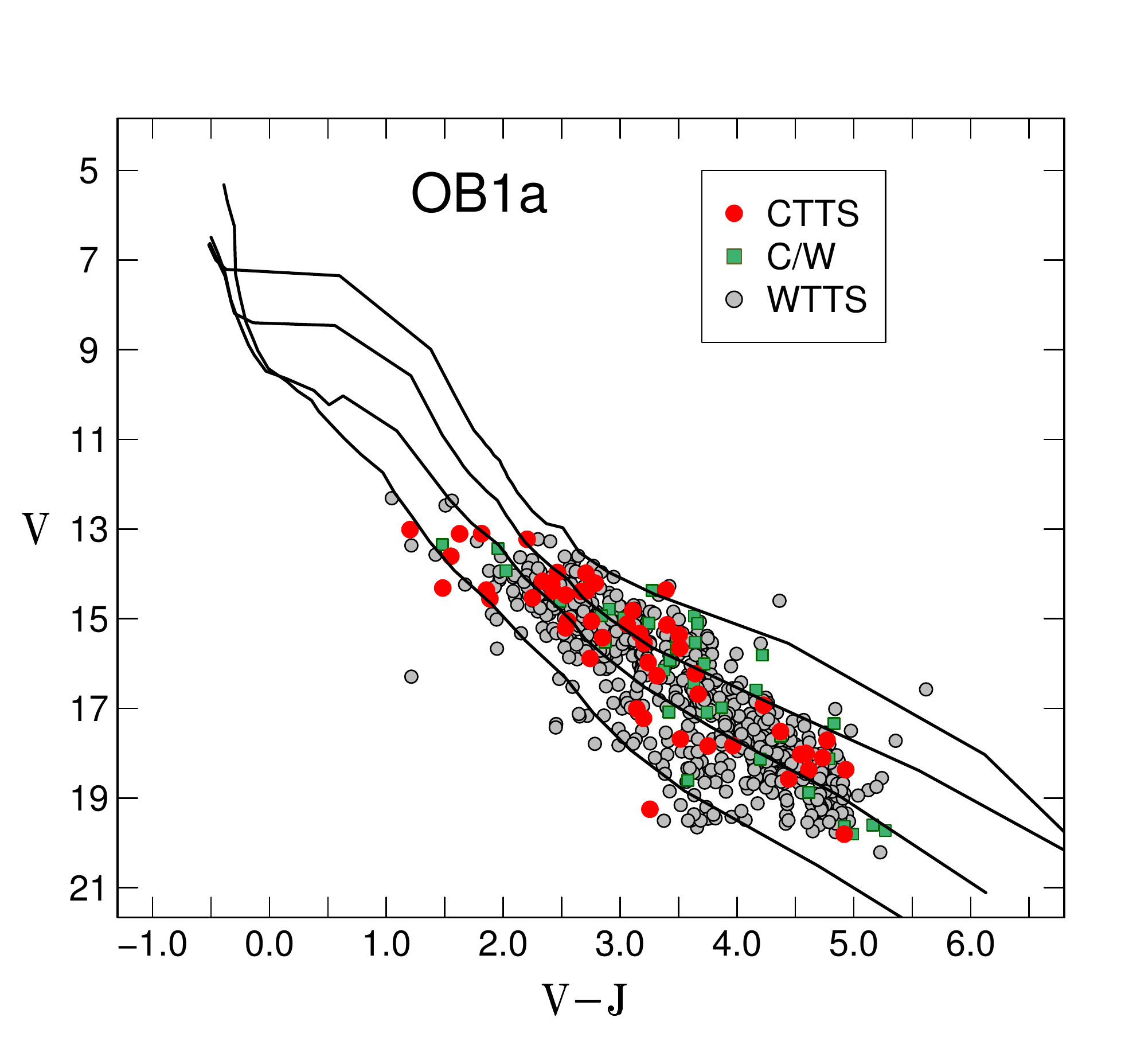} 
\caption{Extinction-corrected color-magnitude diagram for the field population of the OB1a subassociation.
As in previous figures, solid lines
correspond to the 1, 3, 10 and 100 Myr isochrones \citep{siess00}, set at 
the adopted distance of 360 pc, and the different dot symbols mark the CTTS, C/W and WTTS, as indicated in the legend.
}
\label{cmd_ob1a}
\end{figure}

\subsubsection{The PMS population in Orion OB1a}
\label{sec:newgroups_a}

With the distances derived for each group, 
we proceed to plot all the stars in reddening-corrected V-J CMDs, 
along with the \cite{siess00} isochrones
(Figures \ref{cmd_25ori}, \ref{cmd_hr1833} and \ref{cmd_hd35762}). To derive $A_V$ we used preferentially the $V-I_C$ color, which uses the robust mean $V$ and $I_C$ magnitudes from our \textit{CVSO} optical photometry, available for
$\sim 86$\% of the stars. For the remainder of the stars we used $V-J$ or the $R-I_C$ or $I_C-J$ colors. We used the intrinsic colors from 
\cite{kh95} corresponding to the spectral type for the star, and
the \cite{cardelli1989} reddening law with $\rm R_V=3.1$, which is consistent
with recent estimates of extinction toward the Orion region by \cite{schlafly2016}.
It is important to note that because we have extensive variability information in the optical bands for each star, we plot the clipped mean average V magnitudes, so that any spread in luminosity because of variability is minimized. Though ideally we would plot the V-Ic CMDs because they contain
the optical photometry and variability information from our own dataset, 
we instead chose here the V-J color, with the caveat that the 2MASS J-band is not simultaneous with our optical photometry, in order to include the early type stars, which mostly lack I-band data.
The CMDs for each group show a well defined cluster sequence running from the
early type stars all the way to the low-mass PMS members, with a gap in the
range roughly corresponding to spectral types F to late G. Stars in this range
are saturated in our \textit{CVSO} survey, and are hard to discriminate from contaminating field stars based only on the existing information in the
SIMBAD and \cite{kharchenko2005} samples. 
Though accreting members of this class of objects are easy to find using tracers like strong H$\alpha$ emission and IR excess, the WTTS counterparts have been traditionally difficult to pinpoint. Because of their shallow convective outer zones, G and F-type stars do not deplete lithium significantly during their PMS phase. Therefore, in contrast with K and M stars, for these earlier type objects lithium is not an useful youth indicator. Criteria like X-ray activity are not too useful to find PMS stars among solar-type stars, because X-ray emission decays slowly during the first $\sim 100$ Myr in G - early-K stars \citep{briceno97}. This an outstanding issue in which parallaxes
will be crucial to confirm membership, in combination with optical photometry {\sl and} spectroscopy.
The {\sl Gaia} DR2 release will be of great help in filling this region of the CMD.

That the new HR 1833 and HD 35762 TTS clusterings exhibit relatively
well defined sequences in the color-magnitude diagram, 
is a compelling suggestion that they are real stellar aggregates,  less massive siblings of the 25 Ori cluster. In contrast, the CMD for the general field population across 1a (Figure \ref{cmd_ob1a}), is much more spread out in luminosity, as expected
from stars that likely span a broader range of distances and ages compared to the population of the three clusterings.

\subsubsection{The PMS population in Orion OB1b}
\label{sec:newgroups_b}

\begin{figure}[hbt!]
\epsscale{2.43}
\plottwo{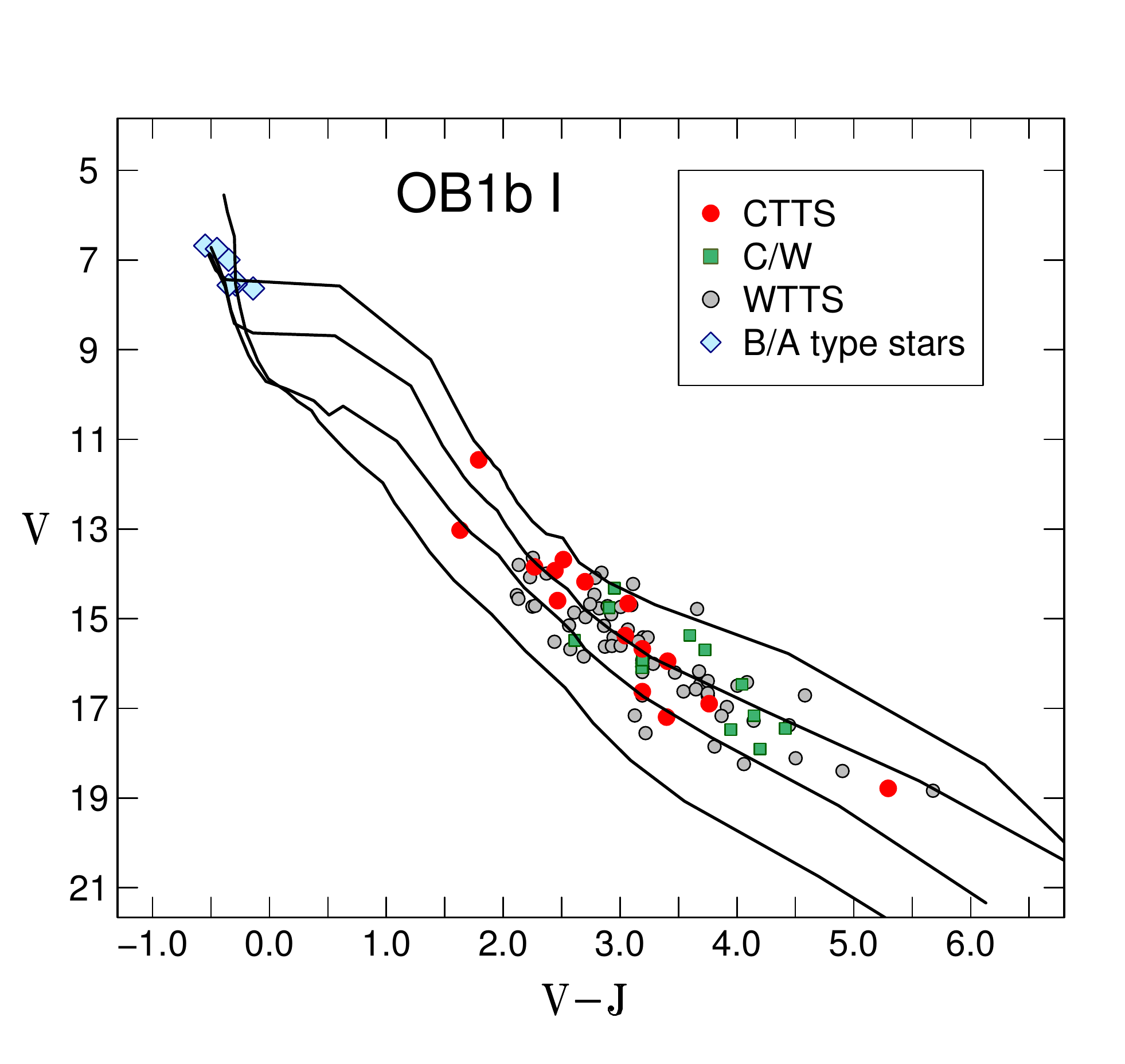}{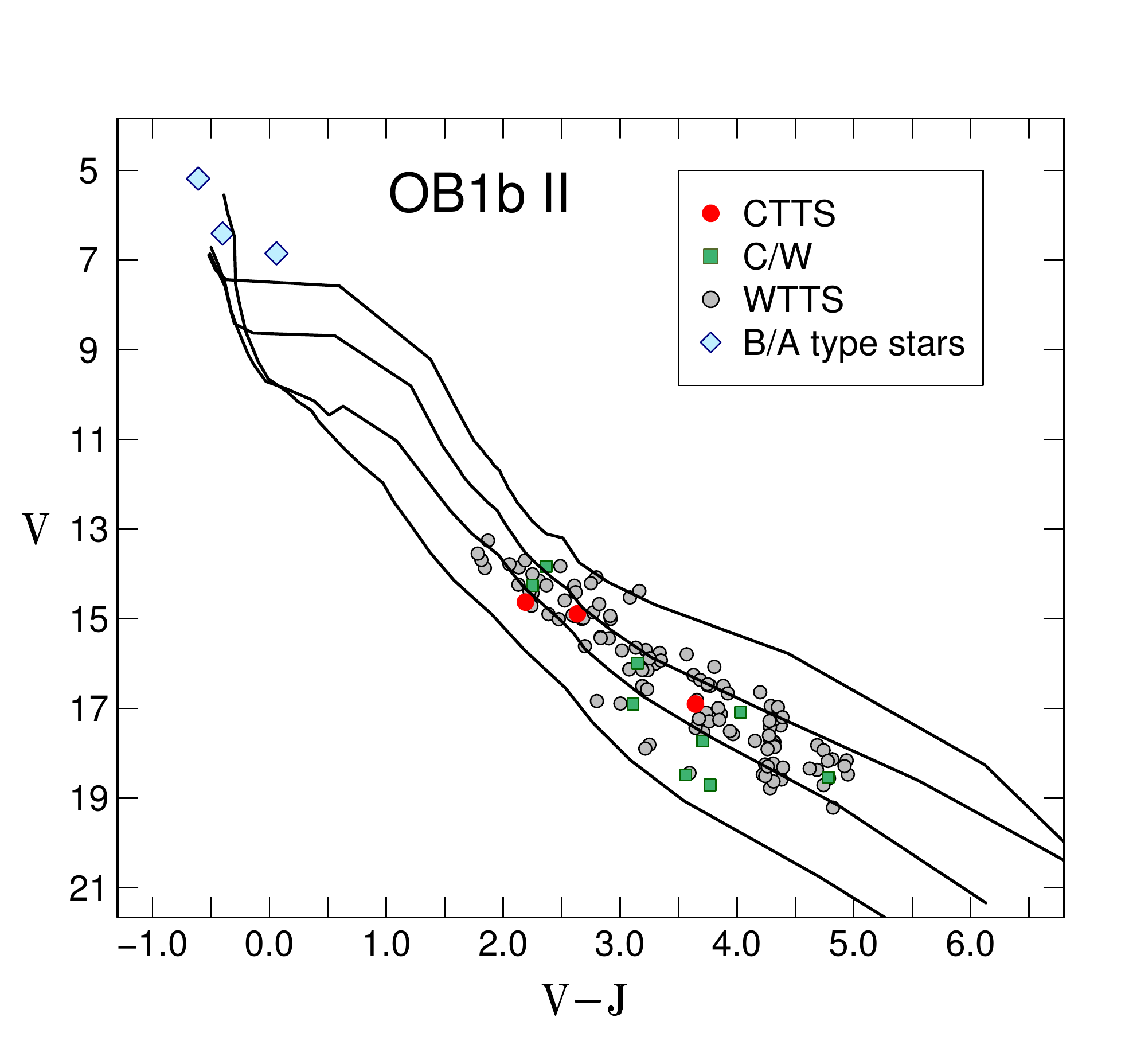} 
\caption{Extinction-corrected color-magnitude diagrams for the population of the two major
overdensities in the OB1b subassociation. Upper panel: the I (a through c) 
overdensity, northwest of $\epsilon$ Ori. Lower panel: the II (a through c) group south of $epsilon$ Ori (Figure \ref{overdensities_1b}).
As in previous figures, solid lines
correspond to the 1, 3, 10 and 100 Myr isochrones \citep{siess00}, set at 
the adopted distance of 380 pc, and the different dot symbols mark the CTTS, C/W and WTTS, and early type B and A stars, as indicated in the legend.
}
\end{figure}
\label{cmds_ob1b}

As discussed in \S \ref{sec:spatial}, we identify two major overdensities
within the Orion OB1b subassociation, which we labeled as I and II, each one
showing hints of further substructure, that we identified as subgroups
a, b and c. However, to simplify our discussion and in order to have
better statistics, here we will consider these
two major structures as two single entities.  In Figure \ref{cmds_ob1b}
we plot the TTS in the V-J CMDs for each of these two groups, together with the early type stars found within each overdensity. We assumed a distance of
390 pc, consistent with the latest determinations for the $\sigma$ Ori
cluster \citep{schaefer2016,hernandez2014,caballero2017}, which is generally assumed to be part of the OB1b region.
There are 231 TTS in these two stellar groups: OB1bI contains 93 TTS, while OB1bII encompasses 138 TTS.
It is interesting that both OB1b groupings exhibit fairly
well defined sequences, and that the early type stars are also roughly located
at their expected location in the CMDs assuming they form part of these two stellar aggregates; though they are clearly more scattered in the CMD compared
with their siblings in the OB1a clusterings, maybe in part because of larger extinction in OB1b compared to OB1a. Also apparent is the fact that OB1bI contains a larger fraction of CTTS than OB1bII, but we defer further discussion to the next section.

In \S \ref{sec:distances} we showed that the OB1b region is composed of two populations separated in distance. In Figure \ref{dist_histo_ob1b} we show the distribution of distances derived from the \textit{Gaia} DR2 parallaxes for each of the two groups of TTS in OB1b.
This plot shows that the bimodality in the distance (or parallax) distribution
in the OB1b subassociation arises from the northern OB1bI group, which has a near ($\sim 355$ pc) component and a far ($\sim 415$ pc) component. In contrast, the majority of the TTS in the southern OB1b II group are at distances within the 
range $d \sim 390 - 450$ pc, with a median distance of $\sim 420$ pc.
As we discuss in \S \ref{sec:distances}, we speculate that this ``near'' component
may be part of the field population of OB1a stars. If so, we would expect these
TTS to share other properties with their OB1a siblings.

\begin{figure}[hbt!]
\includegraphics[scale=0.5]{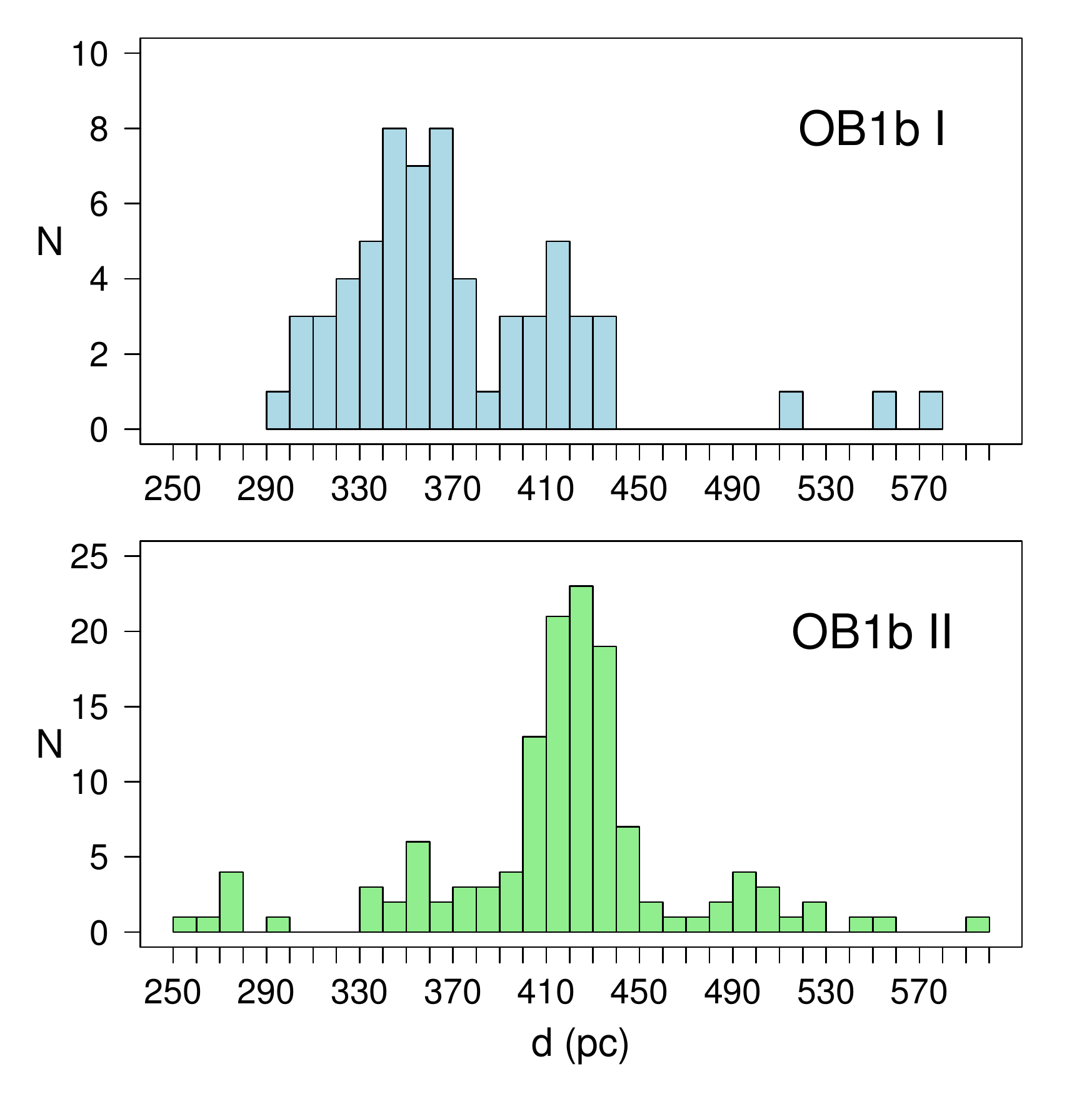} 
\caption{Distribution of \textit{Gaia} DR2 parallaxes for the TTS in groups OB1b I and II.
}
\end{figure}
\label{dist_histo_ob1b}

\subsection{Demographics of young stars in Orion OB1} 
\label{sec:demographics}

\subsubsection{Ages}
\label{sec:ages}

In order to study any evolutionary trends we must first estimate
ages for each region and groups of stars.
The dereddened CMDs provide a useful tool to do this.
For each star we obtained the observed $\rm V-I_C$ color from the \textit{CVSO} robust mean V and I-band magnitudes, 
and then derived the color excess by comparing with the corresponding intrinsic color from Table A5 in \cite{kh95}, provided by our spectral types.
We derived $\rm A_V$ assuming the \cite{cardelli1989} reddening law; we could derive extinctions for over $\sim 88$\% of the \textit{CVSO} stars using $\rm V-I_C$. 
Because $\rm V-I_C$ comes from our quasi-simultaneous \textit{CVSO} photometry,
we adopted this band to perform the reddening correction in preference over over the V-J color. The small fraction of stars dropped because of lack of $\rm V-I_C$ information do not affect in a significant way our final mean age estimates.
We then performed linear a interpolation  within the \cite{baraffe98} and \cite{siess00} model isochrones, using the adopted distances for each group
and region shown Table \ref{tab:distances}.
Finally, we computed clipped mean ages to derive a single age for each region. The values, standard deviation of the fits, and number of stars involved in each calculation, are provided in Table \ref{tab:ages}. 

\begin {deluxetable}{lrrrrrr}[ht!]
\tablewidth{0pt}
\tablecaption{Mean ages for each region\label{tabages}}
\tablehead{
\colhead{Region} & \colhead{Age(B)} & \colhead{$\sigma$(B)} &
\colhead{Age(S)} & \colhead{$\sigma$(S)} & \colhead{$\rm N_*$} & \colhead{Adopted}\\ 
\colhead{}       & \colhead{(Myr)}  &  \colhead{(Myr)}  &  
\colhead{(Myr)} &  \colhead{(Myr)} & \colhead{} & \colhead{Myr}\\
}
\startdata
Cloud B  &  2.6 & 1.3 &  2.3 & 1.0 &  53 & 2.5 \\
Cloud A  &  3.2 & 1.3 &  3.4 & 1.6 &  60 & 3.3 \\
1b       &  4.7 & 1.8 &  5.2 & 2.1 & 489 & 5.0 \\
25 Ori   &  6.1 & 2.4 &  9.1 & 2.3 & 223 & 7.6 \\
HD 35762 &  7.9 & 2.4 &  8.0 & 2.5 &  61 & 8.0 \\
1a       &  9.2 & 2.2 & 12.3 & 2.9 & 811 & 10.8 \\
HR 1833  & 11.8 & 2.5 & 13.9 & 2.7 &  64 & 12.9 \\
\enddata
\tablenotetext{}{Notes: B: \cite{baraffe98}, S: \cite{siess00}}
\end{deluxetable}
\label{tab:ages}


The \cite{siess00} models yield older ages compared to \cite{baraffe98}, but both agree in the sequence of ages, indicating, as expected, that the youngest stars are the ones projected onto the A and B molecular cloud. Within the uncertainties, these two regions can be assumed to be of the same age, $\sim 3$ Myr, a value which is somewhat higher but consistent with estimates for regions like L1641 in the A cloud (\cite{hsu2012}, given the large uncertainties in age determination, particularly for the youngest star-forming regions. We also point out that because our candidate selection was optical, we are reddening-limited toward the molecular clouds, resulting in a modest number of stars compared to the low extinction OB1b and OB1a regions. Therefore we are more susceptible to contamination from  OB1b and OB1a. Such slightly older stars would bias the mean age toward a slightly higher value.
The second oldest population is located in the OB1b region, and the oldest stars are those in the OB1a subassociation, consistent with the historic picture of the progression of ages across Orion OB1.
Within the OB1a region, 25 Ori seems to be the youngest group ($\sim 6-9$ Myr), similar in age to HD 35762, while HR 1833 comes out as the oldest one (12-14 Myr).
The ``field'' population has an average age of $\sim 11$ Myr.

Despite the large uncertainties in the absolute ages of each region, that come from of a complex combination of observational errors that include determination of spectral types, transformation to an effective temperature scale, the adoption of dwarf temperatures and colors for PMS stars, the use of a particular reddening law, the uncertainty in the assumed distances, and finally, but not least, limitations in the theory behind model isochrones and differences between various models, for the purpose of our analysis here onwards, what is important is the age sequence between the various regions. Even with the improved distances that will come with the second release of {\sl Gaia}, differences between models or differing methods, and the various other assumptions mentioned above will likely still introduce significant systematic offsets, in addition to scatter. For example, \cite{bell2013} found ages that are older by a factor $\sim 2$ compared to the most commonly assumed ages for several nearby young regions. Further  discussion of PMS ages and their uncertainties is beyond the scope of this work.

\subsubsection{Pre-main sequence Li depletion}
\label{sec:li_depletion}

The presence of the Li~ I 6707{\AA } line strongly in
absorption is a reliable indicator of youth in K and M-type stars 
(\S \ref{sec:membership}). Because this light element is
burned at temperatures of $\sim 2 \times 10^6\, ^\circ$K,
which roughly corresponds to the temperature at the base of
the convective envelope in a solar-type star at the zero-age main sequence
\citep{siess00}, lithium is steadily depleted during
the PMS phase in low-mass stars, and its surface abundance 
decreases over time, with the known spread caused by effects like rotation and the details of physical processes used in the stellar interior codes \citep{sestito_randich2005,tognelli2012,bouvier2016,bouvier2017}.

Having measured Li equivalent widths (W[Li I]) in a 
homogeneous way for a large number of PMS K-M type stars,
we are in a vantage position to explore from a statistical perspective
the depletion of Li during the PMS phase.
In order to avoid the uncertainties associated with the derivation of Li abundances \citep[see ][]{sestito_randich2005,tognelli2012}, we restrict ourselves to looking at a purely observational measure, W(Li I).  
This approach was considered by \cite{jeffries2014a} in their discussion
of PMS Li depletion.
We adopt the ONC data from \cite{sicilia-aguilar2005b} and the 
Taurus Li measurements from \cite{basri1991}, as representative of
what can be considered as a ``pristine'' or ``undepleted'' Li sample of 
optically visible, $1-2$ Myr old low-mass PMS stars. Both datasets were
obtained with similar spectral resolution as
that of our spectra, so we expect a reasonable comparison with our own measurements, though its important to bear in mind that even in very young regions like Taurus and the ONC there is a considerable spread in Li abundances. Still, past studies indicate that most members in
these regions  have undergone little, if any, Li depletion \cite{palla2005,palla2007}
and \citet{sestito_randich2005}.
Thus we define the ``undepleted'' or ``young'' Li locus in the W(Li I) - $\rm T_{eff}$ diagram, following \cite{briceno07b}, as the region above the line:

$$ \rm W(Li I)_{young} > 1.0591 - (1.52 \times 10^{-4}) \times T_{eff} $$

with W(Li I) in {\AA } and $\rm T_{eff}$ in $^\circ $K. Over 70\% of
Taurus and ONC TTS fall in the young Li locus.

\begin{figure}[hbt!]
\epsscale{2.43}
\plottwo{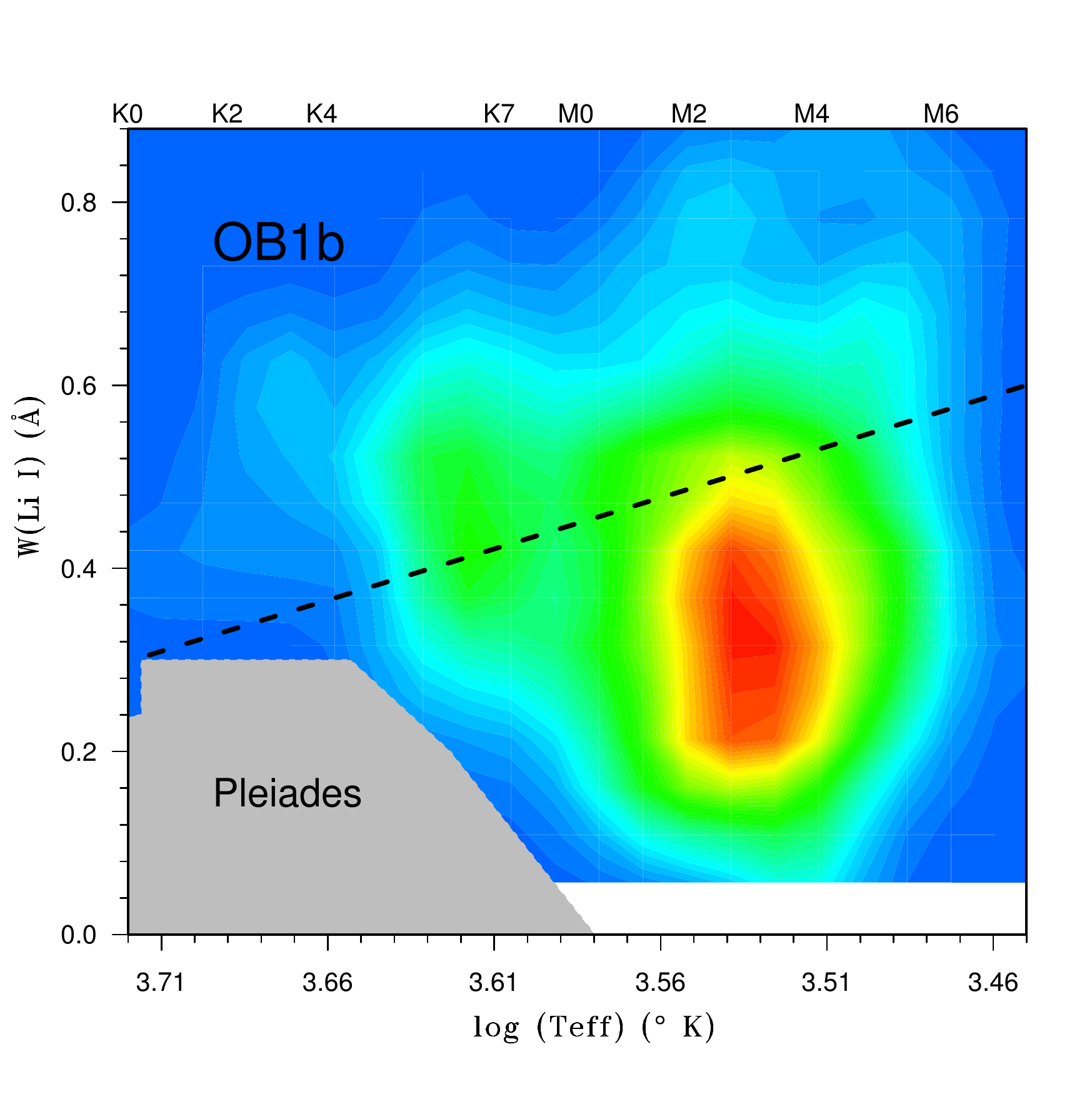}{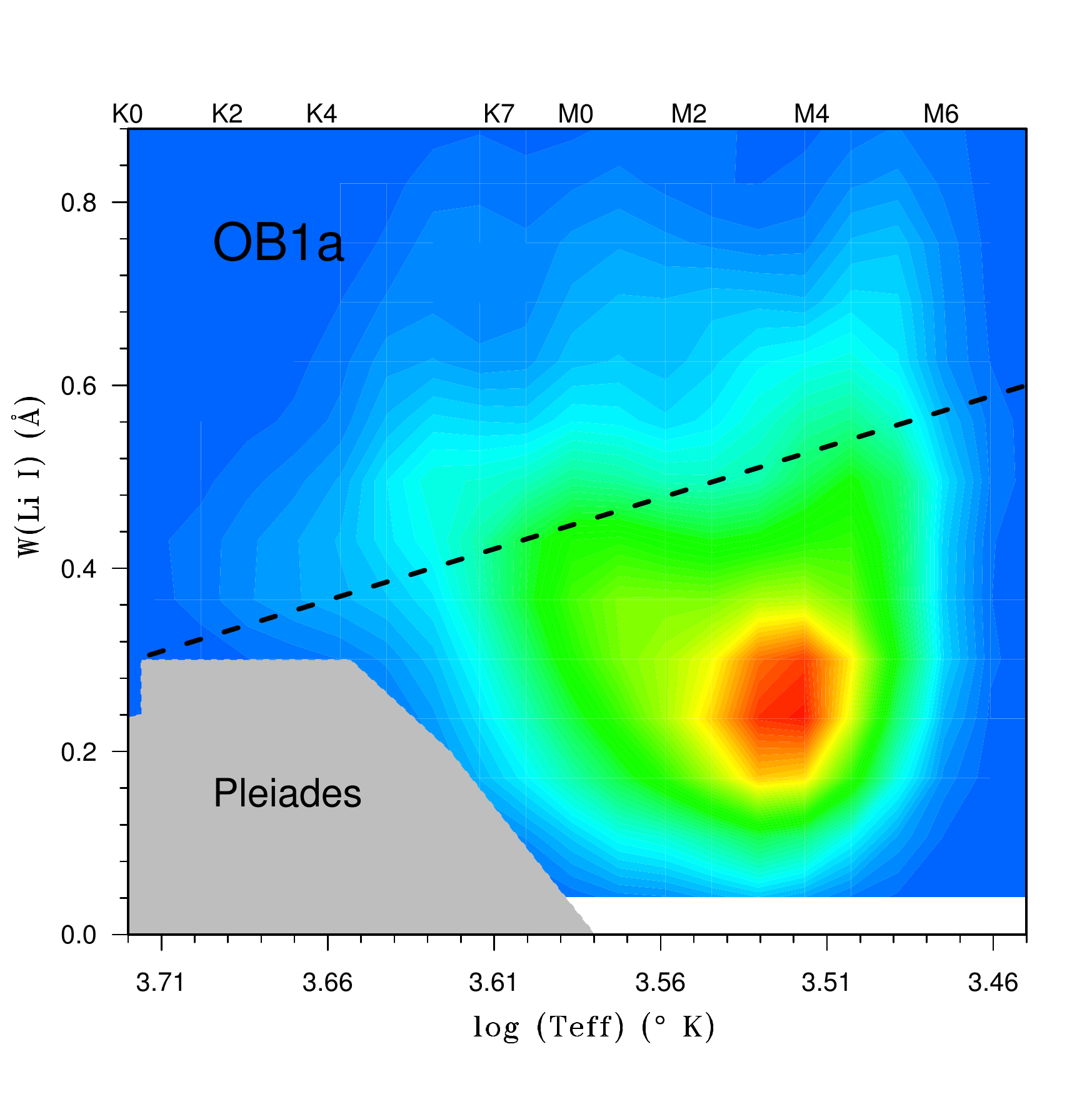} 
\caption{
Comparison of the distribution of Li I $\lambda 6707$ equivalent widths, represented as the surface density of values, for the $\sim 5$ Myr old Ori OB1b TTS, and  the $\sim 11$ Myr old OB1a "field" population. The dashed line is the young Li locus lower boundary (see text), above which are the W(Li I) of most optically visible stars in the ONC and the Taurus clouds.
The locus of Pleiades stars is indicated by the shaded gray region
\citep{soderblom1993,garcia-lopez1994}.
$T_{eff}$ are derived by interpolating our spectral types in Table A5 of \cite{kh95}.
}
\end{figure}
\label{wli_locus}

In Figure \ref{wli_locus} we show the surface density of Li equivalent widths for two populations at different evolutionary stages, the $\sim 5$ Myr old OB1b and the $\sim 11$ Myr old OB1a ``field'' population.  We use these two subsets because they have the largest number of stars, 556 in OB1b and 844 in OB1a, so that statistics are less affected by differences in sample sizes.
We derived the surface density using the Two-Dimensional Kernel Density Estimation recipe in R \cite{venables_ripley2002}. We can see how as we go from a young region to an older one, even allowing for the large spread in W(Li I) at any given $\rm T_{eff}$, in part intrinsic and in part due to measurement errors, a population-wide trend is evident; the K and M stars stars slowly move downwards in the diagram, to smaller values of W(Li I).
This effect was already suggested by us in \cite{briceno07b}, and later shown by \cite{jeffries2014a} in the $\gamma$ Velorum cluster. However, here we provide robust observational evidence of Li depletion with large samples of young populations of low-mass stars, all belonging to the Orion OB1 association.

\begin{deluxetable}{lrrr}[ht!]
\tablewidth{0pt}
\tablecaption{Li Depletion in Orion OB1\label{wli_fractions}}
\tablehead{
\colhead{Region} & \colhead{No. TTS} & \colhead{\# in ``young''} & \colhead{\% in ``young''} \\
\colhead{} & \colhead{} & \colhead{Li locus} & \colhead{Li locus}
}
\startdata
   L1641     & 865  &   352   &   41 \\
   Cloud B   &  68  &    31   &   46 \\
   Cloud A   & 222  &   106   &   48 \\
   1b        & 556  &   165   &   30 \\
   25 Ori    & 223  &    52   &   23 \\
   HD 35762  &  86  &    10   &   12 \\
   1a        & 844  &   183   &   22 \\
   HR 1833   &  65  &    13   &   20 \\
\enddata
\end{deluxetable}
\label{tab:wli_fractions}

In the left panel of Figure \ref{li_acc_evolution},
we show the evolution of the ensemble Li~ I for each population, parametrized by the number of TTS in each group lying within the ``young'' Li locus, 
as indicated in Table \ref{wli_fractions}.  The decrease in the ensemble W(Li I) across different populations can be fit by an exponential decline with
a characteristic timescale $\tau_{Li} = 8.5$ Myr in M-type stars, which constitute the bulk of the samples shown in Figure \ref{wli_locus}. This is in rough agreement,within a factor of $\sim 2$, with theoretical predictions of Li depletion \citep{stahler_palla2005}. For example, \cite{bildsten1997} estimate that a $0.3\, M_{\odot}$ PMS star 
(approximately an M3 TTS) will have depleted its Li by $\sim 50$\% at 16 Myr.
Though few studies similar to ours have been done, our findings seem
consistent with the differences between the overall W(Li I) for M stars
shown by \cite{jeffries2014b} in their comparison of the ONC with $\gamma$ Vel.
The HD 35762 group is an outlier in Figure \ref{li_acc_evolution}, with an unusually low Li-young fraction, that does not fit the trend, though it is only a 1-$\sigma$ deviation. This group merits further investigation in order to determine better its full membership, and derive better parameters for the stellar aggregate.
Large scale spectroscopic studies in increasingly older populations, carried out in a systematic way, could shed light on the evolution of Li in young stellar populations in the little studied 10 - 50 Myr age range. If this behavior could be calibrated, it may prove an useful tool to date PMS populations in the next era of large scale photometric and spectroscopic studies, ushered by the possibility of combining the power of facilities like DECam on the CTIO Blanco telescope, the Zwicky Transient Factory \citep{smith2014}, and LSST \citep{ivezic2008}, with highly multiplexed spectroscopic discovery and characterization machines like DESI \citep{desi2016a,desi2016b}.

\begin{figure}[hbt!]
\includegraphics[scale=0.42]{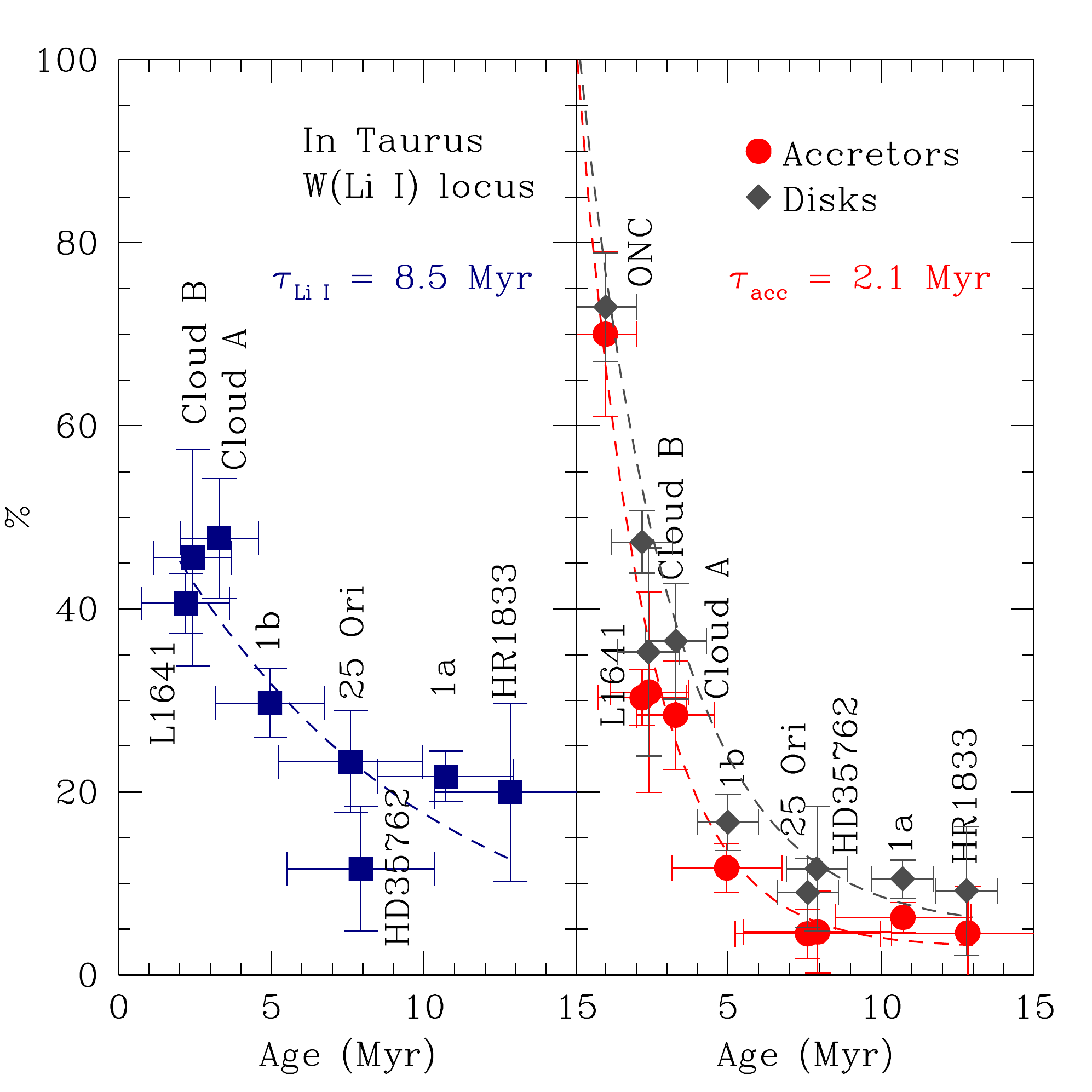} 
\caption{{\bf Left:} Evolution of the fraction of stars with
Li I equivalent widths as large as K and M-type T Tauri 
members of the Taurus star-forming region, 
which we consider as reference of a pristine, optically visible young population. The blue dashed line is a fit to the data, excluding the point for HD 35762.
{\bf Right:} Decline of the accretor fraction as a function of the mean age of each region/group. The red dots indicate the percentage of CTTS in each region, the usual measure of how much of the population is still actively accreting. The red dashed curve is an exponential fit to the data. As a reference, we show with dark diamonds the percentage of disk-bearing stars, determined from the $Ks-W2$ color following Figure 1 of \cite{luhman_mamajek2012}, with an exponential fit indicated by the dashed curve.
}
\end{figure}
\label{li_acc_evolution}

\subsubsection{Evolution of disk accretion}
\label{sec:accretion}

An important indicator related to the evolutionary state of a population of
low-mass PMS stars is the fraction of stars still accreting from their circumstellar disks.
As we have discussed in \S \ref{sec:ctts_classif}, CTTS are low-mass PMS stars considered to be actively accreting from a circumstellar disk. We have introduced an additional class of objects, namely, the C/W stars, which we argue are objects in a stage intermediate between that of the actively accreting CTTS and the largely diskless, chromospherically active WTTS.  These C/W stars are likely undergoing modest and/or variable accretion, in some cases at levels below what can be readily detected in low resolution spectra. Such low-accretors can be diagnosed through high resolution spectroscopy that can resolve the H$\alpha$ line profile, or near-IR spectra that can look at other accretion diagnostics like the He I $\lambda 10830$ line (Thanathibodee et al. 2018, submitted). 
Now we can derive accretor fractions for each region, in order to gauge how accretion is progressing from one population to another.  We determine the percentage of accreting TTS in each region as $\rm \% Accretors = (No.\> CTTS / No.\> TTS) \times 100$.

\begin{deluxetable}{lrrrrr}
\tablewidth{0pt}
\tablecaption{Accretor fractions across Orion OB1 \label{accretor_fractions}}
\tablehead{
\colhead{Region} & \colhead{TTS} & \colhead{CTTS} & \colhead{C/W} & \colhead{\% CTTS} & \colhead{\% C/W}\\ 
}
\startdata
   ONC        & \nodata & \nodata & \nodata & 70(a) & \nodata \\
   L1641      & 865  &   262   &    108  &  30 & 12 \\
   Cloud B    &  68  &    21   &      2  &  31 &  3 \\
   Cloud A    & 222  &    63   &     31  &  28 & 14 \\
   1b         & 556  &    65   &     55  &  12 & 10 \\
   25 Ori     & 223  &    10   &     11  &   5 &  5 \\
   HD 35762   &  86  &     4   &      6  &   5 &  7 \\
   1a         & 844  &    53   &     39  &   6 &  5 \\
   HR 1833    &  65  &     3   &      6  &   5 &  9 \\
\enddata
\tablenotetext{1}{(a) From \cite{hillenbrand98b}}
\end{deluxetable}
\label{tab:acc_fractions}

The percentage of CTTS averaged over our full sample is 11\%, 
but this is a strongly location-dependent value, which is due to the very different ages of each group. It varies from at least 70\% in the $\sim 1$ Myr old ONC \cite{hillenbrand98b}, to $\sim 30$\% in the A and B clouds,
decreasing to 14\% in the OB1b region and finally down to $\sim 5-8$\% once we reach the $\sim 10$ Myr age range of the OB1a subassociation. However, even within the OB1b
region there are considerable differences; the OB1bI group has a much higher
fraction of CTTS ($\sim 22$\%) compared with OB1b II ($3.8$\%).

As disks evolve and dissipate,
less TTS are still actively accreting at a given age, and the fraction of CTTS
diminishes as an increasing fraction of the PMS population tends to be composed by non-accreting WTTS.
This well known trend \citep{hartmann2008}  can now be tested with improved statistics, across populations characterized in a uniform and consistent way.
In the right panel of Figure \ref{li_acc_evolution} we show the decrease of the accretor fraction with age, for all the populations and groups in our sample. As in previous sections, for this analysis we consider OB1b as a single population, in order to provide more robust statistics.
We have included the ONC as reference, taking the value of 70\% provided by \cite{hillenbrand98b}, with the caveat if this numbers not having been derived in the same way as for the rest of the regions and groups. Still, this value is agrees well with the exponential fit to the data, described by the dashed red line:

$$ \% Accretors = C_{acc} \times exp(-t/\tau_acc)$$
with $C_{acc}$= constant, and $T=$ age, and $C_{acc}$ normalized so that the percentage of accretors at and $t= 0$ is 100\%.

This fit gives an accretion e-folding timescale $\tau_{acc}= 2.1$ Myr, consistent with the findings of \cite{fedele2010}, but obtained in a single star forming region, with $\sim 2000$ stars measured in a consistent and uniform way.
As reference, we also plot the disk fractions for each region, derived using the $Ks-W2$ color, by matching the \textit{CVSO} sample with the ALLWISE catalog \citep{wright2010,mainzer2011}. Following \cite{luhman_mamajek2012} (who used $Ks-[4.5]$, from \textit{Spitzer} data), we derived the disc fraction as the number of stars falling above the upper envelope of the distribution of WTTS in a $Ks-W2$ - V-J color-color diagram, divided by the total number of TTS with a $Ks-W2$ measurement in each region. The disk fraction is systematically higher than the accretor fraction
A possibility is that
at every age 
there are systems in which accretion has stopped due to clearing of an inner gap in the disk, but enough dust remains to produce an IR excess. Alternatively, 
there may be a number of 
low accretors that have not yet been
found.
An exponential
fit to the decline in disk-bearing systems as a function of age results in a timescale of 3.2 Myr for the depletion of dust in the innermost parts of the disk. 
Our derived values for $\tau_{acc}$ and $\tau_{disk}$ are in very good agreement with previous studies looking at disk fractions using similar indicators \citep[e.g][]{fedele2010}, the main differences being that, first, we have derived this result for a large sample within the same star-forming region, thus we can expect the stars to have a common origin, and second, all the stars in each of the groups spanning the entire $\sim 3-10$ Myr age range have been selected, confirmed as members, and characterized, in the same way.  We defer further discussion on the ensemble properties of disks in Orion OB1 to a forthcoming paper, in which we carry out an extensive analysis of \textit{Spitzer} and \textit{WISE} infrared data for the \textit{CVSO} sample.

\subsubsection{Variability}
\label{sec:variability}

We now examine the variability among the various groups of stars in our sample.  In Table \ref{variability} we show the mean and median amplitudes of variation in each of the $V$, $R$ and $I_C$ bands, together with the mean and median standard deviation of the variability for each type of TTS: CTTS, C/W and WTTS, as well as for the active and non-active subsets of the field star sample described in \S \ref{sec:ctts_classif}.  
As can be seen in Figure \ref{variability_AmpV}, the amplitude of the mean $V$-band amplitude decreases from the more active CTTS, to the C/W class and finally to the more quiescent WTTS, which are still more variable than the population of active field stars. These active field stars are an important contaminant of photometric surveys like ours, even when folding in variability, because they are active K and M-type stars, significantly variable, so they pass all our selection criteria.  They differ from WTTS only in the lack of Li I $\lambda$ 6707 absorption feature, and their weaker Na I$\lambda 8183,8195$ absorption. In fact, the active group of field stars has a higher $\overline{\Delta (V)} = 0.29$ compared to the non-active group ($\overline{\Delta (V)}= 0.22$). A Welch Two Sample t-test finds that the amplitude of variability in these two groups is different at the 95\% confidence level. So the active group, defined on the basis of H$\alpha$ emission, also has higher photometric variability in the $V$-band, moreover, because these stars are mostly foreground to regions like Orion, they naturally appear above the main sequence in CMDs. These young main sequence dKe and dMe need to be accounted when mapping star-forming regions in future large scale surveys of the galactic disk.

\begin{deluxetable}{lrrrrr}[hbt!]
\tablewidth{0pt}
\tablecaption{Optical Variability in Orion OB1\label{variability}}
\tablehead{
\colhead{} &  \colhead{TTS} & \colhead{CTTS} & \colhead{C/W} & \colhead{WTTS} & \colhead{Field stars} \\
}
\startdata
No. Stars                 &  2064 &   219  &   150   &   1695 &  2418  \\
$\overline{\Delta (V)}$   & 0.403 &  0.722 &  0.474  &  0.356 &  0.240 \\
median $\Delta (V)$       & 0.302 &  0.500 &  0.378  &  0.290 &  0.191 \\
$\overline{\Delta (R)}$   & 0.28  &  0.590 &  0.349  &  0.242 &  0.154 \\
median $\Delta (R)$       & 0.215 &  0.381 &  0.281  &  0.202 &  0.113 \\
$\overline{\Delta (I_C)}$ & 0.244 &  0.419 &  0.261  &  0.225 &  0.160 \\
median $\Delta I_C$       & 0.190 &  0.273 &  0.218  &  0.183 &  0.107 \\
$\overline{\sigma (V)}$   & 0.091 &  0.179 &  0.118  &  0.078 &  0.049 \\
median $\sigma (V)$       & 0.067 &  0.136 &  0.089  &  0.064 &  0.036 \\
$\overline{\sigma (R)}$   & 0.064 &  0.145 &  0.081  &  0.054 &  0.032 \\
median $\sigma (R)$       & 0.048 &  0.095 &  0.064  &  0.045 &  0.025 \\
$\overline{\sigma (I_C)}$ & 0.045 &  0.091 &  0.049  &  0.040 &  0.027 \\
Median $\sigma (I_C)$     & 0.033 &  0.055 &  0.047  &  0.032 &  0.021 \\
\enddata
\end{deluxetable}
\label{tab:variability}

\begin{figure}[hbt!]
\includegraphics[scale=0.42]{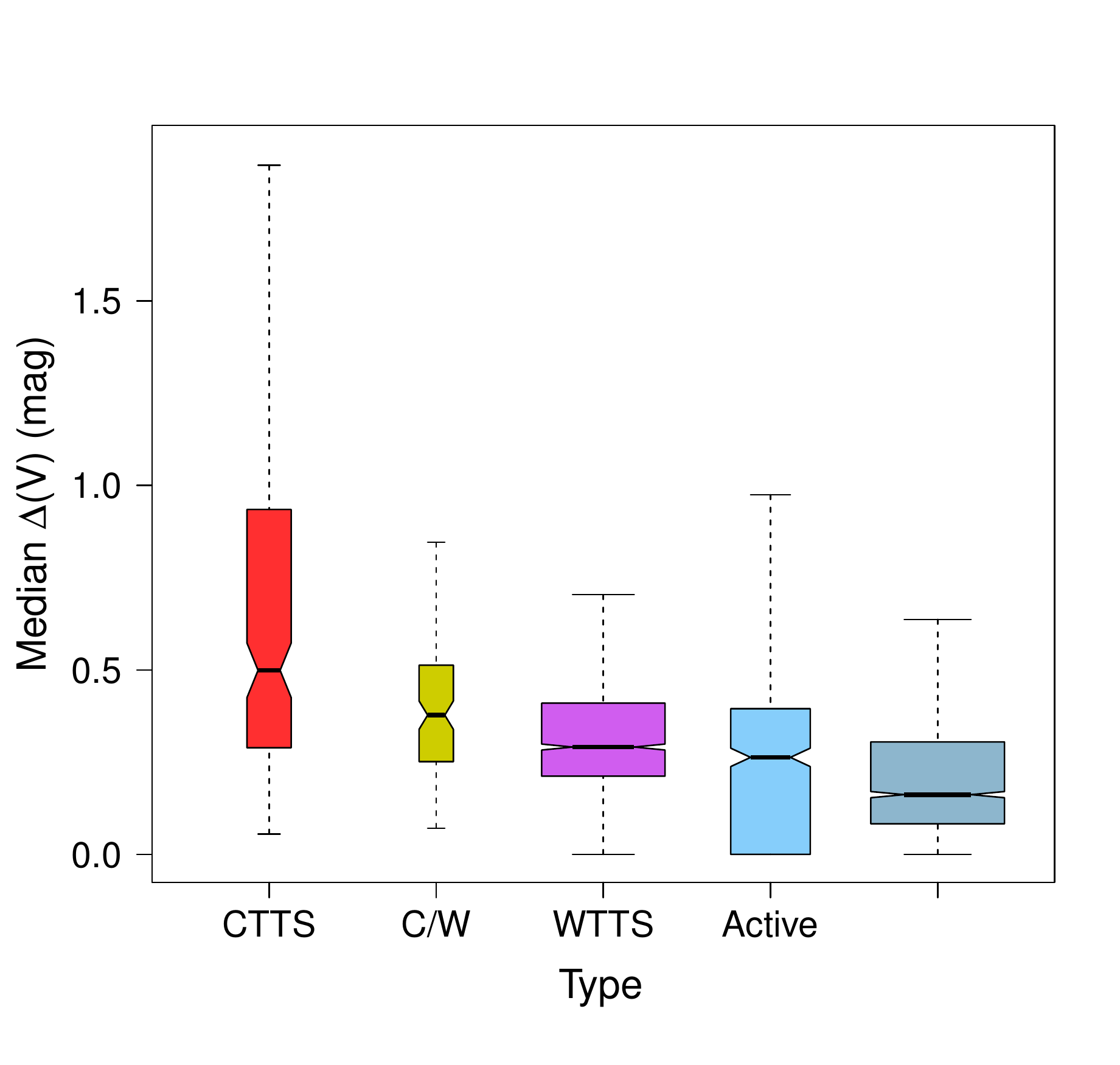} 
\caption{Median amplitude of variability in the V-band for each type of TTS and for field stars. As in Figure \ref{KW1_colors}, this notched box plot shows the first median values together with and third quartiles, and the width of each box is proportional to $\sqrt[]{N}$,
with $N=$ number of stars in each group.
}
\end{figure}
\label{variability_AmpV}

The trend in Figure \ref{variability_AmpV}, combined with that seen in Figure \ref{KW1_colors} provides support to the idea that the C/W class is indeed a transition stage between the more active CTTS and the WTTS, which in turn can be clearly distinguished as a group from their young main sequence counterparts.
Further details on the optical variability of the \textit{CVSO} TTS sample can be found in \cite{karim2016}, where we presented optical light curves for 1974 stars, and derived periods for
564 of them, the large majority WTTS.


\section{Summary and conclusions}
\label{sec:conclusions}

We have carried out an comprehensive synoptic optical photometric survey survey complemented by extensive spectroscopic follow up of the majority of 
the Orion OB1 association, with an emphasis on the off-cloud, 
little extincted and slightly older PMS populations.
We have combined photometric variability with the traditional color-magnitude selection, as in \cite{briceno01,briceno05}, and provide further evidence that the resulting candidate selection is sensitive to the young star population 
in the extended regions beyond the confines of the molecular clouds.

We show the importance of spectroscopic follow up that includes the Li I $\lambda 6707$ line, as the only unambiguous diagnostic of youth in K and M-type PMS stars. With Li I and H$\alpha$ in emission as our main youth diagnostics we confirm 2064 TTS members across Orion OB1, of which 1485 (72\%) are new identifications.
We also provide further evidence, with much larger samples than available before in any single star-forming region, of the usefulness of the Na I $\lambda 8183, 8195$ doublet as a secondary youth indicator, with the caveat that it has significant overlap with the distribution of field K and M type dwarfs.

We provide extensive characterization of the TTS population, 
including cross match with SIMBAD, spectral type for each star, equivalent widths of the H$\alpha \lambda 6563$, Li I $\lambda 6707$ and Na I$\lambda 8183+8195$ lines,
$V$, $R$, $I_C$ robust mean magnitudes, errors and number of observations in each band, amplitude of variability and the standard deviation of the variability of every star in each optical band, the Stetson \cite{stetson96} variability indices and the variability probablity estimates for each band, the 2MASS $J$, $H$ and $K_S$ magnitudes, and the $T_{eff}$, $A_V$, and group or region to which the star is assigned in Orion OB1.  We also compare our study with previous large scale work in Orion, both with the purely survey studies that produced candidate lists, and with spectroscopic
work ending with membership confirmation.

We show that the spatial distribution of the young stars across Orion OB1 is
far from uniform, but rather has a significant degree of substructure.
Beyond the well know clusters in the A and B molecular clouds, which we also
recover, there are a number of clusterings and stellar aggregates in both the
OB1b and the OB1a subassociations. We name the two major stellar overdensities in Orion OB1b, containing 42\% of the 556 stars identified within the OB1b region, as OB1b I with 93 confirmed members, and OB1b II, containing 138 confirmed TTS.  The major groups in OB1a are the 25 Ori cluster, with 223 confirmed members, and the new HR 1833 and HD 35762 groups, with 65 and 86 TTS members respectively. The rest of the 844 TTS widely spread throughout the 
OB1a subassociation is what we call the 1a ``field'' TTS population.
Of the on-cloud sample, 68 TTS are projected onto the B cloud, some of them part of the NGC 2068 and 2071 clusters. In the A cloud we identify 222 stars,
most forming a $\sim 2^\circ$ wide halo around the ONC, the northern part of which
corresponds to the NGC 1977 population, and the southern part to NGC 1980, though this group is truncated at $\delta \sim -6^\circ$ due to the southern limit of our survey. 
These stars likely contaminate many of the ONC censa,
and because they are expected to lie at essentially the same distance as the ONC \citep{kounkel2017b}, and since the proper motions are likely very
similar, and all close to zero, it will be difficult to estimate precisely
by how much they ``contaminate'' existing ONC samples.

We derive distances to the three stellar aggregates in OB1a, 25 Ori, HR 1833 and HD 35762, with three methods: using main sequence fitting of likely B and A-type members, from \textit{Gaia} DR1 parallaxes for these same early type stars, and finally, from \textit{Gaia} DR2 parallaxes for  the TTS members. We find that the different estimates agree at the $1-\sigma$ level, indicating a distance of $\sim 350-360$ pc. With the \textit{Gaia} DR2 data we also derive distances for the other regions. The $\sim 400$ pc distance to the A and B clouds agrees with recent determinations using radio interferometric techniques. We find indications of a bimodal distance distribution within the OB1b region, with ``near'' ($\sim 350$) and ``far'' ($\sim 420$ pc) components, which we suggest may be consequence of contamination by the field population of the OB1a subassociation.

With the distances we use our photometry to place the stars in each group in CMDs, and derive ages by interpolating in \cite{baraffe98} and \cite{siess00} model isochrones. We obtain an age of $\sim 2$ for the B Cloud stars, and $\sim 3$ Myr for the A Cloud stars, this later value consistent with the determinations of \cite{fang2017} and \cite{kounkel2017b}.
The groups in OB1b and OB1a can be ordered in an age sequence that goes from the older OB1a region ($\sim 10$ Myr), containing 25 Ori, HD 35762 and HR 1833, of which the later is the oldest, to the intermediate aged OB1b ($\sim 5$ Myr), to the youngest stars in the A and B molecular clouds.  
This age sequence agrees with
the long standing idea that star formation in Orion started in the OB1a
subassociation $\sim 10-15$ Myr ago, continuing with OB1b, and finalizing 
with the molecular clouds, where the youngest objects and protostars are found today.

We introduce a new class of low-mass PMS stars, the C/W type, which we propose are TTS going through an intermediate evolutionary stage between that of actively accreting CTTS and WTTS, which have stopped accreting at levels detectable in low-resolution spectra. This new classification is based on the  variability of the H$\alpha$ emission line strength observed in multi-epoch low resolution spectra of a subset of our Orion TTS sample. The average disk emission and amplitude of variability of the C/W class fall right between that of the CTTS and WTTS, giving support to the idea that these are TTS at the last stages of their accretion history.

Equipped with an unprecedented large and richly characterized sample of young stars across one of the closest active, massive star forming regions, 
we  carry out a study of the ensemble-wide properties of each stellar group, the demographics of stellar populations in Orion OB1. This has allowed us to observe, for the first time in a single star-forming region, the depletion of Li during the initial $\sim 10 $ Myr of the PMS phase in K and M dwarfs.  By looking at the evolution of the mean W(Li I) for each population, we detect the steady decline of the ensemble-wide strength of the Li I $\lambda 6707$ line. We estimate a timescale of $\sim 8.5$ Myr for this depletion, roughly
in agreement with predictions form theoretical models and with similar work
in other nearby star-forming regions \cite{jeffries2014a}.
We also derive the accretor fractions for each region and fit the decline with an exponential function, which yields a characteristic timescale for the decline of accretion of $2.1$ Myr, consistent with previous findings. Finally, our data provide the opportunity to look at the optical variability properties of each group of stars. The stellar activity gauged by the mean amplitude of the $V$-band variability in each type of TTS, provides new, robust evidence for the known trend of decreasing activity from CTTS, through the C/W class, to the WTTS, through active, likely young main sequence stars, and ending with the levels seen in main sequence, non-active K and M dwarfs.

\vskip 0.1in

\noindent
{\bf Acknowledgments.}
This paper was prepared in part during the sabbatical stay of C. Brice\~no
and A. K. Vivas at the Astronomy Department of the University of Michigan;
we acknowledge the support of the Department and the lively academic atmosphere 
created by 
colleagues and graduated students, 
that was largely conducive to completing this work.
This publication makes use of data products 
from the Two Micron All Sky Survey, which is a joint project 
of the University of Massachusetts and the Infrared 
Processing and Analysis Center/California Institute of 
Technology, funded by the National Aeronautics and Space 
Administration and the National Science Foundation.
This research has made use of the NASA/ IPAC Infrared Science Archive, 
which is operated by the Jet Propulsion Laboratory, California 
Institute of Technology, under contract with the National 
Aeronautics and Space Administration.
This paper uses data products produced by
the OIR Telescope Data Center, supported by the Smithsonian Astrophysical Observatory.
We are grateful to Susan Tokarz at CfA, who is in charge of
the reduction and processing of FAST spectra.
We thank the invaluable assistance of the observers and telescope
operators at the Venezuela Schmidt telescope that made 
possible obtaining the photometric data over these past years,
and of the Telescope Operators and staff at the MMT, WIYN, 
Magellan and SOAR telescopes, without whom we would not have been
able to carry out our extensive spectroscopic follow up work.
We also acknowledge the support from the
CIDA technical staff, and in particular of Gerardo S\'anchez.

\software{IRAF, TOPCAT \citep{taylor2005} }

{\it Facilities:} \facility{Magellan:Clay (M2FS)}, \facility{MMT (Hectospec, Hectochelle)}, \facility{FLWO:1.5m (FAST)}, \facility{CIDA:Stock 1m (CCD Mosaic Camera)}, \facility{SOAR (Goodman spectrograph)}

\bibliography{apj-jour,MyLibrary}

\end{document}

%% file: tabobs_small.tex
\begin{deluxetable*}{crrcccccc}
\tabletypesize{\scriptsize}
\tablewidth{0pt}
\tablecaption{CVSO Observing Log \label{tabobs}}
\tablehead{
\colhead{Year} & \colhead{Month} & \colhead{Day} & \colhead{Obs No.} & \colhead{$\rm DEC_c$(J2000)} &
\colhead{$\rm RA_i$(J2000)} & \colhead{$\rm RA_f$(J2000)} & \colhead{Filters} & \colhead{Hour Angle} \\
\colhead{} & \colhead{} & \colhead{} & \colhead{} & \colhead{(deg)} & \colhead{(deg)} & \colhead{(deg)} &  \colhead{} &  \colhead{(hh:mm:ss)} \\
}
\startdata
  1999  &   12  &   28  &   503  &   +1.0  &   67.5     &  97.5    &   RHIV  & 01:54:20 E \\
  1999  &   12  &   28  &   504  &   +1.0  &   67.5     &  97.5    &   RHIV  & 00:22:26 W \\
  1999  &   12  &   28  &   505  &   +1.0  &   67.5     &  97.5    &   RHIV  & 02:52:39 W \\
  1999  &   12  &   29  &   506  &   +1.0  &   67.5     &  97.5    &   RHIV  & 00:59:53 W \\
  1999  &   12  &   30  &   507  &   +1.0  &   67.5     &  97.5    &   RHIV  & 01:41:07 E \\
  1999  &   12  &   30  &   508  &   +1.0  &   67.5     &  97.5    &   RHIV  & 00:44:12 W \\
  1999  &   12  &   30  &   509  &   +1.0  &   67.5     &  97.5    &   RHIV  & 03:06:26 W \\
  2002  &   10  &   6   &   510  &   +1.0  &   72.5     &  92.98   &   VRIV  & 00:56:00 E \\
  2002  &   10  &   13  &   504  &   +1.0  &   72.5     &  92.5    &   VRIV  & 00:38:00 E \\
  2002  &   11  &   1   &   503  &   +1.0  &   72.5     &  93.15   &   VIRV  & 00:28:18 W \\
\enddata
\tablecomments{Table \ref{tabobs} is published in its entirety in the electronic version
of the Astronomical Journal. A portion is shown here for guidance regarding
its form and contents.}
\end{deluxetable*}

%% file: speclog_short.tex
\begin {deluxetable*}{cllllccl}
\tabletypesize{\scriptsize}
\tablewidth{0pt}
\tablecaption{WIYN-Hydra and MMT-Hectospec Observing Log \label{speclog}}
\tablehead{
\colhead{UT Date} & \colhead{Instrument} & \colhead{Field ID} & \colhead{$\rm RA_c$(J2000)} & \colhead{$\rm DEC_c$(J2000)} & \colhead{$\rm  T_{exp}$} & \colhead{No. objects} & \colhead{Notes} \\
\colhead{(Year Month Day)} & \colhead{} & \colhead{}  & \colhead{(hh:mm:ss.ss)}  & \colhead{(dd:dm:ds.s)} & \colhead{(s)} &  \colhead{} & \colhead{}
}
\startdata
  2000-Nov-26 & Hydra     & W01 & 05:14:00.00 & -00:25:03.0 &    1800 &  57 &                       \\
  2000-Nov-26 & Hydra     & W02 & 05:28:33.10 & -00:39:00.0 &    1800 &  78 &  Field-1a in B05      \\
  2004-Nov-03 & Hectospec & H01 & 05:29:23.28 & -01:25:15.3 &   900x3 & 235 &  decm160\_7 in D08    \\
  2005-Mar-14 & Hectospec & H05 & 05:25:08.04 & +00:20:47.0 &   900x3 & 273 &                       \\                       
  2010-Feb-10 & Hectospec & H30 & 05:39:38.67 & +01:47:54.9 &    3600 & 252 &                       \\
  2013-Nov-30 & M2FS      & M9  & 05:32:09.94 & -02:49:46.8 & 3x600   & 112 &                       \\
  2015-Mar-05 & M2FS      & M1  & 05:21:08.02 & +01:02:09.1 & 3x600+180& 63 &                       \\
\enddata                    
\tablecomments{Table \ref{tabobs} is published in its entirety in the electronic version
of the Astronomical Journal. A portion is shown here for guidance regarding
its form and contents.}
\end{deluxetable*}